\documentclass[reqno,twoside,12pt]{amsart}

\usepackage{amsxtra}            
\usepackage{epsfig}
\usepackage{cite}              
\usepackage{psfrag}
\usepackage{pst-tree}
\usepackage{url}

\newtheorem{result}{Theorem}    

\newtheorem{theorem}{Theorem}[section]
\newtheorem{proposition}[theorem]{Proposition}
\newtheorem{lemma}[theorem]{Lemma}
\newtheorem{corollary}[theorem]{Corollary}

\theoremstyle{definition}
\newtheorem{definition}[theorem]{Definition}
\newtheorem{example}[theorem]{Example}

\theoremstyle{remark}
\newtheorem{remark}[theorem]{Remark}

\numberwithin{equation}{section}

\newcommand{\abs}[1]{\lvert#1\rvert}
\newcommand{\norm}[1]{\lVert#1\rVert}
\newcommand{\normw}[2]{\lVert#1{\rVert}_{#2}}
\newcommand{\supnorm}[1]{\lVert#1{\rVert}_\infty}

\newcommand{\multinorm}[1]{\Vert#1{\rVert}_{\mathcal{L}(^n(\C^{d+1});\C)}}
\newcommand{\piste}{\,\cdot\,}

\newcommand{\average}[1]{\langle#1\rangle}

\newcommand{\eg}{\textit{e.g.}}
\newcommand{\ie}{\textit{i.e.}}
\newcommand{\etc}{\textit{etc.\@}}

\newcommand{\Z}{\mathbb{Z}}
\newcommand{\N}{\mathbb{N}}
\newcommand{\R}{\mathbb{R}}
\newcommand{\C}{\mathbb{C}}

\newcommand{\es}{\mathbb{S}}
\newcommand{\Ham}{\mathcal{H}}

\newcommand{\bigunit}{[-1,1]}
\newcommand{\torus}{{\mathbb{T}^d}}
\newcommand{\T}{\mathcal{T}}
\newcommand{\W}{\mathcal{W}}
\newcommand{\A}{\mathcal{A}}
\newcommand{\B}{\mathcal{B}}
\newcommand{\K}{\mathcal{K}}

\newcommand{\Bphi}[1]{\mathcal{B}^{\Phi}_{#1}}
\newcommand{\Bpsi}[1]{\mathcal{B}^{\Psi}_{#1}}
\newcommand{\el}{\mathcal{L}}
\newcommand{\D}{\mathcal{D}}
\newcommand{\order}[1]{\mathcal{O}(#1)}
\newcommand{\q}{\mathbf{q}}
\newcommand{\qdomain}{{(\Z^d)}^n}
\newcommand{\Phiball}[1]{\bar B^{\Phi}_{#1,r}}
\newcommand{\Psiball}[1]{\bar B^{\Psi}_{#1,r}}

\newcommand{\inv}{^{-1}}

\newcommand{\nonzero}{\setminus\{0\}}
\newcommand{\normalize}{\frac{1}{(2\pi)^d}}
\newcommand{\de}{\partial}

\newcommand{\half}{\tfrac{1}{2}}
\newcommand{\quarter}{\tfrac{1}{4}}

\newcommand{\unpert}{{\rvert}_{\lambda=0}}
\newcommand{\unperte}{{\rvert}_{\epsilon=0}}

\newcommand{\repart}{{\operatorname{\Re\mathfrak{e}}\,}}
\newcommand{\impart}{{\operatorname{\Im\mathfrak{m}}\,}}

\newcommand{\defas}{\mathrel{\raise.095ex\hbox{$:$}\mkern-4.2mu=}}
\newcommand{\defasr}{\mathrel{=\mkern-4.2mu\raise.095ex\hbox{$:$}}}

\newcommand{\mathand}{\quad\text{and}\quad}
\newcommand{\mathwith}{\quad\text{with}\quad}

\newcommand{\beq}{\begin{equation}}
\newcommand{\eeq}{\end{equation}}
\newcommand{\beqn}{\begin{equation*}}
\newcommand{\eeqn}{\end{equation*}}

\DeclareMathAlphabet{\mathfat}{U}{bbold}{m}{n}          
\newcommand{\one}{\mathfat{1}}

\def\XXint#1#2#3{{\setbox0=\hbox{$#1{#2#3}{\int}$}
     \vcenter{\hbox{$#2#3$}}\kern-.5\wd0}}

\begin{document}

\title{Construction of Whiskers for the Quasiperiodically Forced Pendulum}
\author{Mikko Stenlund}
\email{mikko.stenlund@helsinki.fi}
\address{Department of Mathematics and Statistics, University of Helsinki, P.O. Box 68, Fi-00014 University of Helsinki, Finland}


\subjclass[2000]{70K43; Secondary 37J40, 37D10, 70H08, 70K44}


\begin{abstract}
We study a Hamiltonian describing a pendulum coupled with several anisochronous oscillators, giving a simple construction of unstable KAM tori and their stable and unstable manifolds for analytic perturbations.

We extend analytically the solutions of the equations of motion, order by order in the perturbation parameter, to a uniform neighbourhood of the time axis. 
\end{abstract}

\maketitle


\section{Main Concepts and Results}
\subsection{Background and history}
A quasiperiodic motion of a mechanical system is composed of incommensurable periodic motions; the trajectory in phase space winds around on a torus filling its surface densely.
An integrable Hamiltonian system has a great profusion of quasiperiodic motions: if one picks an initial phase point according to a uniform distribution, the trajectory will be quasiperiodic with probability one. The remaining trajectories are periodic.

KAM theory deals with the stability of quasiperiodic motions, or persistence of invariant tori, under small perturbations. Poincar\'e \cite{Poincare1} called this the general problem of dynamics.

In 1954, Kolmogorov \cite{Kolmogorov} outlined a result, made rigorous by Arnold in 1963 \cite{Arnold2}, that quasiperiodic motions are typical also for nearly integrable analytic Hamiltonians under suitable nondegeneracy conditions. Thus, only a small fraction of the tori would be destroyed by the perturbation. Moser managed to prove the same for twist maps \cite{Moser} in 1962, and later for Hamiltonians \cite{MoserRapid1, MoserRapid2}, in the smooth (non-analytic) setting (see also \cite{Moser67}). 

The difficult problem to overcome is the following. Suppose that the Hamiltonian reads $\Ham=\Ham_0+\lambda \Ham_1$, where $\Ham_0$ is integrable and $\lambda$ is considered small. Then one can formally represent a solution to the equations of motion by a power series in $\lambda$, known as the Lindstedt series in this context, conditioned to agree for $\lambda=0$ with a quasiperiodic solution obtained in the integrable case. When one computes the coefficients of the Lindstedt series, however, one encounters expressions containing arbitrarily small denominators. The latter seem to imply that the $k$th coefficient grows like $k!^\alpha$ with a large power $\alpha$. Thus, there is little hope of being able to sum the series and obtain a true solution, unless a miracle occurs.

The proofs mentioned above relied on a rapidly convergent Newton-type iteration scheme, which is interesting in its own right, and yields solutions analytic in $\lambda$. On the other hand, one is then left to wonder why the Lindstedt series \emph{does} converge.

In 1988, an answer was provided by Eliasson \cite{Eliasson}, who managed to identify enormous cancellations among the small denominator contributions and to sum the Lindstedt series ``manually". Gallavotti \cite{GallavottiTwistlessTori,GallavottiTwistless} interpreted the cancellations in a Renormalization Group (RG) framework.  For a review and some extensions, see Gentile and Mastropietro \cite{GentileMastropietro}. The importance of these achievements has to be stressed: they prove the existence of quasiperiodic solutions in an essentially constructive way. 

Motivated by the RG approach of Gallavotti, in the 1999 paper \cite{Kupiainen} Bricmont, ~Gaw{\c{e}}dzki, and Kupiainen identified the cancellations as a consequence of Ward identities (corresponding to a translation invariance of an action functional) in a suitable field theory.

Returning to much earlier works, Moser \cite{Moser67} and Graff \cite{Graff} showed that also hyperbolic tori---tori having local stable and unstable manifolds---would typically persist under small perturbations. In another landmark paper \cite{ArnoldDiffusion}, Arnold had described a mechanism how a chain of such ``whiskered" tori could provide a way of escape for special trajectories, resulting in instability in the system. (As discussed above, a trajectory would typically lie on a torus and therefore stay eternally within a bounded region in phase space.) The latter is often called Arnold mechanism and the general idea of instability goes by the name Arnold diffusion. It is conjectured in \cite{ArnoldAvez} that Arnold diffusion due to Arnold mechanism is present quite generically, among others in the three body problem.

Arnold mechanism is based on Poincar\'e's concept of biasymptotic solutions, discussed in the last chapter of \cite{Poincare3}, that are formed at intersections of whiskers of tori. Following such intersections a trajectory can ``diffuse" in a finite time from a neighbourhood of one torus to a neighbourhood of another, and so on. 

Chirikov's work \cite{Chirikov} is a very nice physical account on Arnold diffusion. Lochak's compendium \cite{LochakCompendium} discusses more recent developments in a readable fashion and is a good point to start learning about diffusion.

The proofs of Moser and Graff mentioned above use the rapidly convergent method of Kolmogorov, but there now exist also constructive proofs in the spirit of Eliasson and Gallavotti. We refer here to Gallavotti \cite{GallavottiTwistless} and Gentile \cite{GentileQuasiflat,GentileExponent}.

\subsection{The model}
We consider the Hamiltonian
\beq
\Ham(I,\phi,A,\psi)=\half I^2+g^2\cos\phi+\half A^2- \lambda f(\phi,\psi)
\label{eq:H}
\eeq
of a pendulum coupled to $d$ rotators, with $\phi\in\es^1\defas\R/2\pi\Z$ and  $I\in\R$ the coordinate and momentum of the pendulum, and $\psi\in\torus\defas(\es^1)^d$ and $A\in\R^d$ the angles and actions of the rotators, respectively. The perturbation $f$ is assumed to be real-valued and real-analytic in its arguments, and $\lambda$ is a (small) real number, whereas the gravitational coupling constant $g$ is taken to be positive. This Hamiltonian is sometimes called \emph{the generalized Arnold model} or the \emph{Thirring model}. It is \emph{the} prototype of a nearly integrable Hamiltonian system close to a simple resonance, as is explained in the introduction of \cite{GentileExponent}. A review of applications can be found in \cite{Chirikov}.

\begin{figure}[!ht]
\epsfig{file=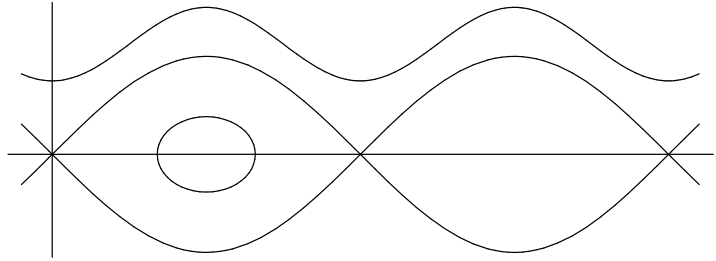,width=0.6\linewidth}
\caption{A $(\phi,I)$ plot showing the unperturbed pendulum separatrix that intersects the $\phi$ axis at integer multiples of $2\pi$---the upright position of the pendulum.}
\label{fig:separatrix}
\end{figure}

The equations of motion are
\beq
\dot\phi=I, \quad \dot\psi=A, \quad 
\dot I=g^2\sin\phi+\lambda\,\de_\phi f,\quad
\dot A=\lambda\,\de_\psi f.
\label{eqm}
\eeq
For the parameter value $\lambda=0$, which is addressed as the unperturbed case, the pendulum and the rotators decouple. The former then has the separatrix flow $\phi:\R\to \es^1$ given by
\beqn
\phi(t)=\Phi^0(e^{gt}),
\eeqn
where
\beqn
\Phi^0(z)=4\arctan z.
\eeqn
By elementary trigonometry, this function possesses the symmetry property
\beq\label{eq:arctan-sym}
\Phi^0(z)=2\pi-\Phi^0(z\inv).
\eeq
It is also odd,
\beqn
\Phi^0(-z)=-\Phi^0(z).
\eeqn
The phase space of the unperturbed pendulum looks as in Figure~\ref{fig:separatrix}, where the separatrix, given by $\Phi^0$, separates closed trajectories (libration) from open ones (rotation).

On the other hand, $\psi:\R\to\torus$ is quasiperiodic:
\beqn
\psi(t)=\psi(0)+\omega t\pmod{2\pi},
\eeqn
such that the vector
\beqn
\omega \defas A(0) \equiv A(t)
\eeqn
satisfies the Diophantine condition
\beq
|\omega\cdot q| > a\,|q|^{-\nu}\quad{\rm for}\quad q\in \Z^d,\;q\neq 0,
\label{Dio}
\eeq
with $a$ and $\nu$ positive.
Thus, at the instability point of the pendulum, the flow possesses the invariant tori
\beqn
\T_0\defas\Bigl\{(\phi,\psi,I,A)=(0,\theta, 0, \omega)\;\Big|\; \theta\in T^d\Bigr\}
\eeqn
indexed by $\omega$, with stable and unstable manifolds ($\W^s_0$ and $\W^u_0$, respectively) coinciding:
\beq
\W^{s,u}_0=\Bigl\{(\phi,\psi,I,A)=\left(\Phi^0(z), \theta, gz\partial_z\Phi^0(z),\omega\right)\;\Big|\; z\in [-\infty,\infty],\;\theta\in T^d\Bigr\}.
\label{unpert}
\eeq
\begin{remark}\label{rem:timescale}
The constant $g$ is the Lyapunov exponent for the unstable fixed point of the pendulum motion; in the limit $s\to-\infty$ two nearby initial angles $\phi(s)$ and $\phi(s+\delta s)$ separate at the exponential rate $e^{gs}$. As $\phi(t)=\Phi^0(e^{t/g^{-1}})$, the Lyapunov exponent fixes a natural time scale of $g^{-1}$ units, characteristic of the pendulum motion in the unperturbed Hamiltonian system \eqref{eq:H}.
\end{remark}

When the perturbation is switched on ($\lambda\neq 0$), we show that some of the invariant tori survive and have stable and unstable manifolds---or ``whiskers'' as Arnold has called them---that may not coincide anymore.

\subsection{Main theorems}
Our approach will be to construct the perturbed manifolds in a form similar to \eqref{unpert} as graphs of analytic functions over a piece of $[-\infty,\infty]\times\torus$. To see how this can be achieved, note that the unperturbed stable and unstable manifolds, $\W^s_0$ and $\W^u_0$, consist of trajectories
\beqn
(\phi(t),\psi(t))=(\Phi^0(e^{gt}),\omega t)
\eeqn
that at time $\pm\infty$ become quasiperiodic, as they wrap tighter and tighter around the invariant torus $\T_0$; indeed $(\phi(t),\psi(t))\sim(0,\omega t)$ in the limit $t\to\pm\infty$.

\begin{figure}[!ht]
\epsfig{file=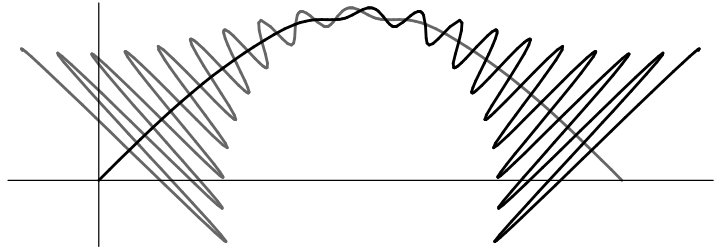,width=0.6\linewidth}
\caption{A schematic $I$-versus-$\phi$ plot, on a section of constant $\psi$ ($d=1$). The stable and unstable manifolds are expected to split, as opposed to coincide. The origin has been shifted for convenience.}
\label{fig:separatrices}
\end{figure}

Analogously, we will find the stable and unstable manifolds of the perturbed tori by looking for solutions of the form
\beq
(\phi(t),\psi(t))=(\Phi(e^{\gamma t},\omega t),\omega t+\Psi(e^{\gamma t},\omega t))=(0,\omega t)+(\Phi,\Psi)(e^{\gamma t},\omega t)
\label{trial}
\eeq
with quasiperiodic behavior in \emph{one} of the two limits $t\to \pm\infty$. Note especially that we anticipate the Lyapunov exponent $\gamma>0$ to depend on $\lambda$, with $\gamma\unpert=g$.
\begin{remark}
One should not assume asymptotic quasiperiodicity in both of the limits $t\to\pm\infty$, as the unstable and stable manifolds, which we denote $\W^u_\lambda$ and $\W^s_\lambda$, are generically expected to depart for nonzero values of the perturbation parameter $\lambda$. Therefore, either the past \emph{or} future asymptotic of a trajectory will evolve so as to ultimately reach the (deformed) invariant torus $\T_\lambda$. The separatrix in Figure~\ref{fig:separatrix} is thus transformed into something like the pair of curves in Figure~\ref{fig:separatrices}.
\end{remark}

Let us denote the total derivative $d/dt$ by $\de_t$ and the complete angular gradient $(\de_\phi,\de_\psi)$ by $\de$ for short. Substituting \eqref{trial} into the equations of motion
\beqn
\de_t^2(\phi,\psi)=(\dot I,\dot A)=(g^2\sin\phi,0)+\lambda\,\de f(\phi,\psi),
\eeqn
we get for $X\defas(\Phi,\Psi)$ the equation
\beqn
(\omega\cdot\de_\theta+\gamma e^{\gamma t}\de_z)^2 X(e^{\gamma t},\omega t)=[(g^2\sin\Phi,0)+\lambda\,\de f(X+(0,\theta))](e^{\gamma t},\omega t),
\eeqn
where $\theta$ stands for the canonical projection $[-\infty,\infty]\times\torus\to\torus$. 

Notice that the partial differential operator
\beqn
\el\defas\omega\cdot\de_\theta+\gamma z\de_z
\label{el}
\eeqn
satisfies the characteristic identity
\beq\label{Lid}
\el F(ze^{\gamma t},\theta+\omega t)
=\de_t F(ze^{\gamma t},\theta+\omega t)
\eeq
for a differentiable map $(z,\theta)\mapsto F(z,\theta)$. Equation \eqref{Lid} simply reflects the time derivative nature of $\el$. In fact, if $T$ is the ``time-reversal map''
\beq\label{eq:time-reversal-map}
T(z,\theta)\equiv (z\inv,-\theta),
\eeq
then, by the chain rule,
\beq\label{eq:time-reversal}
\el (F\circ T)=- (\el F)\circ T.
\eeq

Let us abbreviate
\beq\label{eq:Omega}
\Omega(X) \defas (g^2\sin\Phi,0)+\lambda\,\widetilde\Omega(X)\quad\text{with}\quad
\widetilde\Omega(X)\defas\de f(X+(0,\theta)).
\eeq
As a consequence, we have reduced the equations of motion to the PDE 
\beq\label{xeq}
\el^2 X=\Omega(X)
\eeq
for the map $(z,\theta)\mapsto X(z,\theta)$ in a suitable Banach space of analytic functions, albeit its restriction to the set (``characteristic'')
\beq\label{eq:z,theta}
\left\{(z,\theta)=(e^{\gamma t},\omega t)\;\big|\;t\in\R\right\}
\eeq 
is what one is physically interested in. Our preference of working directly with the invariant manifolds, as opposed to individual trajectories traversing along them, motivates us encoding the time derivative in the operator $\el$. Nevertheless, it will be harmless---and indeed quite informative---for the reader to keep in mind that the objects we deal with originate from \eqref{eq:z,theta} and therefore have a direct physical interpretation.  

The action variables trivially follow from the knowledge of $X(z,\theta)$:
\beqn
(I(t),A(t))=(0,\omega)+Y(e^{\gamma t},\omega t),\quad Y\defas\el X.
\eeqn
The solutions $X$ will provide a parametrization of the deformed tori and their stable and unstable manifolds. 
As hinted below \eqref{trial}, we find two kinds of solutions, $X^u(z,\theta)$ defined for $z\in [-z_0,z_0]\defasr \mathbb{I}^u$ and $X^s(z,\theta)$ defined for $z\in [-\infty,-z_0\inv]\cup [z_0^{-1},\infty]\defasr \mathbb{I}^s$. Here, $z_0>1$. The tori will have the three parametrizations
{\Small
\begin{eqnarray*}
\T_\lambda &= &\left\{(\phi,\psi,I,A)=\left((0,\theta)+X^u(0,\theta), (0,\omega)+Y^u(0,\theta)\right) \;\Big|\; \theta\in T^d\right\}\\
&= &\left\{(\phi,\psi,I,A)=\left((0,\theta)+X^s(\pm\infty,\theta), (0,\omega)+Y^s(\pm\infty,\theta)\right) \;\Big|\; \theta\in T^d\right\},
\end{eqnarray*}
}%
whereas the parametrizations of their stable and unstable manifolds then read
{\Small
\beqn
\W^{s,u}_\lambda=\left\{(\phi,\psi,I,A)=\left((0,\theta)+X^{s,u}(z,\theta), (0,\omega)+Y^{s,u}(z,\theta)\right) \;\Big|\; z\in \mathbb{I}^{s,u},\;\theta\in T^d\right\}.
\label{pert}
\eeqn
}%

In order to enable solving \eqref{xeq}, we need to deal with quantities of the form $(\omega\cdot q)^{-1}$, $q\in\Z^d\nonzero$, stemming from the Fourier representation of the operator $\el$. Here the Diophantine property of the vector $\omega\in\R^d$ stated in \eqref{Dio} steps in. Since $\omega\equiv A\unpert=\dot\psi\unpert$, by rescaling time (and the actions, correspondingly) in the equations of motion \eqref{eqm}, the constant $a$ can be absorbed into $g^2$ and $\lambda$ in the equations of motion, leaving the ratio $\lambda g^{-2}$ unchanged: $(g,\lambda)\mapsto(g/a,\lambda/a^2)$ \footnote{This scaling is responsible for the usual requirement $\lambda=\order{a^2}$ for KAM tori.}. Thus, we may as well take $a$ to be 1 below, transforming the condition on $\omega$ into
\beq\label{eq:Dio2}
\abs{\omega\cdot q}>|q|^{-\nu}\quad{\rm for}\quad q\in \Z^d\nonzero.
\eeq 
We will moreover consider $\lambda$ small in a $g$-dependent fashion, taking
\beq
\epsilon\defas\lambda g^{-2}
\label{eq:epsilon}
\eeq
small. This should be seen as an outreach towards the experimenter, albeit there is a technical wherefore: such a choice is needed for studying the limit $g\to \infty$, which corresponds to rapid forcing; see Remark~\ref{rem:timescale}. The domain we restrict ourselves to is
\beq\label{eq:D}
D\defas\left\{(\epsilon,g)\in \C\times\R\;\big|\;\abs{\epsilon}<\epsilon_0,\; 0<g<g_0\right\},
\eeq
for some positive values of $\epsilon_0$ and $g_0$.

Finally, note that if $X=(\Phi,\Psi)$ solves \eqref{xeq} on some domain $D'\subset [-\infty,\infty]\times\torus$, then so does
\beq
X_{\alpha,\beta}(z,\theta)\defas X(\alpha z,\theta+\beta)+(0,\beta),
\label{xtr}
\eeq
as long as $(\alpha z,\theta+\beta\pmod{2\pi})\in D'$. The aforementioned invariance is a manifestation of the freedom of choosing initial conditions for $(\phi,\psi)$---we may choose the origin of time and the configuration of the physical system there. 

For $\epsilon=0$, the solutions are obtained from
\beq\label{eq:X^0}
X^0(z,\theta)\defas(\Phi^0(z),0)
\eeq
using \eqref{xtr}. In particular, $X^0(1,0)=(\pi,0)$. This will provide us with a natural way of fixing $\alpha$ and $\beta$ below.

We are now ready to state the first of the two main theorems of this article. It is a version of a classical result, and by no means new; earlier treatments include \emph{for instance} \cite{Melnikov63,Moser67, Graff,EliassonBiasymptotic,GallavottiTwistless,GentileQuasiflat,GentileExponent}. However, the interest here lies in the new techniques used in the proof.
\begin{result}[Tori and their whiskers]\label{thm:manifolds}
Let $f$ be real-analytic and even, \ie,
\beqn
f(\phi,\psi)=f(-\phi,-\psi).
\eeqn 
Also, suppose $\omega$ satisfies the Diophantine condition \eqref{eq:Dio2}, and fix $g_0>0$. Then there exist a positive number $\epsilon_0$ and a function $\gamma(\epsilon,g)$ on $D$, analytic in $\epsilon$ with $\abs{\gamma-g}<Cg\abs{\epsilon}$, such that equation~\eqref{xeq} has a solution $X^u$ which is analytic in $\epsilon$ as well as in $(z,\theta)$ in a neighbourhood of $\bigunit\times\torus$ and which satisfies
\beq
X^u(1,0)=(\pi,0),\quad X^u(z,\theta)=X^0(z)+\order{\epsilon}.
\label{norm}
\eeq
Corresponding to the same $\gamma$, there exists a solution $X^s(z,\theta)=X^0(z)+\order{\epsilon}$ which is an analytic function of $(z^{-1},-\theta)$ in a neighbourhood of $\bigunit\times\torus$. The maps
\beq\label{eq:param}
W^{s,u}(z,\theta)=(X^{s,u},Y^{s,u})(z,\theta)+((0,\theta),(0,\omega)),\quad Y^{s,u}\defas\el X^{s,u},
\eeq
provide analytic parametrizations of the stable and unstable manifolds $\W^{s,u}_\lambda$ of the torus $\mathcal{T}_\lambda$.
\end{result}

\begin{remark}\label{rem:manifolds}
The number $\epsilon_0$ above depends on the Diophantine exponent $\nu$ and on $f$. The perturbation $(\phi,\psi)\mapsto f(\phi,\psi)$ is analytic on the compact set $\es^1\times\torus$. By Abel's Lemma (multivariate power series converge on polydisks), it extends to an analytic map on a ``strip'' $\abs{\impart\phi},\abs{\impart\psi}\leq \eta$ ($\eta>0$) around $\es^1\times\torus$. By Theorem~\ref{thm:manifolds}, there exists some $0<\sigma<\eta$ such that each $\theta\mapsto X^{s,u}(\piste,\theta)$ is analytic on $\abs{\impart\theta}\leq \sigma$. 
\end{remark}

An important part of Theorem~\ref{thm:manifolds} is that the domains of $X^u$ and $X^s$ overlap. Namely, if $(z,\theta)\mapsto X(z,\theta)$ solves equation~\eqref{xeq}, then so does $(z,\theta)\mapsto(2\pi,0)-(X\circ T)(z,\theta)$. This is due to \eqref{eq:time-reversal} and the parity of $f$. Consequently, by a simple time-reversal consideration (set $t\mapsto -t$ in \eqref{eq:z,theta}), the stable and unstable manifolds are related through
\beq\label{Xsym}
X^s=(2\pi,0)-X^u\circ T.
\eeq
In particular, as $T(1,0)=(1,0)$,
\beqn
X^s(1,0)=X^u(1,0).
\eeqn
Moreover, the actions $Y^{s,u}=\el X^{s,u}$ satisfy
\beq\label{Ysym}
Y^s=Y^u\circ T,
\eeq
yielding
\beqn
Y^s(1,0)=Y^u(1,0).
\eeqn
In other words, a \emph{homoclinic intersection} of the stable and the unstable manifolds $\W^{s,u}_\lambda$ occurs at $(z,\theta)=(1,0)$, as their parametrizations \eqref{eq:param} coincide at this \emph{homoclinic point}. Since the manifolds $\W^{s,u}_\lambda$ are invariant, there in fact exists a \emph{homoclinic trajectory} on which the parametrizations agree:
\beq\label{eq:homoclinic-trajectory}
W^s(e^{\gamma t},\omega t)\equiv W^u(e^{\gamma t},\omega t).
\eeq

\begin{remark}
Equation~\eqref{Xsym} is what remains of the symmetry $X^0=(2\pi,0)-X^0\circ T$, which is just another way of writing \eqref{eq:arctan-sym}, after the onset of \emph{even} perturbation. This is an instance of \emph{spontaneous symmetry breaking}: The equations of motion, \eqref{xeq}, remain unchanged under the transformation $X\mapsto(2\pi,0)-X\circ T$, but the individual solutions do not respect this symmetry; $X^u\neq X^s=(2\pi,0)-X^u\circ T$, if $\lambda\neq 0$. 
\end{remark}

Coming to the second one of our main results, let us expand
\beqn
X^u=\sum_{\ell=0}^\infty \epsilon^\ell X^{u,\ell}.
\eeqn
In Section~\ref{sec:continuation}, we will show that the common analyticity domain of each $X^{u,\ell}$ in the $z$-variable is in fact much larger than the (small) neighbourhood of $\bigunit$---the corresponding analyticity domain of $X^u$ according to Theorem~\ref{thm:manifolds}; namely it includes the wedgelike region
\beqn
\mathbb{U}_{\tau,\vartheta}\defas \bigl\{\abs{z}\leq \tau\bigr\}\,\bigcup\,\bigl\{\arg{z}\in [-\vartheta,\vartheta]\cup [\pi-\vartheta,\pi+\vartheta]\bigr\}\subset\C
\eeqn
(with some positive $\tau$ and $\vartheta$):

\begin{result}[Analytic continuation]\label{thm:extension}
Each order $X^{u,\ell}$ of the solution extends analytically to a common region $\mathbb{U}_{\tau,\vartheta}\times\{\abs{\impart{\theta}}\leq\sigma\}$. Moreover, if $\psi\mapsto f(\piste,\psi)$ is a trigonometric polynomial of degree $N$, \ie, $N$ is the minimal nonnegative integer such that $\hat f(\piste,q)=0$ whenever $\abs{q}>N$, then $\theta\mapsto X^{u,\ell}(\piste,\theta)$ is a trigonometric polynomial of degree $\ell N$, at most.
\end{result}

\begin{remark}
With $\eta$ and $\sigma$ as in Remark~\ref{rem:manifolds}, the numbers $\tau$ and $\vartheta$ are specified by the following observation: $\Phi^0(z)=4\arctan z$ implies that $\abs{\impart\Phi^0(z)}\leq\eta$ in $\mathbb{U}_{\tau,\vartheta}$ with $\tau$ and $\vartheta$ sufficiently small. By Remark~\ref{rem:manifolds}, $(z,\theta)\mapsto f(\Phi^0(z),\theta)$ is analytic on $\mathbb{U}_{\tau,\vartheta}\times\{\abs{\impart{\theta}}\leq\sigma\}$, which we will use as the basis of the proof.
\end{remark}

In spite of Theorem~\ref{thm:extension}, (a straightforward upper bound on) $X^{u,\ell}$ grows without a limit as $\abs{\repart{z}}\to \infty$, such that there is no reason whatsoever to expect absolute convergence of the series $\sum_{\ell=0}^\infty \epsilon^\ell X^{u,\ell}$ in an unbounded $z$-domain with a fixed $\epsilon$. In fact, it is known that the behavior of the unstable manifold gets extremely complicated for large values of $z$ even with innocent looking Hamiltonian systems.  Still, it seems to us that the possibility of a uniform analytic extension of the coefficients $X^{u,\ell}$ has not been appreciated in the literature. 

Due to \eqref{Xsym}, an analog of Theorem~\ref{thm:extension} and the subsequent discussion are seen to hold for the solution $X^s$, with $z$ replaced by $z\inv$. 

Theorem~\ref{thm:extension} is interesting, because it allows one (at each order in $\epsilon$) to track trajectories $t\mapsto W^{s,u}(e^{\gamma t},\theta+\omega t)$ on the invariant manifolds $\W_\lambda^{s,u}$ for arbitrarily long times in a uniform complex neighbourhood $\abs{\impart{t}}\leq g\inv\vartheta$ of the real line, for arbitrary $\theta\in\torus$. The motivation for doing this stems from studying the splitting of the manifolds $\W_\lambda^{s,u}$ in the vicinity of the homoclinic trajectory \eqref{eq:homoclinic-trajectory}, and is the topic of another article. The general ideology that, being able to extend ``splitting related functions" to a large complex domain yields good estimates, is due to Lazutkin \cite{Lazutkin}, as is emphasized in \cite{LochakMemoirs}.

\subsection{Strategy}
Let us briefly explain how Theorem~\ref{thm:manifolds} will be proved in three steps. Due to \eqref{Xsym}, we may concentrate on studying the unstable manifold. Thus, we write
\beqn
X(z,\theta)\defas X^u(z,\theta)=X_0(\theta)+zX_1(\theta)+\delta_2X(z,\theta).
\eeqn
From \eqref{xeq} we first get an equation for $X_0\defas X^u(0,\piste)$ alone. Second, \emph{given} $X_0$, an equation for $X_1\defas \de_zX^u(0,\piste)$ and $\gamma$ alone is obtained. Third, \emph{given} $X_0$, $X_1$, \emph{and} $\gamma$, an equation for the remainder $\delta_2X$ is obtained.

It turns out that solving for $X_0$ and $X_1$ (together with $\gamma$), \ie, the invariant torus and the linearization of the unstable manifold around it, is difficult. Namely, these problems involve the small denominators of KAM theory. In contrast, solving for $\delta_2X$ amounts to a simple Contraction Mapping argument.

We deduce the existence of $X_0$ from \cite{Kupiainen}. The existence proof of $X_1$ is reminiscent of the RG argument in the latter paper, except that the Lyapunov exponent $\gamma$ has to be fine-tuned to a proper value such that the renormalization flow converges.

At this point we would like to draw the readers attention to the interesting reference \cite{GentileExponent}, where the author takes a different approach. Gentile fixes the perturbed Lyapunov exponent $\gamma$ in advance and replaces $g$ by $\tilde g(\epsilon,\gamma)$ in the Hamiltonian, which is analogous to introducing counterterms in quantum field theory, and finds the corresponding manifolds. One could then solve the implicit equation $\tilde g(\epsilon,\gamma)=g$ and to obtain $\gamma$ as a function of $g$ and $\epsilon$.

\subsection*{Acknowledgements} 
I am indebted to Antti Kupiainen for his help during the course of this work. Guido Gentile, Kari Astala, and Jean Bricmont provided sharp remarks and critical comments on the manuscript that made it more comprehensible and mathematically accurate. I wish to express my gratitude to all of them. I thank Giovanni Gallavotti and Emiliano De Simone for discussions at Rutgers University and University of Helsinki, respectively.

\section{Perturbed Tori}\label{sec:tori}
The perturbed tori will be found by looking for solutions having the general form
\beqn
\phi(t)=\Phi_0(\omega t),\quad\psi(t)=\omega t+\Psi_0(\omega t),
\eeqn
with $\Phi_0:\torus\to\R$ and $\Psi_0:\torus\to\R^d$ satisfying the ``$t\to -\infty$ asymptotics''
\begin{align}
\D^2\Phi_0(\theta)&=g^2\sin\Phi_0(\theta)+\lambda\,\de_\phi f(\Phi_0(\theta),\theta+\Psi_0(\theta))\label{Phi}\\ 
\D^2\Psi_0(\theta)&=\lambda\,\de_\psi f(\Phi_0(\theta),\theta+\Psi_0(\theta)) \label{Psi}
\end{align}
obtained from equation~\eqref{xeq} by putting $z=0$ and $\D=\omega\cdot\de_\theta$. Note that if $X_0=(\Phi_0,\Psi_0)$ is a solution to equations~\eqref{Phi} and \eqref{Psi}, then so is
\beq\label{eq:X0-shift}
\sigma_\beta X_0(\theta)\defas(\Phi_0(\theta+\beta),\Psi_0(\theta+\beta)+\beta)
\eeq
for $\beta\in\torus$. We point out that together \eqref{Phi} and \eqref{Psi} are equivalent to
\beq\label{X0}
\D^2 X_0=\Omega(X_0).
\eeq

\subsection{Spaces of analytic functions}\label{subsec:spaces}
Let us define the spaces we shall be working in. As linear subspaces of $\ell^1$, the Banach spaces
\begin{align*}
\Bphi{\sigma}&\defas\Big\{\Phi:\torus\to\C\;\Big|\;\normw{\Phi}{\sigma}\defas\sum_{q\in\Z^d}\abs{\hat\Phi(q)}e^{\sigma\abs{q}}<\infty\Big\},\\ 
\Bpsi{\sigma}&\defas\Big\{\Psi:\torus\to\C^d\;\Big|\;\normw{\Psi}{\sigma}\defas\sum_{q\in\Z^d}\abs{\hat\Psi(q)}e^{\sigma\abs{q}}<\infty\Big\},
\end{align*}
for any $\sigma\geq 0$, have the advantage that Fourier analysis on their elements is convenient. Furthermore, we are trying to find a solution $X=(\Phi,\Psi)$ analytic on the torus, and, for a suitably small $\sigma$, such a function belongs to $\Bphi{\sigma}\times\Bpsi{\sigma}$ because of the exponential decay of its Fourier coefficients; $\abs{\hat X(q)}<Ce^{-\sigma\abs{q}}$ with some positive constant $C$. Indeed, if $\sigma>0$, the spaces above comprise precisely those functions on the torus that admit an analytic extension to the ``strip'' $\abs{\impart\theta}<\sigma$. We will occasionally write $\B_\sigma$ when referring to either one of $\Bphi{\sigma}$ and $\Bpsi{\sigma}$.

Of course, as our analysis proceeds, the perturbation $f$ will appear all over the place. This, in turn, dictates the analyticity properties of a plethora of maps, in practice introducing the constraint $\sigma\leq\eta$ for the spaces $\B_\sigma$; see Remark~\ref{rem:manifolds}.  

Notice the natural embeddings
\beqn
\B_{\sigma+\alpha}\subset\B_{\sigma},
\eeqn
for $\alpha\geq 0$, due to the inequality
\beq\label{norms}
\normw{\piste}{\sigma}\leq\normw{\piste}{\sigma+\alpha}.
\eeq

Consider the linear operator $\tau_\beta:\B_{\sigma+\alpha}\to\B_\sigma$ defined through setting $\widehat{\tau_\beta X}(q)=e^{iq\cdot\beta}\hat X(q)$, with $\beta\in\C^d$. Whenever $\abs{\impart\beta}\leq\alpha$, $\normw{\tau_\beta}{\mathcal{L}(\B_{\sigma+\alpha};\B_\sigma)}\leq 1$. The realization of $\tau_\beta$ in terms of the variable $\theta$ is just the translation $\Psi(\theta)\mapsto\Psi(\theta+\beta)$. $\tau_\beta$ will serve as a useful device in encoding the real-analyticity of $f$ as an algebraic property into the Fourier series of certain other functions. This is due to the the fact that exponential smallness of $\abs{\widehat{X}(q)}$ in $q$ implies real-analyticity of a function $X$ on the torus, and vice versa. 
 
We shall encounter $n$-linear maps from $\C^{d+1}$ into $\C$. Endowed with the norm
{\Small
\beqn
\multinorm{A}\defas\inf\left\{M\geq 0\;\Big|\;\abs{A(z_1,\dots,z_n)}\leq M\abs{z_1}\dots\abs{z_n}\quad\forall\, z_i\in\C^{d+1}\right\}
\eeqn
}%
they form the Banach space $\mathcal{L}(^n(\C^{d+1});\C)$; see \cite{Chae}.

\subsection[Asymptotics of the solution in the perturbed case]{Past and future asymptotics of the solution in the perturbed case}
This subsection discusses the $t\to\pm\infty$ asymptotics of the solution $X$. In these limits the motion settles onto the ``distorted version'' $\T_\lambda$ of the invariant torus $\T_0$ with the pendulum seizing to swing, but wiggling quasiperiodically about its unstable equilibrium. 

\begin{theorem}\label{thm:tori}
Under the assumptions of Theorem~\ref{thm:manifolds}, there exist positive numbers $r$ and $\epsilon_0$ such that, for $(\epsilon,g)\in D$, equations~\eqref{Phi} and \eqref{Psi} have a unique solution $X_0=(\Phi_0,\Psi_0)$ in the class of those real-analytic functions of $\theta\in\torus$ that satisfy $\normw{\Psi_0}{\ell^1}<r$ and $\average{\Psi_0}=0$ (zero average). The function $X_0$, defined on $\{\abs{\impart{\theta}}\leq\sigma\}\times D$ for some $\sigma>0$, is analytic and uniformly bounded by $(C\abs{\epsilon},Cg^2\abs{\epsilon})$. Moreover, it is $\R\times\R^d$-valued on $\torus$ for $\epsilon$ real. Thus, any real-analytic solution $X_0'=(\Phi_0',\Psi_0')$ with $\average{\Psi_0'}=\beta\in\R^d$ and $\normw{\Psi_0'-\beta}{\ell^1}<r$ must be the one given by
\beqn
X_0'(\theta)\equiv X_0(\theta+\beta)+(0,\beta),
\eeqn
\ie, $X_0'=\sigma_\beta X_0$, using the notation of \eqref{eq:X0-shift}.
\end{theorem}
\begin{remark}
Remark~\ref{rem:manifolds} below Theorem~\ref{thm:manifolds} holds true. Recall that we have defined $\epsilon:=\lambda g^{-2}$ in \eqref{eq:epsilon} and the domain $D$ in \eqref{eq:D}. This is a version of the KAM Theorem. Notice that $X_0\in \Bphi{\sigma}\times\Bpsi{\sigma}$.
\end{remark}
\begin{proof}
The proof is a reduction to the one given in \cite{Kupiainen}. Here we systematically omit the subindex $0$ of $\Phi_0$, $\Psi_0$, and $X_0$. Let us concentrate on the pendulum part, equation~\eqref{Phi}, first. We expect $\Phi$ to be close to its unperturbed value, zero, and it pays to cancel the leading term of $g^2\sin\Phi(\theta)$ on the right-hand side by subtracting $g^2\Phi(\theta)$ from both sides. We then have
\beq\label{Phi1}
(\D^2-g^2)\Phi=U(\Phi,\Psi)\defasr U(X)
\eeq
with
\beq\label{U}
U(X)(\theta)\defas g^2(\sin\Phi(\theta)-\Phi(\theta))+\lambda\,\de_\phi f(\Phi(\theta),\theta+\Psi(\theta)).
\eeq

Pay attention to the fact that $U(X)(\theta)$ depends locally on $X$---only through $X(\theta)$, that is. Abusing notation, we shall use $U(X)(\theta)$, $U(X,\theta)$, $U(X(\theta),\theta)$, \etc\@, in the same meaning, whichever is the most convenient form. Now, $U(\chi,\theta)$ is analytic in the vector argument $\chi=(\chi_\phi,\chi_\psi)$ in the region $\abs{\chi_\phi},\abs{\chi_\psi}\leq \eta$, where $\eta>0$ depends on the analyticity domain of $f$; see Remark~\ref{rem:manifolds} on page \pageref{rem:manifolds}. 

Let us now write down the Fourier--Taylor expansion
\begin{align}
U(X(\theta),\theta)&=\sum_{n=0}^\infty\frac{1}{n!}D^nU(0,\theta)\,(X(\theta),\dots,X(\theta))\nonumber\\
&=\sum_{n=0}^\infty\frac{1}{n!}\sum_{\substack{\q=(q_1,\dots,q_n) \\ q_i\in\Z^d}} e^{i\theta\cdot\sum_i q_i}\,D^nU(0,\theta)\,(\hat X(q_1),\dots,\hat X(q_n)),\label{Taylor}
\end{align}
where $D^nU(0,\theta)\in\mathcal{L}(^n(\C^{d+1});\C)$ is the $n{\text{th}}$ Fr\'echet derivative of the map $U(\piste,\theta):\C^{d+1}\to\C:\chi\mapsto U(\chi,\theta)$.

The map $\theta\mapsto U_n(\theta)\defas \frac{1}{n!}D^nU(0,\theta)$ is analytic in the same domain as $\theta\mapsto U(0,\theta)=\lambda\,\de_\phi f(0,\theta)$, \ie, $\abs{\impart \theta}\leq \eta$. Its Fourier representation $U_n(\theta)=\sum_{q\in\Z^d} e^{iq\cdot\theta}u_n(q)$ has coefficients
\beq\label{kernel}
u_n(q)=\normalize\int_\torus  e^{-iq\cdot \theta}\,\frac{1}{n!}D^nU(0,\theta)\,d\theta
\eeq
in $\mathcal{L}(^n(\C^{d+1});\C)$. Using this notation, we translate \eqref{Taylor} into the Fourier language;
\begin{align}\label{TaylorF}
\widehat {U(X)}(q)=\sum_{n=0}^\infty\sum_{\q\in\qdomain}u_n(q-\sum_{i=1}^n q_i)\,(\hat X(q_1),\dots,\hat X(q_n)).
\end{align}

The right-hand side of equation~\eqref{TaylorF} is a power series in $\hat X$, converging whenever $\hat X$ is sufficiently close to zero. Namely, we have
\begin{lemma}\label{decay}
The multilinear maps $u_n(q)$ obey the bound
\beq\label{bound}
\multinorm{u_n(q)}\leq Cg^2(r_0^3+\abs{\epsilon})(r_0/e)^{-n}e^{-\rho|q|},
\eeq
where $\rho$ and $r_0$ is any pair of positive numbers satisfying $\rho+r_0=\eta$, $\eta>0$ being the width of the analyticity domain of $f$ as explained in Remark~\ref{rem:manifolds}.
\end{lemma}
The proof of Lemma~\ref{decay} is straightforward, but, for the sake of continuity, is given in Subsection~\ref{subsec:proof.lem1and2} below.

Considering the closed origin-centered balls of radius $r<r_0/2$ in $\Bphi{\sigma}$ and $\Bpsi{\sigma}$---$\Phiball{\sigma}$ and $\Psiball{\sigma}$, respectively---we next study $U_\beta:\Phiball{\sigma}\times\Psiball{\sigma}\to \Bphi{\sigma}:(\Phi,\Psi)\mapsto \tau_\beta U(\tau_{-\beta}\Phi,\tau_{-\beta}\Psi)$. By equation~\eqref{U},
\beq\label{Ubeta}
U_\beta(\Phi(\theta),\Psi(\theta),\theta)=U(\Phi(\theta),\theta+\beta+\Psi(\theta)), 
\eeq
when $\beta\in\R^d$. The right-hand side is analytic in $\beta$, and extends to $\abs{\impart\beta}+\sigma+r<\eta$ through the same expression, leaving $U_\beta$ analytic with respect to $X$. 

More quantitatively, one checks using the bound \eqref{bound} that the power series 
\begin{align}
\widehat{U_{\beta}(X)}(q) = \sum_{n=0}^\infty\sum_{\q\in\qdomain}e^{i\beta\cdot(q-\sum_iq_i)}\,u_n(q-\sum_{i=0}^n q_i)\,(\hat X(q_1),\dots,\hat X(q_n)),
\label{Ubeta2}
\end{align}
converges uniformly with respect to $X$ and $\beta$, even if the latter has a small imaginary part. In fact, $\normw{U_\beta(X)}{\sigma}$ obeys the upper bound 
{\small
\begin{align*}
&\sum_{n=0}^\infty\sum_{\q\in\qdomain}\sum_{q\in\Z^d}e^{\sigma\abs{q}} \Bigl | e^{i\beta\cdot(q-\sum_iq_i)}\,u_n(q-\sum_{i=0}^n q_i)\,(\hat X(q_1),\dots,\hat X(q_n)) \Bigr | \\
&\leq\sum_{n=0}^\infty\left(\sum_{q\in\Z^d}e^{(\abs{\impart\beta}+\sigma)\abs{q}}\multinorm{u_n(q)}\right)\,\sum_{\q\in\qdomain}\prod_{i=1}^n\,\abs{\hat X(q_i)}e^{\sigma\abs{q_i}}\\
&\leq Cg^2(r_0^3+\abs{\epsilon})\sum_{n=0}^\infty\sum_{q\in\Z^d}e^{(\abs{\impart\beta}+\sigma-\rho)\abs{q}}\,(r_0/e)^{-n}\,\normw{X}{\sigma}^n\leq Cg^2(r_0^3+\abs{\epsilon}),
\end{align*}
}%
if we choose $\abs{\impart\beta}+\sigma<\rho=\eta-r_0$ and $r<r_0/2e$, since $\normw{X}{\sigma}\leq 2r$.
Thus, fixing $r=r_0/6$, say, we obtain
\beq\label{Ubound}
\sup_{X\in\Phiball{\sigma}\times\Psiball{\sigma}}\normw{U_\beta(X)}{\sigma}\leq Cg^2(r^3+\abs{\epsilon})
\eeq
whenever 
\beq\label{eq:betacond}
\abs{\impart\beta}+\sigma+6r<\eta.
\eeq

\begin{lemma}\label{lemma:phi}
Suppose \eqref{eq:betacond} holds, and $\Psi\in\Psiball{\sigma}$. Then, for $r$ and $\epsilon_0$ small enough,
\beqn
(\D^2-g^2)\Phi=U_\beta(\Phi,\Psi)
\eeqn
has a solution $\Phi_\beta(\Psi)\in\Phiball{\sigma}$, real-valued provided $\beta$, $\epsilon$, and $\Psi$ are, and there are no other solutions in the $\ell^1$-ball $\Phiball{0}\supset\Phiball{\sigma}$. In fact, $\Phi_\beta(\Psi)=\tau_\beta\Phi_0(\tau_{-\beta}\Psi)$. The map $\Psi\mapsto\Phi_\beta(\Psi)$ is analytic on $\Psiball{\sigma}$. $\Phi_\beta(\Psi)$ also depends analytically on $\beta$ as well as on $(\epsilon,g)\in D$ (see \eqref{eq:D}), and obeys the bound
\beq\label{estimate}
\normw{\Phi_\beta(\Psi)}{\sigma}\leq C\abs{\epsilon}
\eeq
uniformly in $\Psi$, $\beta$, and $g$.
\end{lemma}

\begin{remark}
The smallness condition is $C(r^3+\epsilon_0)\leq r$, where $C$ is the same constant as in \eqref{Ubound} and contains the norm of the perturbation $f$.
\end{remark}

The standard but lengthy proof of Lemma~\ref{lemma:phi} may be found in Subsection~\ref{subsec:proof.lem1and2}.

Let us come back to equation~\eqref{Psi}, whose right-hand side may now be written solely in terms of $\Psi\in\Psiball{\sigma}$, amounting to
\beq\label{PsiV}
\D^2\Psi=V(\Psi)
\eeq
with $V(\Psi)(\theta)\equiv \lambda\,\de_\psi f(\Phi(\Psi)(\theta),\theta+\Psi(\theta))$. Consider then $V_\beta(\Psi)\defas\tau_\beta V(\tau_{-\beta}\Psi)$. By Lemma~\ref{lemma:phi}, it reads
\beqn
V_\beta(\Psi)(\theta)\equiv V(\tau_{-\beta}\Psi)(\theta+\beta)\equiv\lambda\,\de_\psi f(\Phi_\beta(\Psi)(\theta),\theta+\beta+\Psi(\theta))
\eeqn 
and is analytic in the domain 
\beq\label{eq:Vdomain}
\Psiball{\sigma}\times D\times \{\abs{\impart\theta}\leq\sigma\}\times\{\beta\,|\;\abs{\impart\beta}+\sigma+6r<\eta\}
\eeq
 with the uniform bound
\beqn
\normw{V_\beta(\Psi)}{\sigma}\leq\sup_{\abs{\impart\phi},\abs{\impart\psi}\leq \eta}\,\abs{\lambda\,\de_\psi f(\phi,\psi)}\leq Cg^2\abs{\epsilon},
\eeqn
provided $C\abs{\epsilon}\leq \eta$ (see \eqref{estimate}).

Equation~\eqref{PsiV} is the variational equation corresponding to the action functional
\beqn
S:\Psiball{\sigma}\to\R:\Psi\mapsto S(\Psi)=\int_\torus s(\Psi,\theta)\,d\theta
\eeqn
given by the integrand
\beqn 
s(\Psi,\theta)=\half(\Phi\D^2\Phi+\Psi\cdot\D^2\Psi)+g^2\cos\Phi-\lambda f(\Phi,\theta+\Psi),
\eeqn
where $\Phi=\Phi(\Psi)$. $S$ is invariant under the $\torus$-action $\Psi(\theta)\mapsto\Psi_\beta(\theta)\defas\Psi(\theta+\beta)+\beta$, $\beta\in\R^d$. Hence, ${\de_\beta S(\Psi_\beta)\rvert}_{\beta=0}=0$ yields the \emph{Ward identity}
\beq\label{Ward}
\int_\torus\frac{\delta S(\Psi)}{\delta\Psi^i(\theta)}\,d\theta=\int_\torus\Psi(\theta)\cdot\de_{\theta^i}\frac{\delta S(\Psi)}{\delta\Psi(\theta)}\,d\theta\quad (i=1,\dots,d)
\eeq
of the symmetry in the functional derivative notation. In fact,
\beqn
\frac{\delta S(\Psi)}{\delta\Psi(\theta)}=(\D^2\Psi-V(\Psi))(\theta).
\eeqn
Integrating by parts three times one sees that
\beqn
\int_\torus\Psi(\theta)\cdot\de_{\theta^i}\D^2\Psi(\theta)\,d\theta=-\int_\torus\Psi(\theta)\cdot\de_{\theta^i}\D^2\Psi(\theta)\,d\theta=0.
\eeqn
The general identity \eqref{Ward} therefore reduces to the identity
\beq\label{WardV}
\int_\torus V^i(\Psi,\theta)\,d\theta=\int_\torus\Psi(\theta)\cdot\de_{\theta^i} V(\Psi,\theta)\,d\theta
\eeq
for the map $V$.

In conclusion, we have the KAM-type small denominator problem \eqref{PsiV} with $V_\beta(\Psi,\theta)$ analytic in the domain \eqref{eq:Vdomain} and bounded there by $C\abs{\lambda}$, together with the Ward identity \eqref{WardV} stemming from a translation symmetry of the action that generates the equation. Furthermore, $V_\beta(\Psi,\theta)$ is real-valued whenever $\beta$, $\epsilon$, and $\Psi$ are. For $0<\sigma<\eta-6r$---so that we may choose $\impart\beta\neq0$---this is precisely the setup in \cite{Kupiainen}, where the authors devise a method for dealing with such problems using a Renormalization approach. 

The subtle analysis in \cite{Kupiainen} yields a unique solution $\Psi\in\Psiball{\sigma}$ to \eqref{PsiV} with zero average and analytic in $(\epsilon,g)\in D$. The inevitable loss of analyticity takes place in the domain of $\beta$. The map $\theta\mapsto\Psi(\theta)$ is $\R^d$-valued on the torus for real $\epsilon$ and satisfies $\normw{\Psi}{\sigma}\leq C\abs{\lambda}=Cg^2\abs{\epsilon}$.  

Denote by $\Psi_n$, $n\in\Z_+$, the unique solution to \eqref{PsiV} in the ball $\Psiball{\sigma/n}$. Since $\Psiball{\sigma}\subset\Psiball{\sigma/n}$, $\Psi$ has to coincide with $\Psi_n$. Hence, $\Psi$ is the unique solution in 
\beqn
\bigcup_{n=1}^\infty\Psiball{\sigma/n}\supset\left\{\Psi:\torus\to\R^d\;\Big|\;\Psi\;\text{real-analytic and}\;\normw{\Psi}{\ell^1}< r\right\}.
\eeqn
Indeed, assuming the map $\theta\mapsto\Psi(\theta)$ is real-analytic, $\normw{\Psi}{\sigma/n}<\infty$ for some $n$, and we have that $\normw{\Psi}{\sigma/n}\searrow\normw{\Psi}{0}\equiv\normw{\Psi}{\ell^1}$ as $n\to\infty$. Thus, if $\normw{\Psi}{\ell^1}<r$, we gather that $\normw{\Psi}{\sigma/n}<r$ for sufficiently large values of $n$.

This concludes the proof of Theorem~\ref{thm:tori}.
\end{proof}

\subsection{Proofs of Lemmata~\ref{decay}\label{subsec:proof.lem1and2} and \ref{lemma:phi}}

\begin{proof}[Proof of Lemma~\ref{decay}]
Write $\norm{\piste}=\multinorm{\cdot}$ for short. From \eqref{kernel} and the Cauchy Integral Theorem,
\begin{align*}
\norm{u_n(q)}&= \Bigl \| \normalize\int_\torus  e^{-iq\cdot (\theta+i\beta)}\,\frac{1}{n!}D^nU(0,\theta+i\beta)\,d\theta \Bigr \| \\
&\leq e^{q\cdot\beta}\frac{1}{n!}\sup_{\theta\in\torus}\norm{D^nU(0,\theta+i\beta)},
\end{align*}
for $\beta\in\R^{d}$ and $\abs{\beta}\leq\eta$. Take $0<\rho<\eta$ and choose $\beta=-\rho q/\abs{q}$. We compute the standard norm of $n$-homogeneous polynomials,
\beqn
{\norm{D^nU(0,\theta+i\beta)}}_{\mathcal{P}^n(\C^{d+1};\C)}\defas\sup_{\abs{z}\leq 1}\abs{D^nU(0,\theta+i\beta)\,(z,\dots,z)},
\eeqn
which, using the Cauchy Integral Formula, turns into
{\small
\beqn
\sup_{\abs{z}\leq 1} \Bigl | \frac{n!}{2\pi i}\oint_{\de\mathbb{D}(0,r_0)}\frac{U(\zeta z,\theta+i\beta)\, d\zeta}{\zeta^{n+1}} \Bigr |
 \leq n!\,r_0^{-n}\sup_{\abs{z}\leq 1}\sup_{\zeta\in\de\mathbb{D}(0,r_0)}\,\abs{U(\zeta z,\theta+i\beta)}.
\eeqn
}%
Here $\mathbb{D}(0,r_0)$ is the origin-centered circle of radius $r_0$ in the complex plane, with the constraint $r_0+\rho\leq \eta$. For $\abs{z}\leq r_0$ and  $\abs{\impart\theta}\leq\rho$ we estimate 
\beqn
\abs{U(z,\theta)}\leq Cg^2(r_0^3+\abs{\epsilon});
\eeqn
see equation~\eqref{U}. Here we have singled out $\lambda g^{-2}=\epsilon$, and $C$ is independent of $g$. We stress that $U(z,\theta)$ simply stands for the expression obtained from the expression of $U(X,\theta)$ in \eqref{U} by replacing $X(\theta)$ by $z\in\C^{d+1}$.

Symmetric multilinear maps are fully determined by their diagonal---the corresponding homogeneous polynomial, that is---which is explicitly confirmed by the Polarization Formula \cite{Chae,Dineen}. Hence, in order to obtain the estimate in \eqref{bound}, we multiply the corresponding polynomial estimate by the factor $n^n/n!\sim e^n/\sqrt{2\pi n}$.
\end{proof}

\begin{proof}[Proof of Lemma~\ref{lemma:phi}]
The proof is a simple application of the Banach Fixed Point Theorem. We fix $\Psi\in\Psiball{\sigma}$ and study the operator $F(\Phi)\defas (\D^2-g^2)^{-1}U_\beta(\Phi,\Psi)$.

First, $(\D^2-g^2)^{-1}$ is a linear operator bounded in norm by $g^{-2}$. From \eqref{Ubound}, 
\beqn
\normw{F(\Phi)}{\sigma}\leq g^{-2}\normw{U_\beta(\Phi,\Psi)}{\sigma}\leq C(r^3+\abs{\epsilon})\leq r
\eeqn
for sufficiently small $r$ and $\epsilon$, which means that $F(\Phiball{\sigma})\subset\Phiball{\sigma}$. Proving contractiveness resembles estimating the norm of $U_\beta$ in the proof of Theorem~\ref{thm:tori}, and is omitted. The existence and uniqueness of the solution $\Phi(\Psi,\beta)\in\Phiball{\sigma}$ now follow. 

For $\beta$, $\epsilon$, and $\Psi$ real, $F$ maps the closed subset of real-valued functions $\Phi\in\Phiball{\sigma}$ into itself and is a contraction there, so $\Phi(\Psi,\beta)$ is real-valued by uniqueness.

The operator $F$ depends analytically on the parameter $\Psi$ in $\Psiball{\sigma}$. Consider the sequence ${(F^k(0))}_{k\in\N}$
of successive substitutions. Each element $F^k(0)$ is analytic in $\Psi\in\Psiball{\sigma}$. Furthermore, the Banach Fixed Point Theorem guarantees that such a sequence converges to the fixed point $\Phi(\Psi,\beta)$ in geometric progression;
\beqn
\normw{F^k(0)-\Phi(\Psi,\beta)}{\sigma}\leq\frac{\mu^n}{1-\mu}\,\normw{F(0)}{\sigma}<\frac{r\mu^n}{1-\mu}.
\eeqn
Consequently, $\Phi(\Psi,\beta)$ is the uniform limit of a sequence of analytic functions, and, as such, analytic itself. The same argument goes for $(\epsilon,g)\in D$ (see \eqref{eq:D}), as well as for $\beta$ in the domain specified by \eqref{eq:betacond}.

Because \eqref{norms} implies $\Psi\in\Psiball{\sigma}\subset\Psiball{0}$, in fact $\Phi(\Psi,\beta)$ is the unique solution in $\Phiball{0}$.

Let us denote $\Phi(\Psi)=\Phi(\Psi,0)$. If $\Psi\in\Psiball{\sigma}$ and $\abs{\impart{\beta}}\leq \sigma/2$, then $\tau_{-\beta}\Psi\in\Psiball{\sigma/2}$, such that $\Phi=\Phi(\tau_{-\beta}\Psi)$ is the unique element in $\Phiball{\sigma/2}$ solving $\Phi=(\D^2-g^2)^{-1}U(\Phi,\tau_{-\beta}\Psi)$. The diagonality of $\tau_\beta$ and $\D$ yields
\beqn
\Phi_\beta(\Psi)=(\D^2-g^2)^{-1}U_\beta(\Phi_\beta(\Psi),\Psi),
\eeqn
where $\Phi_\beta(\Psi)=\tau_\beta\Phi(\tau_{-\beta}\Psi)\in\Phiball{0}$. But $\Phi(\Psi,\beta)$ was the unique solution in $\Phiball{0}$, such that $\Phi_\beta(\Psi)=\Phi(\Psi,\beta)\in\Phiball{\sigma}$. For larger $\abs{\impart{\beta}}$ one obtains an analytic continuation.

\textit{A priori}, we know that $\normw{\Phi_\beta(\Psi)}{\sigma}=\normw{F(\Phi_\beta(\Psi))}{\sigma}\leq r$. On the other hand, we know that $\Phi_\beta(\Psi)\unperte=0$ by uniqueness, whence the estimate \eqref{estimate} follows.
\end{proof}

\section{Lyapunov Exponent---Linearizing the Unstable Manifold}
In this section we study the motion in the immediate vicinity of the torus $\T_\lambda$ corresponding to the solution $X_0(\theta)$ of Theorem~\ref{thm:tori}. To that end, suppose $X(z,\theta)$ is an analytic solution to equation~\eqref{xeq} with $X(0,\theta)=X_0(\theta)$. Then $X_1(\theta)\defas\de_zX(0,\theta)$ should satisfy the equation
\beq\label{X1}
(\D+\gamma)^2X_1=D\Omega(X_0)X_1,
\eeq
as $\Omega(X)(z,\theta)$ depends on $z$ only through $X$ evaluated at $(z,\theta)$.

Note that \eqref{X1} is a problem of ``eigenvalue type''; recalling $\gamma\unperte=g$, we will strive to choose $\gamma=\gamma(\epsilon,g)$ in a $g$-dependent neighbourhood, say
\beq\label{eq:gammacondition}
\abs{\gamma-g}<g/2,
\eeq
of its unperturbed value $g$, such that \eqref{X1} has a nontrivial solution. That we succeed is the content of Theorem~\ref{thm:linearization}. Consequently, our $\gamma$ will depend analytically on $\epsilon$, nicely controlled by $\abs{\gamma-g}<Cg\abs{\epsilon}$.

The subtlety of proving Theorem~\ref{thm:linearization} lies in solving a small denominator problem. We go about dealing with it using a Renormalization Group method, treating such small denominators scale by scale. Here we show that the framework of \cite{Kupiainen} is applicable. The proof, though, is self-contained.

First, view the map $X\mapsto\Omega(X)$ as the map that takes the pair $(\Phi,\Psi)$ to \linebreak $(\Omega_\Phi(\Phi,\Psi),\Omega_\Psi(\Phi,\Psi))$ with the components $\Omega_\Phi(\Phi,\Psi)=g^2\sin\Phi+\lambda\,\de_\phi f(\Phi,\theta+\Psi)$ and $\Omega_\Psi(\Phi,\Psi)=\lambda\,\de_\psi f(\Phi,\theta+\Psi)$. Then  the component form of \eqref{X1} reads
\beq\label{X1mat}
(\D+\gamma)^2\begin{pmatrix}\Phi_1 \\ \Psi_1\end{pmatrix}=
\begin{pmatrix}
g^2\cos\Phi_0+\lambda f_{\phi,\phi} & \lambda f_{\phi,\psi} \\
\lambda f_{\psi,\phi} & \lambda f_{\psi,\psi}
\end{pmatrix}\begin{pmatrix}\Phi_1 \\ \Psi_1\end{pmatrix}.
\eeq
In each entry, $f_{a,b}$ stands for the matrix $(\de_b\de_a f)(\Phi_0,\theta+\Psi_0)$.

From \eqref{X1mat} we get for $\Psi_1$ the equation
\beq\label{eq:psi1}
\Psi_1=\left[(\D+\gamma)^2-\lambda f_{\psi,\psi}\right]^{-1}(\lambda f_{\psi,\phi}\Phi_1)\defasr J\Phi_1,
\eeq
Here $J$ is a well-defined bounded linear operator from $\Bphi{\sigma}$ to $\Bpsi{\sigma}$, provided that $\epsilon_0$ is small. Checking this is straightforward implementation of Neumann series and the fact that the operator $(\D+\gamma)^{-2}$ has the diagonal Fourier kernel
\beq\label{eq:Oker1}
(\D+\gamma)^{-2}(p,q)=\delta_{p,q}(i\omega\cdot q+\gamma)^{-2},\quad p,q\in\Z^d.
\eeq
Using \eqref{eq:gammacondition}, one obtains the bound
\beq\label{eq:Jbound}
\normw{J}{\mathcal{L}(\Bphi{\sigma};\Bpsi{\sigma})}\leq C\abs{\epsilon}.
\eeq

\begin{remark}\label{rem:epsilon}
The definition of $J$ is an instance where demanding smallness of $\epsilon\defas \lambda g^{-2}$ is natural, indeed necessary.
\end{remark}

Consequently, using \eqref{eq:psi1}, we get for $\Phi_1$ the equation
\beq\label{eq:phi1}
[(\D+\gamma)^2-g^2]\Phi_1=g^2(\cos\Phi_0-1)\Phi_1+\lambda f_{\phi,\phi}\Phi_1+\lambda f_{\phi,\psi}J\Phi_1\defasr H\Phi_1.
\eeq
Recall that $\Phi_0\unperte=0$ by Lemma~\ref{lemma:phi}. Therefore $H\unperte=0$, and $\Phi_1\unperte=4$ (due to $\Phi^0(z)=4\arctan z$) is a physically motivated nontrivial solution to \eqref{eq:phi1}. In other words, the differential operator $(\D+g)^2-g^2$ is \emph{singular}. On the other hand, when $\epsilon\neq 0$ is small, we know that $\Phi_0$ remains close to zero, making the whole right-hand side in \eqref{eq:phi1} small. We then hope to find a Lyapunov exponent $\gamma$, close to $g$, such that  $(\D+\gamma)^2-g^2-H$ stays singular and the equation still admits a nontrivial solution close to the constant function 4.

It follows from \eqref{eq:Jbound} that the operator $H$ appearing in \eqref{eq:phi1}, which lies in $\mathcal{L}(\Bphi{\sigma})\equiv\mathcal{L}(\Bphi{\sigma};\Bphi{\sigma})$, has the useful properties below. The proof comprises Subsection~\ref{subsec:proof.lem:H}.

\begin{lemma}\label{lem:H}
Denote the kernel of $H\in\mathcal{L}(\Bphi{\sigma})$ by $H(p,q)$, $(p,q)\in\Z^d\times\Z^d$. For $\abs{\impart{\kappa}}\leq g/3$, there exists an operator $H(\kappa)\in\mathcal{L}(\Bphi{\sigma})$ related to $H$ by
\beqn
(t_sH)(p,q)\defas H(p+s,q+s)=H(\omega\cdot s;p,q),\quad s\in\Z^d.
\eeqn
Let $0<\sigma'<\sigma$. The kernel $H(\kappa;p,q)$ is analytic on 
\beqn
\left\{(\kappa,\epsilon,g,\gamma)\;\big|\;\abs{\impart{\kappa}}\leq g/3,\, (\epsilon,g)\in D,\,\abs{\gamma-g}<g/2\right\}
\eeqn
and it satisfies the bound
\beqn
\abs{H(\kappa;p,q)}\leq Cg^2\abs{\epsilon}\,e^{-\sigma'\abs{p-q}}
\eeqn
with $C=C(\sigma')$. As for the $\kappa$-derivatives,
\beqn
\abs{H^{(k)}(\kappa;p,q)}\leq Ck!\,(g/3-\abs{\impart{\kappa}})^{-k}g^2\abs{\epsilon}^2\,e^{-\sigma'\abs{p-q}},\quad k\geq 1.
\eeqn
Moreover,
\beqn
\left|\frac{\de}{\de\gamma} H(0;0,0)\right|\leq C\abs{\epsilon}^2g\,\frac{1}{1-2\abs{\gamma-g}/g}.
\eeqn
\end{lemma}

\subsection{Proof of Lemma~\ref{lem:H}}\label{subsec:proof.lem:H}
\begin{proof}
To simplify notations, we decompose
\beqn
H=H_1+H_2\quad\text{with}\quad H_2=\lambda f_{\phi,\psi}J. 
\eeqn
Let $\Phi$ and $\Psi$ be arbitrary functions in the spaces $\Bphi{\sigma}$ and $\Bpsi{\sigma}$, respectively.

$H_1$ acts as ordinary multiplication: $H_1\Phi(\theta)=H_1(\theta)\Phi(\theta)$ with $H_1(\theta)\in\C$. We write $\widehat H_1$ for the Fourier transform of the map $\theta\mapsto H_1(\theta)$. Denoting a kernel element of the operator $H_1$ by $H_1(p,q)$, we have $H_1(p,q)\equiv\widehat H_1(p-q)$.
We gather that 
\beq\label{eq:t_sH_1}
t_sH_1=H_1 
\eeq
holds, and that the kernel of $H_1$ satisfies
\beqn
\abs{H_1(p,q)}\leq C\abs{\lambda}\,e^{-\sigma\abs{p-q}},\quad p,q\in\Z^d.
\eeqn
Here $\sigma>0$ is the width of the analyticity strip around the real $\torus$ of the map $\theta\mapsto H_1(\theta)$, \ie, of $X_0$. Since, by Theorem~\ref{thm:tori}, $X_0$ is analytic with respect to $(\epsilon,g)\in D$, so is $H_1(p,q)$. 

Observe that the expression defining $J$ in \eqref{eq:psi1} may be cast as  
\beqn
J\Phi=\left[\one-(\D+\gamma)^{-2}(\lambda f_{\psi,\psi})\right]^{-1}(\D+\gamma)^{-2}(\lambda f_{\psi,\phi}\Phi)=B\Lambda O\Phi,
\eeqn
where $B$, $\Lambda$, and $O$ stand for $\left[\one-(\D+\gamma)^{-2}(\lambda f_{\psi,\psi})\right]^{-1}$, $(\D+\gamma)^{-2}$, and $\lambda f_{\psi,\phi}$, respectively. Assuming each index $a$ and $b$ in $f_{a,b}$ stands either for $\phi$ or $\psi$, the reader should bear in mind that $f_{a,b}$ refers to the multiplication operator corresponding to the Jacobian matrix $(\de_b\de_af)(\Phi_0,\theta+\Psi_0)$. Its Fourier kernel reads $f_{a,b}(p,q)=\hat f_{a,b}(p-q)$,
whence the translation invariance
\beq\label{eq:t_sf=f}
t_s f_{a,b}= f_{a,b}.
\eeq

Denoting $\Lambda(q)\equiv \Lambda(q,q)\equiv (i\omega\cdot q+\gamma)^{-2}$, we are interested in the kernel
\beq\label{eq:Jker}
J(p,q)=\sum_{r\in\Z^d}B(p,r)\Lambda(r)O(r,q),\quad p,q\in\Z^d,
\eeq
of $J$. We shall also need the ``shifted version'' of $\Lambda(q)$,
\beq\label{eq:Lambdak}
\Lambda(\kappa;q)\defas(i\omega\cdot q+i\kappa+\gamma)^{-2},\quad\kappa\in\C.
\eeq
It is related to $\Lambda(q)$ by the property
\beq\label{eq:t_sLambda}
t_s\Lambda(q)=\Lambda(\omega\cdot s;q).
\eeq
Further, $\Lambda(\kappa;q)$ is analytic on $\{\kappa\;|\;\abs{\impart\kappa}\leq g/3\}\times\{\gamma\;|\;\abs{\gamma-g}<g/2\}$ and satisfies
\beq\label{eq:Lambdakbound}
\abs{\Lambda(\kappa;q)}\leq 36g^{-2}.
\eeq

Equation~\eqref{eq:Lambdakbound} also means that the operator $\Lambda(\kappa)$ corresponding to the kernel in \eqref{eq:Lambdak} belongs to $\mathcal{L}(\B_\sigma)$ with $\normw{\Lambda(\kappa)}{\mathcal{L}(\B_\sigma)}\leq 36g^{-2}$. Interpreting $f_{a,b}$ as a multiplication operator, $\normw{f_{a,b}}{\mathcal{L}(\B_\sigma)}\leq \normw{f_{a,b}}{\sigma}$
shows that $B,O\in\mathcal{L}(\B_\sigma)$.

As in the case of $H_1$, $O$ acts as multiplication by a real-analytic function whose modulus is bounded by $C\abs{\lambda}$. Thus, we estimate
\beq\label{eq:Oker}
\abs{O(p,q)}\leq C\abs{\lambda}\,e^{-\sigma\abs{p-q}}\mathand\abs{\Lambda(p)O(p,q)}\leq C\abs{\epsilon}\,e^{-\sigma\abs{p-q}}.
\eeq

Bounding the kernel of $B$ calls for the Neumann series
\beq\label{eq:Bneumann}
B=\sum_{k=0}^\infty B_k,\quad\text{with}\quad B_k\defas\left(\lambda\,\Lambda f_{\psi,\psi}\right)^k.
\eeq
Clearly $\normw{B_k}{\mathcal{L}(\B_\sigma)}\leq (C\abs{\epsilon})^k$ and $\abs{B_k(p,q)}\leq (C\abs{\epsilon})^k$ such that, by Fubini's Theorem,
\beq\label{eq:Bker}
\widehat{B\Psi}(p)=\sum_{k=0}^\infty \widehat{B_k\Psi}(p)=\sum_{q\in\Z^d}\sum_{k=0}^\infty B_k(p,q)\hat\Psi(q).
\eeq

The expression of $B_k$ contains $k-1$ products of the operator $\lambda\,\Lambda f_{\psi,\psi}$ with itself,
which appear as convolutions in terms of Fourier transforms. Explicitly, 
\begin{align}
B_k(p,q)=\lambda^k \sum_{q_i\in\Z^d}\Lambda(p) \hat f_{\psi,\psi}(p-q_1)\cdots \Lambda(q_{k-1}) \hat f_{\psi,\psi}(q_{k-1}-q).\label{eq:B_k}
\end{align}
Using the bound $\abs{\Lambda(p)\hat f_{\psi,\psi}(q)}\leq Cg^{-2}\, e^{-\sigma\abs{q}}$ 
we see that, for $0<\sigma'<\sigma$,
\begin{align*}
\abs{B_k(p,q)}  \leq \left(Cg^{-2}\abs{\lambda}\right)^k e^{-\sigma'\abs{p-q}}\sum_{q_i\in\Z^d}e^{-(\sigma-\sigma')(\abs{p-q_1}+\dots+\abs{q_{k-1}-q})} \leq \left(C\abs{\epsilon}\right)^k e^{-\sigma'\abs{p-q}}.
\end{align*}
Thus, choosing $\epsilon$ appropriately small we make the geometric series arising in \eqref{eq:Bker} convergent and obtain
\beqn
\abs{B(p,q)},\,\abs{J(p,q)}\leq C\,e^{-\sigma'\abs{p-q}}
\eeqn
with the aid of \eqref{eq:Oker} in \eqref{eq:Jker}.
Finally, 
\beq\label{eq:H_2bound}
\abs{H_2(p,q)}\leq C g^{2}\abs{\epsilon}^2e^{-\sigma'\abs{p-q}}.
\eeq

Exploiting \eqref{eq:t_sf=f}, we compute
\beq\label{eq:t_sH_2}
t_sH_2=\lambda\,t_s( f_{\phi,\psi}J)=\lambda\,f_{\phi,\psi}\,t_sJ=\lambda\,f_{\phi,\psi}\,(t_sB)(t_s\Lambda)O.
\eeq
With the aid of \eqref{eq:Bneumann} and \eqref{eq:t_sLambda}, $t_sB_k=\lambda^k \left[\Lambda(\omega\cdot s) f_{\psi,\psi}\right]^k$.
Thus, $(t_sB_k)(p,q)$ depends on $s$ only through $\omega\cdot s$. Moreover, the dependence on $\omega\cdot s$ is analytic in a neighbourhood of the real line:
Consider the shifted quantity
\begin{align*}
B_k(\kappa;p,q)\defas\lambda^k \sum_{q_i\in\Z^d}\Lambda(\kappa;p) \hat f_{\psi,\psi}(p-q_1)\cdots \Lambda(\kappa;q_{k-1}) \hat f_{\psi,\psi}(q_{k-1}-q),
\end{align*}
which for $\kappa=\omega\cdot s$ becomes $(t_sB_k)(p,q)$. The summand above is analytic on 
\beqn
D_g\defas\{\epsilon\;|\;\abs{\epsilon}<\epsilon_0\}\times\{\kappa\;|\;\abs{\impart\kappa}\leq g/3\}\times\{\gamma\;|\;\abs{\gamma-g}<g/2\},
\eeqn
and the sum converges uniformly, as is readily observed after recalling the bound \eqref{eq:Lambdakbound} on $\Lambda(\kappa;q)$ and looking at the estimation of $\abs{B_k(p,q)}$. Thus, $B_k(\kappa;p,q)$ is analytic. But the Neumann series $\sum_{k=0}^\infty B_k(\kappa;p,q)$ also converges uniformly, making the limit $B(\kappa;p,q)$ analytic on  $D_g$. Evidently,
$
(t_sB)(p,q)=B(\omega\cdot s;p,q).
$
The kernel $B_k(\kappa;p,q)$ defines an operator $B(\kappa)$. Motivated by equation~\eqref{eq:t_sH_2}, we extend the definition of $H_2$ and set $H_2(\kappa)\defas \lambda f_{\phi,\psi}B(\kappa)\Lambda(\kappa)O$.
Using \eqref{eq:Lambdakbound}, a straightforward computation shows that also $H_2(\kappa;p,q)$ obeys \eqref{eq:H_2bound} and is analytic on $D_g$. Furthermore,
\beqn
(t_sH_2)(p,q)=H_2(\omega\cdot s;p,q).
\eeqn

Recalling the translation invariance \eqref{eq:t_sH_1} of $H_1$, we simply take $H(\kappa)\defas H_1+H_2(\kappa)$.

The bound on the derivative $H^{(k)}(\kappa;p,q)$ is achieved by a Cauchy estimate. To that end, one observes
$
H'(\kappa)=H_2'(\kappa) 
$
and uses the bound \eqref{eq:H_2bound} on $D_g$. Similarly, because $X_0$ is independent of $\gamma$, $\de H/\de\gamma=\de H_2/\de\gamma$, and we get the bound on $\de H(0;0,0)/\de\gamma$.

The constants above are independent of $g$, as long as $0<g<g_0$. That is to say, the estimates hold on $\bigcup_{0<g<g_0}D_g = \bigl\{(\kappa,\epsilon,g,\gamma)\;\big|\;\abs{\impart{\kappa}}\leq g/3,\, (\epsilon,g)\in D,\,\abs{\gamma-g}<g/2\bigr\}$. 
\end{proof}

\subsection[Linearized invariant manifolds]{Linearized invariant manifolds: rudiments of renormalization}
We now proceed to stating the main theorem of this section, discussing the linearization $X_1$. Our proof is based on a Renormalization Group (RG) technique we present below.

\begin{theorem}\label{thm:linearization}
Under the assumptions of Theorem~\ref{thm:manifolds}, there exist a number $\epsilon_0$ and a map $\gamma=\gamma(\epsilon,g)$ on $D$, analytic in $\epsilon$, with $\abs{\gamma-g}\leq Cg\abs{\epsilon}$, such that equation~\eqref{X1} has a nontrivial solution $X_1$ which is
\begin{enumerate}
\item analytic in $\abs{\epsilon}<\epsilon_0$ and
\item analytic in $\theta$ in a complex neighbourhood $\,\mathcal{U}$ of $\,\torus$,
\end{enumerate}
and satisfies the physical constraint
\beqn
\Phi_1\unperte\equiv 4=\average{\Phi_1}.
\eeqn
Furthermore, it is real-valued if $\epsilon$ and $\theta$ are real, and
\beqn
\sup_{\theta\in \mathcal{U}}\,\abs{\Psi_1(\theta)}\leq C\abs{\epsilon}\mathand\sup_{\theta\in \mathcal{U}}\,\abs{\Phi_1(\theta)-4}\leq Cg\abs{\epsilon}.
\eeqn
The map $\gamma$ is independent of $\average{\Psi_0}$. If $X_1$ and $X_1'$ correspond to $X_0$ and $X_0'$ of Theorem~\ref{thm:tori}, respectively, with $\average{\Psi_0}=0$ and $\average{\Psi_0'}=\beta\in\R^d$, then
\beqn
X_1'(\theta)\equiv X_1(\theta+\beta).
\eeqn
\end{theorem}
\begin{remark}
We chose the normalization $4$, because $X^0(z,\theta) =(4\arctan z,0)$ is the unperturbed solution (separatrix) and $\arctan{z}=z+\order{z^3}$. 
\end{remark}
\begin{remark}\label{rem:uniqueness}
The pair $(\gamma,X_1)$ of Theorem~\ref{thm:linearization} is unique in the sense that it is the only one making \emph{our construction} work, which is manifested by Lemma~\ref{lem:flowcontrol} below. We do not prove the uniqueness of $X_1$. However, for a given solution $X_1$ the value of $\gamma$ is unique: If $\gamma'$ is another one, \eqref{X1} yields $(\D+\gamma)^2X_1=(\D+\gamma')^2X_1$ because $D\Omega(X_0)$ is independent of $\gamma$. This shows that $\gamma'=\gamma$, because $\hat \Phi_1(0)=4\neq 0$. 
\end{remark}

Let us commence sketching the backbone of Theorem~\ref{thm:linearization} by recalling equation~\eqref{eq:phi1}. We expand the square on the left-hand side and obtain
\beq\label{eq:phi1mod}
(\D^2+2\gamma\D)\Phi_1=(H+g^2-\gamma^2)\Phi_1.
\eeq

For a small $\epsilon\neq 0$, $\Phi_1$ should remain close to the unperturbed value $4=\de_z\Phi^0(0)$. Due to the linearity of \eqref{eq:phi1mod} such a solution may be normalized as $\average{\Phi_1}=4$. Thus, we set
\beq\label{eq:phi=4+xi}
\Phi_1(\theta)=4+\xi(\theta),
\eeq
where we \emph{demand} the function $\xi:\torus\to\R$ to vanish on the average, \ie,
\beq\label{eq:xi(0)=0}
\hat\xi(0)=0.
\eeq
Plugging \eqref{eq:phi=4+xi} into \eqref{eq:phi1mod} results in
\beqn
(\D^2+2\gamma\D)\xi=\pi_0(\xi+4),\quad\text{where}\quad\pi_0\defas H+g^2-\gamma^2.
\eeqn
After switching into Fourier representation, this reads
\begin{align}
\hat\xi(q) &= G(q)\Bigl[\sum_{p\in\Z^d}\pi_0(q,p)\hat\xi(p)+\hat\rho_0(q)\Bigr]\quad\text{if $q\in\Z^d\nonzero$},\label{eq:xifourier} \\
0 &= \sum_{p\in\Z^d}\pi_0(0,p)\hat\xi(p)+\hat\rho_0(0),\label{eq:xifourier0}
\end{align}
where $\rho_0$ is a function defined through its Fourier transform by setting
\beq\label{eq:rhozero}
\hat\rho_0(q)\defas 4\pi_0(q,0).
\eeq
The symbol $G(q)$ stands for the diagonal element $G(q,q)$ of the operator $G$ whose Fourier kernel is given by
\beq\label{eq:Gker}
G(p,q)\defas\delta_{p,q}
\begin{cases}
\left(2i\gamma\,\omega\cdot q -(\omega\cdot q)^2\right)^{-1} & \text{if $q\in\Z^d\nonzero$}, \\
0 & \text{if $q=0$}.
\end{cases}
\eeq

The matter of the fact is that, in terms of our new notations, any solution $\xi$ of 
\beq\label{eq:xi}
\xi=G(\pi_0\xi+\rho_0)
\eeq
also solves \eqref{eq:xifourier}; only the zero mode constraint \eqref{eq:xi(0)=0} has been included here. After finding such a $\xi$, we go on to show that it is a solution to \eqref{eq:xifourier0}, as well.

As is apparent from the definition of $G$, this problem involves arbitrarily small denominators $\omega\cdot q$. Our strategy is to recursively decompose $G$ into parts, each of which corresponds to denominators up to a given order of magnitude. We then end up solving ``partial problems'' of \eqref{eq:xi} scale by scale, and show that these solutions converge to a true solution of \eqref{eq:xi} as the recursion proceeds and smaller and smaller denominators become dealt with.

\begin{figure}[!ht]
\psfrag{x}{$\chi$}
\psfrag{y}{}
\epsfig{file=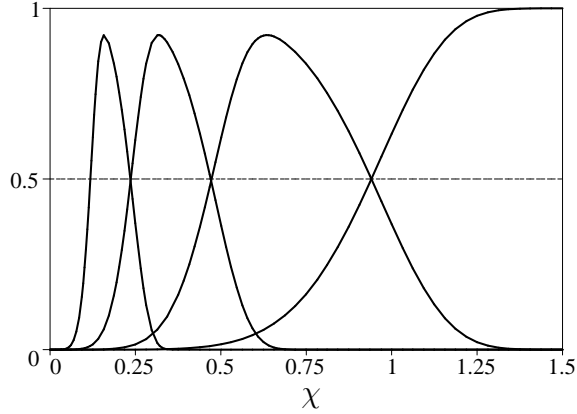,width=0.5\linewidth}
\caption{Graphs of $\chi_n-\chi_{n+1}$ with $n=0,1,2,3$, and  $\aleph=\half$. The maxima are located roughly at $\aleph^{n}$.}
\label{fig:pofunity}
\end{figure}

Leaving the all-important scaling parameter $\aleph\in{]0,1[}$ to be decided later\footnote{Aleph, $\aleph$, is the first letter in the Hebrew alphabet.}, we shall need the entire functions
\beqn
\chi_n:\C\to\C:\chi_n(\kappa)=
\begin{cases}
e^{-(\aleph^{-n}\kappa)^6} & \text{if $n\in\Z_+$}, \\
1 & \text{if $n=0$}.
\end{cases}
\eeqn
Their importance lies in the fact that the sequence $(\chi_n-\chi_{n+1})_{n\in\N}$ of functions is an analytic partition of unity on $\R\nonzero$; on this set
$
0\leq 1-\chi_{N}\nearrow 1$ pointwise, as $
\quad N\to\infty.
$
Some of the first members of the sequence appear plotted in Figure~\ref{fig:pofunity}.
The number 6 in the exponent is a choice of convenience; it is the one used in \cite{Kupiainen}. 

Let us now introduce the diagonal operators $G_n$ and $\Gamma_n$, $n\in\N$, defined by
\beqn
G_n(q)\defas\chi_n(\omega\cdot q)G(q)\mathand\Gamma_n\defas G_n-G_{n+1},
\eeqn
respectively. Observe that $G_0=G$ and $G_n(0)=0$. The point here is that in $\Gamma_n(q)$ the functions $\chi_n(\omega\cdot q)-\chi_{n+1}(\omega\cdot q)$ act as cutoffs for the values of $\omega\cdot q$. Each $\Gamma_n$ deals with the denominators $\omega\cdot q$ that are roughly of order $\aleph^n$ and, intuitively, 
\beq\label{eq:Gamma<n}
\Gamma_{<n}\defas\sum_{k=0}^{n-1}\Gamma_k=G-G_n
\eeq
gets closer and closer to $G$ as $n$ tends to infinity. Instead of the full equation \eqref{eq:xi}, consider the easier, approximate problem
\beq\label{eq:xiappr}
 x_n=\Gamma_{<n}(\pi_0x_n+\rho_0),
\eeq
obtained by replacing $G$ with $\Gamma_{<n}$. It is easier since $\Gamma_{<n}$ discards the most dangerous ones of the small denominators. However, its solution should become a better and better approximation of the solution of \eqref{eq:xi} with increasing $n$.

Having $G_0=G_1+\Gamma_0$, we decompose $\xi=\xi_1+\eta_0$ and assume that $\eta_0=\eta_0(\xi_1)$ solves the ``large denominator problem''
\beq\label{eq:eta_apriori}
\eta_0=\Gamma_0(\pi_0(\xi_1+\eta_0)+\rho_0).
\eeq
Then, solving the original problem \eqref{eq:xi} for $\xi$ amounts to solving
\beq\label{eq:xi1temp}
\xi_1=G_1(\pi_0(\xi_1+\eta_0)+\rho_0)
\eeq
for $\xi_1$.

Assuming $\one-\Gamma_0\pi_0$ is invertible\footnote{Think of $\Gamma_0$ as comprising only large denominators and $\pi_0$ being proportional to $\epsilon$.}, we can extract $\eta_0$ out of \eqref{eq:eta_apriori} and get
\beq\label{eq:eta}
\eta_0=(\one-\Gamma_0\pi_0)^{-1}\Gamma_0(\pi_0\xi_1+\rho_0).
\eeq
Therefore, \eqref{eq:xi1temp} transforms into
\beqn
\xi_1=G_1(\one-\pi_0\Gamma_0)^{-1}(\pi_0\xi_1+\rho_0)
\eeqn
with the aid of the identities
\beqn
\pi_0(\one-\Gamma_0\pi_0)^{-1}=(\one-\pi_0\Gamma_0)^{-1}\pi_0
\mathand
(\one-\pi_0\Gamma_0)^{-1}\pi_0\Gamma_0=(\one-\pi_0\Gamma_0)^{-1}-\one.
\eeqn
Thus, defining the new objects
\beqn
\pi_1\defas(\one-\pi_0\Gamma_0)^{-1}\pi_0\mathand
\rho_1\defas(\one-\pi_0\Gamma_0)^{-1}\rho_0,
\eeqn
we obtain
\beqn
\eta_0=\Gamma_0(\pi_1\xi_1+\rho_1)
\eeqn
and
\beq\label{eq:xi1}
\xi_1=G_1(\pi_1\xi_1+\rho_1).
\eeq

Indeed, equation~\eqref{eq:xi1} has \emph{precisely the same form} as the original problem \eqref{eq:xi}. Now, relaxing the assumption that $\eta_0$ be \textit{a priori} known, suppose we are able to solve \eqref{eq:xi1}, and take \eqref{eq:eta} as the definition of $\eta_0$, instead. Then the solution of the full problem is recovered using the simple relation
\begin{align*}
\xi & =\xi_1+\eta_0=(\one-\Gamma_0\pi_0)^{-1}(\xi_1+\Gamma_0\rho_0).
\end{align*}

Owing to the aforementioned formal covariance between equations \eqref{eq:xi} and \eqref{eq:xi1}, we may iterate the construction above. Thus, in general, solving
\beq\label{eq:xi_general}
\xi_{n+1}=G_{n+1}(\pi_{n+1}\xi_{n+1}+\rho_{n+1})
\eeq
for $\xi_{n+1}$ with the definitions
\begin{eqnarray}
\pi_{n+1} & \defas & (\one-\pi_{n}\Gamma_{n})^{-1}\pi_{n},\label{eq:pirec} \\
\rho_{n+1} & \defas & (\one-\pi_{n}\Gamma_{n})^{-1}\rho_{n},\label{eq:rhorec}\\
\eta_{n} & \defas & \Gamma_n(\pi_{n+1}\xi_{n+1}+\rho_{n+1}),\label{eq:etarec}
\end{eqnarray}
produces $\xi_n=\xi_{n+1}+\eta_n$, or
\beq\label{eq:previous_xi}
\xi_n=(\one-\Gamma_n\pi_n)^{-1}\left(\xi_{n+1}+\Gamma_n\rho_{n}\right)
\eeq
for the solution of $\xi_n=G_n(\pi_n\xi_n+\rho_n)$.

Equations \eqref{eq:previous_xi} and \eqref{eq:rhorec} reveal through
\begin{align*}
\pi_n\xi_n+\rho_n&=\pi_n\left[(\one-\Gamma_n\pi_n)^{-1}\xi_{n+1}+\Gamma_n\rho_{n+1}\right]+(\one-\pi_n\Gamma_n)\rho_{n+1} 
\end{align*}
the recursion invariance
\beq\label{eq:recinv}
\pi_0\xi_0+\rho_0=\pi_1\xi_1+\rho_1=\dots=\pi_n\xi_n+\rho_n=\cdots
\eeq
in our construction.

Let us tidy up the notation by giving the definitions
\beq\label{eq:v&f}
v_n(y)\equiv\pi_n y+\rho_n\mathand f_n\defas\one+\Gamma_{<n}v_n\mathwith\Gamma_{<0}=0.
\eeq
In particular, \eqref{eq:recinv} takes the form $v_n(\xi_n)=v_0(\xi_0)$. We also set
\beq\label{eq:Xidef}
\Xi_n(y)\equiv(\one-\Gamma_n\pi_n)\inv\left(y+\Gamma_n\rho_n\right),
\eeq
such that \eqref{eq:previous_xi} reads $\xi_n=\Xi_n(\xi_{n+1})$, and \eqref{eq:recinv} reduces to
\beq\label{eq:vrec}
v_{n+1}=v_n\circ\Xi_n.
\eeq
The latter is a convenient way of writing
$
v_{n+1}=(\one-\pi_n\Gamma_n)\inv v_n.
$
Notice also that $\Xi_n$ is formally invertible. 

One easily verifies
\beq\label{eq:Xi}
\Xi_n=\one+\Gamma_n v_{n+1}.
\eeq
As a consequence,
\beq\label{eq:frec}
f_{n+1}=f_n\circ\Xi_n.
\eeq
Since $f_0=\one$, we have the cumulative formula
\beq\label{eq:fcum}
f_n=\,\Xi_0\circ\Xi_1\circ\dots\circ\Xi_{n-1}.
\eeq
Hence, a similar expansion of \eqref{eq:vrec} implies
\beqn
v_n=v_0\circ f_n.
\eeqn
Inserting here the definition of $f_n$, we get
\beq\label{eq:vfp}
v_n=v_0\circ(\one+\Gamma_{<n}v_n).
\eeq

\begin{proposition}\label{prop:xi}
Let $\xi_n=\Xi_n(\xi_{n+1})$. If $\xi_{n+1}$ satisfies $\xi_{n+1}=G_{n+1}(\pi_{n+1}\xi_{n+1}+\rho_{n+1})$, then $\xi_n$ satisfies $\xi_n=G_n(\pi_n\xi_n+\rho_n)$, and vice versa.
\end{proposition}
\begin{proof}
Suppose $\xi_{n+1}=G_{n+1}v_{n+1}(\xi_{n+1})$. By $G_n=G_{n+1}+\Gamma_n$ and \eqref{eq:vrec},
\beqn
G_nv_n\circ\Xi_n=G_{n+1}v_{n+1}-\one+\one+\Gamma_nv_{n+1}.
\eeqn
But, with the aid of \eqref{eq:Xi}, this transforms into
\beqn
(G_nv_n-\one)\circ\Xi_n=G_{n+1}v_{n+1}-\one.
\eeqn
As $\Xi_n$ is invertible with $\xi_n=\Xi_n(\xi_{n+1})$, the identity above proves the formal equivalence of the small denominator problems \eqref{eq:xi_general}, or $G_nv_n(\xi_n)=\xi_n$, with differing $n$.
\end{proof}

Recalling \eqref{eq:fcum}, we immediately arrive at

\begin{corollary}\label{cor:xi}
If $\xi_n=G_n(\pi_n\xi_n+\rho_n)$, then 
\beqn
\xi_0\defas f_n(\xi_n)=\xi_n+\Gamma_{<n}v_n(\xi_n) 
\eeqn
solves the complete problem: $\xi_0=G_0(\pi_0\xi_0+\rho_0)$.
\end{corollary}

\begin{remark}
The solution $\xi_0$ above comprises two terms having clear interpretations. The first term, $\xi_n$, solves the small denominator problem, namely $\xi_n=G_n(\pi_n\xi_n+\rho_n)$, at the $n\text{th}$ step. The second term, $\Gamma_{<n}v_n(\xi_n)$, on the other hand, consists of the sum
\beqn
\eta_{<n}(\xi_n)\defas\sum_{k=0}^{n-1}\eta_k(\xi_{k+1})\mathwith\xi_{k+1}=\left(\Xi_{k+1}\circ\dots\circ\Xi_{n-1}\right)(\xi_n),
\eeqn
where $\eta_k=\eta_k(\xi_{k+1})$ solves the large denominator problem $\eta_k=\Gamma_kv_k(\xi_{k+1}+\eta_k)$ in analogy with \eqref{eq:eta_apriori}. Indeed, $\Gamma_kv_k(\xi_{k+1}+\eta_k)=\Gamma_k v_k(\xi_k)=\Gamma_k v_{k+1}(\xi_{k+1})=\eta_k$.
\end{remark}

Finally, we make a crucial observation. If we operate on \eqref{eq:vfp} by $\Gamma_{<n}$ and set 
\beq\label{eq:x_n}
x_n\defas{f_n(0)}=\Gamma_{<n}v_n(0),
\eeq
we solve the approximate problem \eqref{eq:xiappr}:
\beqn
x_n=\Gamma_{<n}(\pi_0x_n+\rho_0).
\eeqn
We shall demonstrate that the approximate solutions $x_n$ form a Cauchy sequence in a simple Banach space, and that their limit
\beq\label{eq:xidef}
\xi\defas\lim_{n\to\infty}x_n
\eeq
solves the original equation \eqref{eq:xi}.

Just to motivate the above discussion, think of an abstract map $\mathcal{R}_n$ that takes $(\pi_n,\rho_n,G_n)$ to $(\pi_{n+1},\rho_{n+1},G_{n+1})$. The recursion scheme
{\small
\beqn
\xi=G(\pi_0\xi+\rho_0)\overset{\mathcal{R}_0}{\mapsto}\xi_1=G_1(\pi_1\xi_1+\rho_1)\overset{\mathcal{R}_1}{\mapsto}\cdots\overset{\mathcal{R}_{n-1}}{\mapsto}\xi_n=G_n(\pi_n\xi_n+\rho_n)\overset{\mathcal{R}_n}{\mapsto}\cdots
\eeqn
}%
is called \emph{renormalization} of the problem, and $\mathcal{R}_n$ is the corresponding renormalization transformation. Then, in view of Proposition~\ref{prop:xi}, it remains for one to demonstrate that this process ``converges'', in order to be able to solve the original equation $\xi=G(\pi_0\xi+\rho_0)$. That is to say, one wishes that the \emph{renormalization flow} of the triplet $(\pi_0,\rho_0,G_0)$, $(\pi_n,\rho_n,G_n)=(\prod_{k=0}^{n-1}\mathcal{R}_k)(\pi_0,\rho_0,G_0)$, in a sense tends to a fixed point $(\pi^*,\rho^*,G^*)$ of some limiting operator ``$\mathcal{R}_\infty=\lim_{k\to\infty}\mathcal{R}_k$'' as $n\to\infty$, and that the equation
\beq\label{eq:xi_star}
\xi^*=G^*(\pi^*\xi^*+\rho^*)
\eeq
is well-defined and solvable. 

In our case $G^*\rho^*=0$, such that the equation is linear and possesses the trivial solution $\xi^*=0$. Corollary~\ref{cor:xi} then throws light on why \eqref{eq:xidef} should solve \eqref{eq:xi}; $f_n(\xi_n)$ solves it, and $\xi_n$ approaches zero with increasing $n$. Therefore, it is fair to expect that also $\lim_{n\to\infty}f_n(0)$ is a solution.

\subsection{Banach spaces}\label{subsec:rigorousRG}
Technically speaking, we need to control the renormalization flow \eqref{eq:pirec}--\eqref{eq:etarec} by estimating the kernel elements of $\Gamma_n$ and $\pi_n$, for the operators $\one-\pi_n\Gamma_n$ and $\one-\Gamma_n\pi_n$ had better be invertible between suitable spaces. Such Banach spaces will be defined in this subsection.

We begin by analyzing the properties of the operators $\Gamma_n$. \emph{A priori}, one expects the most significant contribution to arise from such $q$'s that $\omega\cdot q=\order{\aleph^{n}}$, due to the cutoff $\chi_n-\chi_{n+1}$ in the definition of these operators. Therefore, \eqref{eq:Gker} implies
\beq\label{eq:gamma_heur}
\abs{\Gamma_n(q)}=\order{g\inv \aleph^{-n}}.
\eeq
More accurately, it is fairly easy to obtain
\beq\label{eq:chibound}
\abs{\chi_n(\kappa)-\chi_{n+1}(\kappa)}\leq C\abs{\aleph^{-n}\kappa}^\ell
\begin{cases}
e^{-\half\abs{\aleph^{-n}\kappa}^6} & \text{if $n\geq 1$},\\
1 \, & \text{if $n=0$},
\end{cases}
\eeq
for $\ell=0,1,\dots,6$, in a strip $\abs{\impart\kappa}<\aleph^{n}b$.  $C$ only depends on the parameter $b$. Pay attention to the fact that $G(q)$, which was defined in \eqref{eq:Gker}, only depends on $q$ through $\omega\cdot q$. Therefore, it is handy to introduce the analytic function
\beqn
\iota:\C\setminus\left\{0,2i\gamma\right\}\to\C:\iota(\kappa)=(2i\gamma\kappa-\kappa^2)^{-1}.
\eeqn
In particular, $\iota(\omega\cdot q)=G(q)$ for $q\neq 0$. This motivates the further definition 
\beqn
\Gamma_n(\kappa;p,q)\defas\delta_{p,q}
{
\begin{cases}
[\chi_n(\omega\cdot q+\kappa)-\chi_{n+1}(\omega\cdot q+\kappa)]\iota(\omega\cdot q+\kappa) & \text{if $\omega\cdot q+\kappa\neq 0$},\\
0 & \text{if $\omega\cdot q+\kappa=0$}.
\end{cases}
}%
\eeqn
The importance of the resulting operator $\Gamma_n(\kappa)$ is based on the possibility of viewing $\omega\cdot q$ as a complex ``variable'':
\beqn
\Gamma_n(q,q)=\Gamma_n(0;q,q)=\Gamma_n(\omega\cdot q;0,0).
\eeqn
We shall often write $\Gamma_n(\kappa;q)$ instead of the complete $\Gamma_n(\kappa;q,q)$.

Imposing the condition $b\leq g/2$ on $b$ we get within $\abs{\impart\kappa}<\aleph^{n}b$ that
{
\begin{align}
\abs{\Gamma_n(\kappa;q)} = 
\aleph^{-n}\frac{\abs{\chi_n(\omega\cdot q+\kappa)-\chi_{n+1}(\omega\cdot q+\kappa)}}{\abs{\aleph^{-n}(\omega\cdot q+\kappa)}}\,\abs{(\omega\cdot q+\kappa)\,\iota(\omega\cdot q+\kappa)} 
\notag \\
\leq C_\Gamma\,g\inv \aleph^{-n}\min(1,\abs{\aleph^{-n}(\omega\cdot q+\kappa)}^5)
\begin{cases}
e^{-\half\abs{\aleph^{-n}(\omega\cdot q+\kappa)}^6} & \text{if $n\geq 1$},\\
1\, & \text{if $n=0$},
\end{cases}\label{eq:Gammakernel}
\end{align}
}%
making use of \eqref{eq:chibound}. In particular, we have confirmed the heuristic estimate \eqref{eq:gamma_heur}.

Now to the spaces promised. Ultimately the solution of \eqref{eq:xi}, namely $\xi$ (and therefore $\Phi_1$) will live in the space $\Bphi{\alpha^*}\subset\ell^1(\Z^d;\C)$ for a sufficiently small width $\alpha^*$ of the analyticity strip---see Subsection~\ref{subsec:spaces}. The following weights will come in handy:
\beq\label{eq:weights}
w_n(q)\defas 
\begin{cases}
e^{\aleph^{-n}\abs{\omega\cdot q}} & \text{if $n\geq 1$},\\
1 & \text{if $n=0$}.
\end{cases}
\eeq
We extend these to negative indices by setting
$
w_{-n}(q)\equiv w_n(q)\inv.
$

\begin{definition}[Spaces $h_n$]
For $n\in\Z$, let
\beqn
\normw{\xi}{n}\defas\sum_{q\in\Z^d}\abs{\hat\xi(q)}w_n(q).
\eeqn
These norms induce the Banach spaces $h_n$. Observe that $h_0$ is the space $\ell^1(\Z^d;\C)$.
\end{definition}

Notice that our weights satisfy
\beq\label{eq:weightproperty}
w_{n+1}(q)^{\aleph}=w_n(q)\mathand w_n(q)\geq 1\qquad(n\geq 1).
\eeq
The spaces at hand thus realize the embedding hierarchy
\beqn
h_{n+1}\subset h_n\qquad(n\in\Z)
\eeqn
due to the trivial inequalities
\beq\label{eq:normineq}
\normw{\piste}{n}\leq\normw{\piste}{n+1}\qquad(n\in\Z).
\eeq

Operator norms $\normw{\piste}{\el(h_n;h_m)}$ between such spaces $h_n$ and $h_m$ will be denoted by $\normw{\piste}{n;m}$ for short. We actually have, 
\begin{align}\label{eq:operatornorm}
\normw{L}{n;m}  =\sup_{q\in\Z^d}\sum_{p\in\Z^d}\abs{L(p,q)}w_m(p)w_{-n}(q)\qquad(m,n\in\Z).
\end{align}
Either from this or from \eqref{eq:normineq} by the Schwarz inequality one infers
\beqn\label{eq:opnorms}
\normw{\piste}{n+1;m}\leq\normw{\piste}{n;m}\leq\normw{\piste}{n;m+1}\qquad(m,n\in\Z)
\eeqn
such that the operator spaces satisfy
\beqn
\el(h_{n};h_{m+1})\subset\el(h_{n};h_{m})\subset\el(h_{n+1};h_{m})\qquad(m,n\in\Z).
\eeqn
Moreover, the Schwarz inequality implies the useful bounds
\beq\label{eq:normofproduct}
\normw{L_1L_2}{n;m}\leq\normw{L_1}{l;m}\normw{L_2}{n;l}\qquad(l,m,n\in\Z).
\eeq

From now on, $n$ will always assume \emph{nonnegative} values. Define the domain
\beq\label{eq:D_n}
D_n\defas\left\{\kappa\in\C\,\big|\,\abs{\kappa}<\aleph^n b\right\},
\eeq
recalling that $b\leq g/2$. Then \eqref{eq:Gammakernel} easily validates the bounds
\beq\label{eq:Gamma_n}
\normw{\Gamma_{n}(\kappa)}{-n;n}\leq C_\Gamma\,g\inv\aleph^{-n}
\eeq
for $\kappa\in D_{n}$, where the (new) constant $C_\Gamma$ is independent of $\kappa$ and $g$ as long as $g<g_0$.
This shows, in particular, that
\beqn
\Gamma_{n}(\kappa)\in\el(h_{-n};h_n)\subset\el(h_0;h_0).
\eeqn

\begin{remark}\label{rem:weights}
The weights $w_n(q)$ arise as follows. The diagonal kernel of $\Gamma_n$ is strongly concentrated around small denominators $\omega\cdot q$ of order $\aleph^{n}$; for large $\omega\cdot q$ the value of $\Gamma_n(q)$ is very close to zero, \emph{but not quite equal to zero}. Therefore, in an expression such as $\widehat{\Gamma_n\xi}(q)=\Gamma_n(q)\hat\xi(q)$ we cannot let $\abs{\hat\xi(q)}$ be \emph{arbitrarily} large for large values of $\omega\cdot q$. This ``tail'' can be of the order of $w_n(q)=e^{\aleph^{-n}\abs{\omega\cdot q}}$, say, which amounts to $\xi\in h_{-n}$. 

It has to be emphasized that having the same power of $\aleph^{-n}$ and $\abs{\omega\cdot q}$ in $w_n(q)$ is crucial, which can be read off from \eqref{eq:Gammakernel}. This way $\omega\cdot q$ ``scales'' as $\aleph^n$ in all estimates in the $n$th step of the iteration.

The motivation for introducing the spaces $h_n$, on the other hand, comes from the fact that in the recursion \eqref{eq:pirec} the domain of $\pi_n$ will shrink. So, in the norms $\normw{\piste}{n}$ we incorporate a weight that increases as $n$ grows. It is a matter of convenience to use the inverse of the weight $w_n(q)^{-1}$ appearing in $\normw{\piste}{-n}$.
\end{remark}

\subsection[Renormalization: estimates and the Lyapunov exponent]{Renormalization made rigorous: estimates and the Lyapunov exponent}\label{subsec:RGestimates}
The rest of this section is devoted to demonstrating that the renormalization flow of $\pi_n$ in \eqref{eq:pirec} is controlled in the norms $\normw{\piste}{n;-n}$ such that the products $\normw{\pi_n}{n;-n}\normw{\Gamma_n}{-n;n}$ are small, so as to make the recursion formulae \eqref{eq:pirec}--\eqref{eq:etarec} well-defined through Neumann series. Recalling \eqref{eq:Gamma_n}, the task roughly amounts to making sure that $\normw{\pi_n}{n;-n}$ decays at least as rapidly as $\aleph^{n}$ with increasing $n$.

According to Lemma~\ref{lem:H}, $\pi_0=H+g^2-\gamma^2\in\el(\Bphi{\sigma})$ can be written as
\beqn
\pi_0(p,q)=p_0(\omega\cdot q)\delta_{p,q}+\tilde\pi_0(p,q), 
\eeqn
where $\tilde\pi_0$ vanishes on the diagonal, and in the first term
\beqn
p_0(\kappa)\defas \delta_0+\bar p_0(\kappa),\qquad \bar p_0(0)=0,
\eeqn
depends analytically on $\kappa$, as long as $\abs{\impart\kappa}\leq g/3$; explicitly $\delta_0=H(0;0,0)+g^2-\gamma^2$ and $\bar p_0(\kappa)=H(\kappa;0,0)-H(0;0,0)$. 

Similarly, we split $\pi_n$ into its diagonal and off-diagonal parts:
\beqn
\pi_n(p,q)=p_n(\omega\cdot q)\delta_{p,q}+\tilde\pi_n(p,q), \qquad \tilde\pi_n(q,q)=0,
\eeqn
with
\beqn
p_n(\kappa)=\delta_n+\bar p_n(\kappa),\qquad\bar p_n(0)=0.
\eeqn
The possibility of doing this follows from the computation
\begin{align*}
t_s\pi_0=t_sH+g^2-\gamma^2=H(\omega\cdot s)+g^2-\gamma^2\defasr\pi_0(\omega\cdot s)
\end{align*}
and its recursive consequence
\begin{align*}
t_s\pi_{n+1}=\left(\one-\pi_n(\omega\cdot s)\, \Gamma_n(\omega\cdot s)\right)^{-1}\pi_n(\omega\cdot s)
\defasr\pi_{n+1}(\omega\cdot s).
\end{align*}

Motivated by the computation above, let us inductively define the maps
\beq\label{eq:pirec2}
\pi_{n+1,\beta}(\kappa)\defas\left(\one-\pi_{n\beta}(\kappa)\, \Gamma_{n}(\kappa)\right)^{-1}\pi_{n\beta}(\kappa),\qquad \kappa\in D_n,\quad\abs{\impart\beta}<\alpha_n,
\eeq
starting at 
\beqn
\pi_{0\beta}(\kappa)\defas P_0(\kappa)+\tilde\pi_{0\beta}(\kappa),\qquad \kappa\in D_0,\quad\abs{\impart{\beta}}<\alpha_0,
\eeqn
by setting $b\leq g/3$ in \eqref{eq:D_n}. Here
\begin{align*}
P_0(\kappa;p,q)&\defas\, p_0(\kappa+\omega\cdot q)\delta_{p,q},\\
\tilde\pi_{0\beta}(\kappa;p,q)&\defas\, e^{i\beta\cdot(p-q)}H(\kappa;p,q)(1-\delta_{p,q}),
\end{align*}
and, with $\sigma'$ coming from Lemma~\ref{lem:H},
\beq\label{eq:alpha_n}
\alpha_{n+1}\defas\left(1-\frac{4}{(n+3)^2}\right)\alpha_{n},\qquad\alpha_0< \sigma'.
\eeq
In particular, Eric Weisstein's World of Mathematics \cite{Weisstein} tells us that
\beq\label{eq:alpha_lim}
\alpha_n\searrow\,\alpha_0\cdot\prod_{k=3}^\infty\left(1-\frac{4}{k^2}\right)=\frac{\alpha_0}{6}>0\qquad\text{as}\qquad n\to\infty.
\eeq

As far as notation is concerned, we may omit $\beta$ if it equals zero: $\pi_n(\kappa)\equiv \pi_{n0}(\kappa)$, and so forth. By a straightforward induction argument, 
\beqn
\pi_{n\beta}(\kappa;p,q)\defas e^{i\beta\cdot(p-q)}\pi_n(\kappa;p,q),
\eeqn
such that $\beta$ does not enter the diagonal of $\pi_{n\beta}$. Of course,
\beq\label{eq:betakernelest}
\abs{\pi_{n\beta}(\kappa;p,q)}= e^{-\impart\beta\cdot (p-q)}\abs{\pi_{n}(\kappa;p,q)}.
\eeq

For clarity, set
\beqn
P_n(\kappa)\defas\delta_n\one+\bar P_n(\kappa)\quad\text{with}\quad \bar P_n(\kappa;p,q)\defas \bar p_n(\kappa+\omega\cdot q)\delta_{p,q},
\eeqn
so that we may express the operator $\pi_{n\beta}(\kappa)$ itself, without reference to its kernel, as
\beqn
\pi_{n\beta}(\kappa)=P_n(\kappa)+\tilde\pi_{n\beta}(\kappa)=\delta_n+\bar P_n(\kappa)+\tilde\pi_{n\beta}(\kappa),\qquad \delta_n\equiv\delta_n\one,
\eeqn
for short. This decomposition satisfies
\begin{align}
\normw{\pi_{n\beta}(\kappa)}{n;-n} 
& \leq\abs{\delta_n}+\normw{\bar P_n(\kappa)}{n;-n}+\normw{\tilde\pi_{n\beta}(\kappa)}{n;-n}.\label{eq:pidecomp}
\end{align}
It will turn out that the sum in \eqref{eq:pidecomp} is finite if $\kappa\in D_n$ and $\abs{\impart\beta}<\alpha_n$---indeed very small, as we are trying to prove---meaning that $\pi_{n\beta}(\kappa)\in\el(h_n;h_{-n})$.

The crux of analyzing the renormalization flow is the following lemma, for which we provide an inductive proof later on in this section. The reader is advised to take the result as granted for now.
\begin{lemma}[Modified Lyapunov exponent controls the flow]\label{lem:flowcontrol}
Set $b= g/3$ and $\aleph=\min(\frac{1}{8},b^2)$. There exist constants $c_\gamma>0$, $C>0$, $c>0$, $\mu>1$, and a unique Lyapunov exponent $\gamma$ satisfying
\beq\label{eq:gammabound}
\abs{\gamma-g}< c_\gamma\abs{\epsilon}g
\eeq
such that, for any $n\in\N$, the bounds
{\small\begin{align}
\normw{\tilde\pi_{n\beta}(\kappa)}{n;-n} & \leq  C\abs{\epsilon}g
\begin{cases} 
g & \text{if $n=0$}, \\
\aleph^n e^{- c\mu^n} & \text{if $n\geq 1$},
\end{cases}\label{eq:tildepi} \\
\normw{\bar P_n(\kappa)}{n;-n}  & \leq  C\abs{\epsilon}g\, 
\begin{cases} 
g & \text{if $n=0$}, \\
\aleph^n & \text{if $n\geq 1$},
\end{cases}\label{eq:tildep}\\
\abs{\delta_n} & \leq C\abs{\epsilon}g\, 
\begin{cases} 
g & \text{if $n=0$}, \\
\aleph^{2n} & \text{if $n\geq 1$},
\end{cases}\label{eq:deltabound}
\end{align}}
hold true for $(\epsilon,g)\in D$, $\kappa\in D_n$ and $\abs{\impart\beta}<\alpha_n$. Moreover, $c$ is bounded away from zero and $\mu\to\infty$ in the limit $g\to 0$.
\end{lemma}

\begin{remark}
The sole purpose of introducing the complex variable $\kappa$ is to go about proving the bound \eqref{eq:tildep} on the \emph{diagonal} part of $\pi_n$. We use analyticity in $\kappa$ and restrict the latter to a domain of ever decreasing size. 

The \emph{possibility} of including the complex parameter $\beta$ in the analysis, on the other hand, facilitates proving exponential decay of $\pi_n(\kappa;p,q)$ in the quantity $\abs{p-q}$. This is sufficiently rapid for obtaining the bound \eqref{eq:tildepi} on the \emph{off-diagonal} part of $\pi_n$. Also the analyticity strip of $\beta$ around $\R$ is taken narrower and narrower upon iteration, but no narrower than a certain limit ($\alpha_0/6$).
\end{remark}

\begin{corollary}\label{cor:pi}
The bounds of Lemma~\ref{lem:flowcontrol} imply
{\small\beqn
\normw{\pi_{n\beta}(\kappa)}{n;-n}\leq C\abs{\epsilon}g
\begin{cases} 
g & \text{if $n=0$} \\
\aleph^n & \text{if $n\geq 1$},
\end{cases}
\eeqn}%
\end{corollary}

The caveat to get around in the proof of Lemma~\ref{lem:flowcontrol} is that $\delta_n$ is reluctant to go to zero along the recursion. To change the state of affairs, we fine-tune the Lyapunov exponent $\gamma$ such that also $\delta_n\to 0$ as $n\to\infty$. As stated in the lemma, there turns out to exist precisely one such value of $\gamma$. This is what ultimately enables us to prove the convergence of our renormalization scheme, consequently validating Theorem~\ref{thm:linearization} discussing the linearized solution $X_1$. For the sake of continuity, we first give the simple proof of Theorem~\ref{thm:linearization} and only then prove Lemma~\ref{lem:flowcontrol}. 

\subsection{Proof of Theorem~\ref{thm:linearization}}
With $x_n$ as in \eqref{eq:x_n}, the task is to show that the limiting function $\xi$---see \eqref{eq:xidef}---is an analytic solution to \eqref{eq:xi}.

Given the formal definition $y_\beta\defas\tau_\beta y$, \eqref{eq:x_n} implies
\beqn
x_{n\beta}= f_{n\beta}(0).
\eeqn
Recalling \eqref{eq:frec} and \eqref{eq:fcum}, one clearly has
\beqn 
f_{n+1,\beta}=f_{n\beta}\circ\Xi_{n\beta}\mathand f_{n\beta}=\Xi_{0\beta}\circ\Xi_{1\beta}\circ\dots\circ\Xi_{n-1,\beta}.
\eeqn
Hence, the recursion relation 
\beqn
x_{n+1,\beta}=x_{n\beta}+\bigl(f_{n\beta}(\Xi_{n\beta}(0))-f_{n\beta}(0)\bigr)
\eeqn
follows. Here \eqref{eq:Xidef} extends to
\beq\label{eq:Xidef2}
\Xi_{n\beta}(y)\equiv (\one-\Gamma_n\pi_{n\beta})\inv(y+\Gamma_n\rho_{n\beta}).
\eeq

Notice that the flows of $\rho_{n\beta}$ and $4\pi_{n\beta}(\piste,0)$ \footnote{$\pi_{n\beta}(\piste,0)$ is shorthand for the function $\theta\mapsto \sum_{q}e^{iq\cdot \theta}\pi_{n\beta}(q,0)$.} are identical. Furthermore, the initial conditions agree according to \eqref{eq:rhozero}, such that
\beqn
\hat\rho_{n\beta}(q)\equiv 4\pi_{n\beta}(q,0).
\eeqn

Because
$
D\Xi_{n\beta}(y)\equiv(\one-\Gamma_n\pi_{n\beta})\inv,
$
the chain rule reveals
\beqn
Df_{n\beta}(y)\equiv(\one-\Gamma_0\pi_{0\beta})\inv(\one-\Gamma_1\pi_{1\beta})\inv\cdots(\one-\Gamma_{n-1}\pi_{n-1,\beta})\inv.
\eeqn
Recursive implementation of Corollary~\ref{cor:pi} in the form
\beqn
\normw{(\one-\Gamma_{n}\pi_{n\beta})\inv}{n;n-1}\leq \normw{(\one-\Gamma_{n}\pi_{n\beta})\inv}{n;n}\leq 2
\eeqn
implies that $Df_{n\beta}(y)\in\el(h_{n-1};h_0)$ with
$
\sup_{y\in h_n}\normw{Df_{n\beta}(y)}{n-1;0}\leq 2^n.
$
By the Mean-Value Theorem we go on to estimate
\beq\label{eq:xdiff}
\normw{x_{n+1,\beta}-x_{n\beta}}{0}\leq 2^n\normw{\Xi_{n\beta}(0)}{n}
\eeq
with the aid of the inequality $\normw{\piste}{n-1}\leq\normw{\piste}{n}$.

\begin{lemma}\label{lem:Xi}
For parameters as in Lemma~\ref{lem:flowcontrol} and $\epsilon_0$ small, we may perceive $\Xi_{n\beta}$ as an analytic map from $h_n$ to $h_n\subset h_{n-1}$ with
{\normalsize\beqn
\normw{\Xi_{n\beta}(0)}{n}\leq C\abs{\epsilon}
\begin{cases}
g & \text{if $n=0$},\\
e^{-c\mu^n} & \text{if $n\geq 1$}.
\end{cases}
\eeqn}%
\end{lemma}

\begin{proof}
Since $\Gamma_n$ annihilates the zero mode ($\Gamma_n(0)=0$), $\Gamma_n\rho_{n\beta}=4(\Gamma_n\tilde\pi_{n\beta})(\piste,0)$, which is super-exponentially small in the norm $\normw{\piste}{n}$ by $\normw{\tilde\pi_{n\beta}(\piste,0)}{-n}\leq\normw{\tilde\pi_{n\beta}}{n;-n}$ and Lemma \ref{lem:flowcontrol}. According to \eqref{eq:Xidef2}, Lemma~\ref{lem:Xi} clearly holds if we take $\epsilon$ small enough so as to validate $\normw{\Gamma_n}{-n;n}\normw{\pi_{n\beta}}{n;-n}\leq \half$, say, for each $n$. For the bounds on $\pi_{n\beta}$ and $\Gamma_n$ we refer the reader to Corollary~\ref{cor:pi} and \eqref{eq:Gamma_n}, respectively. 
\end{proof}

By Lemma~\ref{lem:Xi}, $f_{n\beta}$ maps $h_{n-1}$ to $h_0$, confirming that $x_{n\beta}\in h_0$ for each $n$. Coming back to \eqref{eq:xdiff} and taking $\abs{\impart{\beta}}<\alpha^*\defas\alpha_0/6$ (see \eqref{eq:alpha_lim}), the sequence $(x_{n\beta})_{n\in\N}$ is Cauchy in the Banach space $h_0$.
Moreover, $x_{0\beta}=0$ gives us
\beqn
\normw{\xi_\beta}{0} \leq\sum_{n=0}^\infty\,\normw{x_{n+1,\beta}-x_{n\beta}}{0}\leq C \abs{\epsilon}g,
\eeqn
where the factor $g$ is due to the last statement in Lemma~\ref{lem:flowcontrol}. It is implied that
\beqn
\abs{\hat \xi(q)}\leq C\abs{\epsilon}g\,e^{-\alpha^*\abs{q}}\qquad (q\in \Z^d).
\eeqn
We infer that $\xi$ is real-analytic on $\torus$. 

Recalling that $\lim_{n\to\infty}\Gamma_{<n}(q)=G(q)$ for each $q\in\Z^d$, let us take the \emph{pointwise} limit $n\to\infty$ of \eqref{eq:xi} in the Fourier representation: $\hat\xi(q)$ equals
\begin{align*}
\lim_{n\to\infty}\Gamma_{<n}(q)\left(\widehat{\pi_0x_n}(q)+\hat\rho_0(q)\right)=G(q)\lim_{n\to\infty}\left(\widehat{\pi_0x_n}(q)+\hat\rho_0(q)\right) 
 = G(q)\bigl(\widehat{\pi_0\xi}+\hat\rho_0\bigr)(q),
\end{align*}
because $\pi_0$ is a continuous operator on $h_0$. Indeed, $\xi$ solves \eqref{eq:xi}!

Out of curiosity, we conclude by the recursion invariance \eqref{eq:recinv} that
\beq\label{eq:xi_n2}
\xi_{n}=G_n(\pi_n\xi_n+\rho_n)=G_n(\pi_0\xi+\rho_0)
\eeq
converges to $\xi^*= 0$, pointwise in terms of the Fourier representation. Hence, equation \eqref{eq:xi_general} really trivializes in the large-$n$-limit. Another way of seeing this is the pointwise bound
$|\widehat{G_n\rho_n}(q)|\leq C\abs{G(q)}\normw{\tilde\pi_n}{n;-n}$,
which tends to zero and paraphrases $G^*\rho^*=0$ below \eqref{eq:xi_star}.

We still need to demonstrate that the solution $\xi$ of \eqref{eq:xifourier} also solves \eqref{eq:xifourier0}, \ie, that
\beqn
\bigl(\widehat{\pi_0\xi}+\hat\rho_0\bigr)(0)=\pi_0(0,\piste)\hat\xi+\hat\rho_0(0)=\sum_{p\in\Z^d}\pi_0(0,p)\hat\xi(p)+\hat\rho_0(0)=0.
\eeqn
From \eqref{eq:xi_n2}, $G_n(q)\defas\chi_n(\omega\cdot q)G(q)$, $G(0)=0$,  and \eqref{eq:xi},
\begin{align*}
\bigl|\widehat{\pi_n\xi_n}(0)\bigr| & \leq\sum_{q\in\Z^d}\abs{\tilde\pi_n(0,q)}\chi_n(\omega\cdot q)\, \bigl|G(q)\bigl(\widehat{\pi_0\xi}+\hat\rho_0\bigr)(q)\bigr| \\
& \leq \normw{\xi}{0} \sup_{q\in\Z^d}\abs{\tilde\pi_n(0,q)}\chi_n(\omega\cdot q) \\ & \leq\normw{\xi}{0}\,\normw{\tilde\pi_n}{n;-n}\sup_{q\in\Z^d\nonzero}w_n(q)\chi_n(\omega\cdot q) \\
& \leq C\normw{\xi}{0}\,\normw{\tilde\pi_n}{n;-n}\quad\longrightarrow\quad 0\quad\text{as}\quad n\to\infty,
\end{align*}
Thus,
\beqn
\bigl(\widehat{\pi_0\xi}+\hat\rho_0\bigr)(0)=\lim_{n\to\infty}\bigl(\widehat{\pi_n\xi_n}+\hat\rho_n\bigr)(0)=\lim_{n\to\infty}{\hat\rho_n}(0).
\eeqn
But
\beqn
\lim_{n\to\infty}{\hat\rho_n}(0)=4\lim_{n\to\infty}\pi_n(0,0)=4\lim_{n\to\infty}\delta_n=0,
\eeqn
and we are done with the construction of $(\gamma,X_1)$ under the assumption $\average{\Psi_0}=0.$

{\bf The case $\average{\Psi_0}\neq 0$.} If $X_1$ solves \eqref{X1}, it is a matter of applying the translation $\tau_\beta$ on both sides of the equation to get $(\D+\gamma)^2 X_1'=D\Omega(X_0')X_1'$, where $X_0'=\tau_\beta X_0+(0,\beta)$ and $X_1'=\tau_\beta X_1$. In other words, the translation property in the formulation of the theorem holds, and the value of $\gamma$ does not change under such translations. \qed

\subsection{Proof of Lemma~\ref{lem:flowcontrol}}
We begin by deriving several identities that are easy to refer to below. To this end, let us look at the flow \eqref{eq:pirec2} more closely, observing that we may formally split
\begin{align*}
\bigl(\one  -\pi_{n\beta}(\kappa)\Gamma_n(\kappa)\bigr)^{-1}  =  \left(\one-P_n(\kappa)\Gamma_n(\kappa)\right)^{-1}+r_{n\beta}(\kappa).
\end{align*}
The remainder $r_{n\beta}(\kappa)$ reads explicitly
\beqn
r_{n\beta}(\kappa)\defas\left(\one-\pi_{n\beta}(\kappa)\Gamma_n(\kappa)\right)^{-1}\,\tilde\pi_{n\beta}(\kappa)\Gamma_n(\kappa)\,\left(\one-P_n(\kappa)\Gamma_n(\kappa)\right)^{-1}.
\eeqn
In fact, this quantity is asymptotically \emph{very small} in $\el(h_{-n};h_{-n})$ due to the explicit factor $\tilde\pi_{n\beta}$; given the bounds \eqref{eq:tildepi}--\eqref{eq:deltabound} for some particular value of $n$,
\beq\label{eq:r_n}
\normw{r_{n\beta}(\kappa)}{-n;-n}\leq C\abs{\epsilon} e^{- c\mu^n}.
\eeq

Continuing abstractly, \eqref{eq:pirec2} becomes
\beqn
\pi_{n+1,\beta}(\kappa)=\left(\one-P_n(\kappa)\Gamma_n(\kappa)\right)^{-1}P_n(\kappa)+s_n(\kappa)+\tilde s_{n\beta}(\kappa),
\eeqn
where $s_n(\kappa)$ is the diagonal and $\tilde s_{n\beta}(\kappa)$ the off-diagonal part of the small remainder term $r_{n\beta}(\kappa)\pi_{n\beta}(\kappa)$, respectively. Therefore, the diagonal $P_n(\kappa)$---containing the problematic $\delta_n$---and the off-diagonal $\tilde \pi_{n\beta}(\kappa)$ iterate according to the rules
\beq\label{eq:pflow}
\begin{cases}
P_{n+1}(\kappa)=\left(\one-P_n(\kappa)\Gamma_n(\kappa)\right)^{-1}P_n(\kappa)+s_n(\kappa),\\
\tilde\pi_{n+1,\beta}(\kappa)=\tilde s_{n\beta}(\kappa).
\end{cases}
\eeq
Notice that $s_n(\kappa)$ is indeed free of $\beta$, because each $P_n(\kappa)$ is.

By construction, $\delta_n=\pi_{n\beta}(0;0,0)=P_n(0;0)$ for each $n$, such that the diagonality of $(\one-P_n(0)\Gamma_n(0))^{-1}$ with $\Gamma_n(0;0)=0$ implies that changes in $\delta_n$ upon iteration only arise from the small term $s_n$ in \eqref{eq:pflow}:
\beq\label{eq:deltaflow}
\delta_{n+1}=\delta_n+d_n,\qquad d_n\defas s_n(0;0).
\eeq
But $r_{n\beta}(0;0,0)=0$, again because $\Gamma_n(0;0)=0$, such that
\begin{align}
d_n & =s_n(0;0)=\left(r_{n}\pi_{n}\right)(0;0,0) 
 =(r_{n}\tilde\pi_{n})(0;0,0).\label{eq:d=rtildepi}
\end{align}
We remind the reader of our convention of dropping one of the kernel indices of diagonal operators. For instance, $s_n(\kappa;q)\equiv s_n(\kappa;q,q)$. 

It is convenient to spell out a consequence of \eqref{eq:pflow}:
\beq\label{eq:barpflow}
\bar P_{n+1}(\kappa)=\bar P_n(\kappa)+P_n(\kappa)\Gamma_n(\kappa)(\one-P_n(\kappa)\Gamma_n(\kappa))^{-1}  P_n(\kappa)+(s_n(\kappa)-d_n).
\eeq

\begin{proof}[Proof of Lemma~\ref{lem:flowcontrol}]
Here we finally prove that the bounds \eqref{eq:tildepi}--\eqref{eq:deltabound}, such that \eqref{eq:pflow}---and indeed everything above---becomes not only formally justified. To this end, we proceed by induction. As iterating \eqref{eq:tildepi} and \eqref{eq:tildep} is rather easy, the proof boils down to \emph{choosing the value of our free parameter}, the Lyapunov exponent $\gamma$, so as to guarantee that $\delta_n$ satisfies \eqref{eq:deltabound} at each step.

{\bf (i) Case $n=0$.} Consider $\kappa$ restricted to $D_0$ with $b\leq g/3$. Lemma~\ref{lem:H} and $\bar P_0(\kappa;q)=H(\kappa;q,q)-H(0;0,0)$ readily imply
\beqn
\normw{\bar P_0(\kappa)}{0;0}\leq C_0\abs{\lambda}.
\eeqn
Furthermore, increasing $C_0$ and employing \eqref{eq:betakernelest} with $\abs{\impart\beta}<\alpha_0<\sigma'$,
\beqn
\normw{\tilde\pi_{0\beta}(\kappa)}{0;0}\leq C_0\abs{\lambda}.
\eeqn
The leading Taylor coefficient $\bar p_0'(0)=H'(0;0,0)$ and the corresponding remainder of the function $\bar p_0=H(\piste;0,0)-H(0;0,0)$ satisfy
\beqn
\abs{\bar p_0'(0)}\leq \quarter C_0\abs{\epsilon}^2 g\mathand\abs{\bar p_0(\kappa)-\bar p_0'(0)\kappa}\leq \half C_0\abs{\epsilon}^2\abs{\kappa}^2, 
\eeqn
taking $C_0$ large enough.

Assume that $\gamma$ lies in the open $g$-centered disk of radius $c\abs{\epsilon}g$:
\beq\label{eq:I_gamma}
\gamma\in I_\gamma\defas \mathbb{D}(g,c_\gamma\abs{\epsilon}g).
\eeq
Recall that $\delta_0=\epsilon g^2 u(\epsilon,g,\gamma)+g^2-\gamma^2$, where $\epsilon g^2 u(\epsilon,g,\gamma)=H(0;0,0)$. If $\delta_0(\gamma_1)=\delta_0(\gamma_2)$ and we denote $\gamma_i=g(1+x_i)$, the Mean-Value Theorem yields
\beqn
\abs{\gamma_1-\gamma_2}\leq \half \bigl(\abs{x_1+x_2}+\abs{\epsilon} g\supnorm{\de_\gamma u}\bigr)\abs{\gamma_1-\gamma_2}. 
\eeqn
By Lemma~\ref{lem:H}, $\supnorm{\de_\gamma u}\leq C\abs{\epsilon}g\inv/(1-2c_\gamma\abs{\epsilon})$, and $\abs{x_1+x_2}<2c_\gamma\abs{\epsilon}$. For a sufficiently small $\abs{\epsilon}$, we gather $\gamma_1=\gamma_2$, such that $\gamma\mapsto\delta_0$ is one-to-one on $I_\gamma$. Moreover, the image of the disk $I_\gamma$ contains the disk $\mathbb{D}\bigl(0,(2c_\gamma-c_\gamma^2\abs{\epsilon}-\supnorm{u})\abs{\epsilon}g^2\bigr)$. Thus, for a sufficiently large value of $c_\gamma$ and small value of $\epsilon$, there exists a \emph{closed} set $J_0\subset I_\gamma$ which $\gamma\mapsto \delta_0$ maps analytically and \emph{bijectively} onto the closed disk
\beqn
I_0\defas \mathbb{\bar D}(0,C_0\abs{\epsilon}g^2).
\eeqn
We are about to prove below that a correct choice of $\gamma$ leads to
\beqn
\delta_n\in I_n \defas\mathbb{\bar D}(0,C_0\abs{\epsilon}g\,\aleph^{2n})
\eeqn
for each and every $n\in\Z_+$.

{\bf (ii) Induction step: hypotheses.} Fix $n\in \N$. Suppose
\beqn
\normw{\tilde\pi_{n\beta}(\kappa)}{n;-n} \leq  C_{n}\abs{\epsilon}g\, \aleph^n \begin{cases}
g & \text{if $n=0$},\\
e^{-c\mu^n} & \text{if $n\geq 1$}, 
\end{cases}
\eeqn
for some constants $c>0$ and $\mu>1$---to be fixed later---and
\beqn
\normw{\bar P_{n}(\kappa)}{n;-n}  \leq  C_{n}\abs{\epsilon}g\, \aleph^{n}
\begin{cases}
g & \text{if $n=0$},\\
1 & \text{if $n\geq 1$}, 
\end{cases}
\eeqn
hold true for $\abs{\epsilon}<\epsilon_{n}$, $\abs{\impart{\beta}}<\alpha_{n}$, and $\kappa\in D_{n}$. Suppose there exists a closed set $J_n\subset I_\gamma$ and a bijective analytic map $\Delta_n:J_n\to I_n:\gamma\mapsto\delta_n$.

Further, let the kernel elements of these operators be analytic in $D_{n}$ and continuous in the closure $\bar D_{n}$. Also the estimates
\beq\label{eq:taylor_it}
\abs{\bar p_{n}'(0)}\leq\left(1-\frac{1}{n+2}\right)\frac{1}{2}C_n g
\begin{cases}
\abs{\epsilon}^{2} & \text{if $n=0$},\\
\abs{\epsilon}^{3/2} & \text{if $n\geq 1$}, 
\end{cases}
\eeq
and
\beq\label{eq:taylor_it2}
\abs{\bar p_{n}(\kappa)-\bar p_{n}'(0)\kappa}\leq \frac{1}{2} C_{n}\aleph^{-n}\abs{\kappa}^2
\begin{cases}
\abs{\epsilon}^{2} & \text{if $n=0$},\\
\abs{\epsilon}^{3/2} & \text{if $n\geq 1$}, 
\end{cases},
\eeq
which facilitate dealing with the Taylor expansion of $\bar p_n$, are supposed to be satisfied.

In particular, it follows from \eqref{eq:pidecomp}, $b\leq g/3$, and the inductive hypotheses that
\beq\label{eq:phyp}
\abs{p_{n}(\kappa)}\leq B_nC_{n}\abs{\epsilon}g\,\aleph^{n} \mathwith B_n\defas
\begin{cases}
b\abs{\epsilon} + g & \text{if $n=0$}, \\
b\abs{\epsilon}^{1/2}+\aleph^n & \text{if $n\geq 1$},
\end{cases}
\eeq
and
\beq\label{eq:pihyp}
\normw{\pi_{n\beta}(\kappa)}{n;-n} \leq  A_nC_{n}\abs{\epsilon}g\,\aleph^{n},
\eeq
where
\beq\label{eq:A_nprop}
A_n\defas 
\begin{cases}
g & \text{if $n=0$}, \\
1+\aleph^n+e^{-c\mu^n} & \text{if $n\geq 1$}.
\end{cases}
\eeq

The strategy is to iterate the above hypotheses and prove that, in the bitter end, $C_n$ and $\epsilon_n$ can be chosen in an $n$-independent fashion, \emph{uniformly in $g$}.

{\bf (ii a) The off-diagonal $\tilde\pi_{n+1,\beta}(\kappa)$.} If $\tilde\beta\in\C^d$, then
\beqn
\abs{\tilde\pi_{n+1,\tilde\beta}(\kappa;p,q)}\,e^{-\impart(\beta-\tilde\beta)\cdot(p-q)}w_{n}(p)^{-1}w_{n}(q)^{-1} \leq \normw{\tilde\pi_{n+1,\beta}(\kappa)}{n;-n}.
\eeqn
But with a modification of \eqref{eq:r_n},
\beq\label{eq:r_n_mod}
\normw{r_{n\beta}(\kappa)}{-n;-n}\leq 4C_\Gamma C_n\abs{\epsilon}\tilde B_n
\quad\text{where}\quad
\tilde B_n\defas
\begin{cases}
g & \text{if $n=0$},\\
e^{-c\mu^n} & \text{if $n\geq 1$}, 
\end{cases}
\eeq
such that
\beqn
\normw{\tilde\pi_{n+1,\beta}(\kappa)}{n;-n}\leq \normw{r_{n\beta}(\kappa)\pi_{n\beta}(\kappa)}{n;-n}\leq 2C_n\abs{\epsilon}g\,\aleph^n
\tilde B_n
\eeqn
both provided $\epsilon$ meets the condition
\beq\label{eq:lambda_n_cond}
\abs{\epsilon}\leq\epsilon_{n+1}\defas\max{\big(\epsilon_n,\half(A_nC_nC_\Gamma)\inv\big)}.
\eeq
Hence, if $\abs{\impart\beta}<\alpha_{n}$,
\beqn
\abs{\tilde\pi_{n+1,\tilde\beta}(\kappa;p,q)}\leq 2C_{n}\abs{\epsilon}g\,\aleph^n \tilde B_n\cdot (1-\delta_{p,q})\,e^{\impart(\beta-\tilde\beta)\cdot(p-q)}w_{n}(p)w_{n}(q).
\eeqn

Now assume $\abs{\impart{\tilde\beta}}<\alpha_{n+1}$ and, fixing $p$ and $q$, take
\beqn
\beta=\tilde\beta+i(\alpha_{n}-\alpha_{n+1})\frac{p-q}{\abs{p-q}}.
\eeqn
Obviously $\abs{\impart{\beta}}<\alpha_{n}$. What we get this way is
\beqn
\abs{\tilde\pi_{n+1,\tilde\beta}(\kappa;p,q)}\leq 2C_{n}\abs{\epsilon}g\,\aleph^n \tilde B_n \cdot(1-\delta_{p,q})\,e^{-(\alpha_{n}-\alpha_{n+1})\abs{p-q}}w_{n}(p)w_{n}(q)
\eeqn
for each pair $(p,q)\in\Z^d\times\Z^d$. Thus, from the expression \eqref{eq:operatornorm} for the norm,
\begin{align}
&\normw{\tilde\pi_{n+1,\tilde\beta}(\kappa)}{n+1;-(n+1)} \notag \\ 
&\qquad\leq 2C_{n}\abs{\epsilon}g\,\aleph^n  \tilde B_n\sup_{q\in\Z^d}\sum_{p\in\Z^d\setminus\{q\}}e^{-4 (n+3)^{-2}\alpha_{n}\abs{p-q}}\frac{w_{n}(p)w_{n}(q)}{w_{n+1}(p)w_{n+1}(q)} \notag \\
&\qquad\leq 2C_{n}\abs{\epsilon}g\,\aleph^n  \tilde B_n \sum_{p\in\Z^d\nonzero}e^{-4 (n+3)^{-2}\alpha_{n}\abs{p}}w_{n+1}(p)^{-(1-\aleph)}.\label{eq:tildepi1}
\end{align}
After \eqref{eq:weightproperty}, the second inequality follows from shifting $p$ to $p+q$. We control the above bound by treating the cases $\abs{\omega\cdot p}\leq \aleph^{(n+1)/2}$ and $\abs{\omega\cdot p}>\aleph^{(n+1)/2}$ separately. In fact, if $\abs{\omega\cdot p}\leq \aleph^{(n+1)/2}$, then $\abs{p}>\aleph^{-(n+1)/2\nu}$ follows from \eqref{eq:Dio2}, and
\beqn
e^{-4(n+3)^{-2}\alpha_{n}\abs{p}}< e^{-2n^{-2}\alpha_{n}\abs{p}}\cdot e^{-2(n+1)^{-2}\alpha_{n}\aleph^{-(n+1)/2\nu}},\quad w_{n+1}(p)^{-(1-\aleph)}<1,
\eeqn
whereas
\beqn
\abs{\omega\cdot p}>\aleph^{(n+1)/2}\quad\Longrightarrow\quad w_{n+1}(p)^{-(1-\aleph)}<e^{-(1-\aleph)\aleph^{-(n+1)/2}}.
\eeqn
Since $\alpha_n>\alpha_0/6$ by \eqref{eq:alpha_lim} and, for $a>1$ and $m>0$,
$
m^{-2}a^m\geq \frac{e^2}{4}(\ln{a})^2,
$
we have
\beqn
e^{-2(n+1)^{-2}\alpha_{n}\aleph^{-(n+1)/2\nu}}\leq e^{-\frac{1}{12} e^2\alpha_0\ln(\aleph^{-1/4\nu})\,\aleph^{-(n+1)/4\nu}}.
\eeqn

The remaining $d$-dimensional geometric series satisfies 
\beqn
\sum_{p\in\Z^d\nonzero}e^{-2 (n+1)^{-2}\alpha_0\abs{p}}\leq C(d)\biggl(\frac{(n+1)^2}{\alpha_0}\biggr)^d.
\eeqn
Hence, we infer that if $\abs{\impart\beta}<\alpha_{n+1}$ and $\kappa\in D_{n}$, then
\beqn
\normw{\tilde\pi_{n+1,\beta}(\kappa)}{n+1;-(n+1)} \leq C_{n+1}\abs{\epsilon}g\,\aleph^{n+1} e^{-c\mu^{n+1}},
\eeqn
where we finally pin down the values of the previously free parameters 
\beqn
c\defas\half\min{\left(\tfrac{1}{12} e^2\alpha_0\ln(\aleph^{-1/4\nu}),1-\aleph\right)}>0\!\mathand \!\mu\defas \aleph^{-1/\max{(4\nu,2)}}>1,
\eeqn
and take
\beq\label{eq:C_n1}
C_{n+1}\geq 2C(d)\,\aleph\inv e^{-c\mu^{n+1}} \tilde B_n \biggl(\frac{(n+1)^2}{\alpha_0}\biggr)^d  C_n.
\eeq

{\bf (ii b.1) The non-constant part $\bar P_{n+1}(\kappa)$ of the diagonal.} If $\kappa\in D_{n+1}$ and $\abs{\omega\cdot q}<\aleph^{n}(1-\aleph) b$, then $\kappa+\omega\cdot q\in D_n$. So, by \eqref{eq:weightproperty},
{\Small
\begin{align*}
\normw{&\bar P_{n+1}(\kappa)}{n+1;-(n+1)}  =\sup_{q\in\Z^d}\abs{\bar P_{n+1}(\kappa;q)}w_{n+1}(q)^{-2}\\
&\quad\leq\, \max\bigg\{\sup_{\substack{\abs{\omega\cdot q}<\aleph^{n}(1-\aleph) b}}\frac{\abs{\bar p_{n+1}(\kappa+\omega\cdot q)}}{w_{n+1}(q)^{2}} ,\sup_{\substack{\abs{\omega\cdot q}\geq\aleph^{n}(1-\aleph) b}}\abs{\bar P_{n+1}(\kappa;q)}w_{n}(q)^{-2/\aleph}\bigg\}\\
&\quad\leq\,\max\bigg\{\sup_{\substack{\abs{\omega\cdot q}<\aleph^{n}(1-\aleph) b}}\frac{\abs{\bar p_{n+1}(\kappa+\omega\cdot q)}}{w_{n+1}(q)^{2}} , \quad e^{-2b\aleph^{-1}(1-\aleph)^2}\normw{\bar P_{n+1}(\kappa)}{n;-n}\bigg\}.
\end{align*}
}%
But we know that the relations $\normw{\bar P_{n+1}(\kappa)}{n;-n}\leq \normw{P_{n+1}(\kappa)}{n;-n}+\abs{\delta_{n+1}}$ and $\abs{\delta_{n+1}}=\abs{P_{n+1}(0;0)}\leq\normw{P_{n+1}(0)}{n;-n}$ hold. Moreover, \eqref{eq:pihyp} and \eqref{eq:pirec2} yield
\beqn
\normw{P_{n+1}(\kappa)}{n;-n}\leq\normw{\pi_{n+1,\beta}(\kappa)}{n;-n}\leq 2\normw{\pi_{n\beta}(\kappa)}{n;-n} \leq 2A_nC_n\abs{\epsilon}g\,\aleph^n,
\eeqn
assuming \eqref{eq:lambda_n_cond} and $\kappa\in D_{n}\supset D_{n+1}$ hold. Observe that, for positive $x$ and $p$, $x\inv e^{-x^{-p}/(ep)}\leq 1$. Consequently, if we demand that
\beq\label{eq:aleph_b_g}
\aleph\leq\min{\left(\tfrac{1}{8},b^2\right)},
\eeq 
say, and
\beq\label{eq:C_n2}
C_{n+1}\geq A_n C_n,
\eeq
it remains to be proven that
\beq\label{eq:bar_p}
\sup_{\substack{\kappa\in D_{n+1} \\ \abs{\omega\cdot q}<\aleph^{n}(1-\aleph) b}}\frac{\abs{\bar p_{n+1}(\kappa+\omega\cdot q)}}{w_{n+1}(q)^{2}}\leq C_{n+1}\abs{\epsilon}g\,\aleph^{n+1}.
\eeq
Notice that the rather arbitrary \eqref{eq:aleph_b_g} imposes an interrelation between $\aleph$ and $g$, which is needed in the limit $b\leq g/3\to 0$; since we cannot take $b$ large, we have to take $\aleph=o(b)$ in order to guarantee $e^{-2b\aleph^{-1}(1-\aleph)^2}\leq \aleph/4$ above. 

In order to verify \eqref{eq:bar_p}, we use the recursion formula 
\beq\label{eq:bar_p-flow}
\bar p_{n+1}-\bar p_{n} \,=\,  p_{n}\,\gamma_{n}\,a_n\, p_{n}+s_{n}(\piste;0)-s_{n}(0;0)
\eeq
subject to
\beqn
a_n \defas  (1-p_n\gamma_n)\inv \mathand
\gamma_n(\kappa) \defas \Gamma_n(\kappa;0),
\eeqn
which is an advocate of \eqref{eq:barpflow}. The bound \eqref{eq:Gammakernel} yields 
\beq\label{eq:gamma_n-bound}
\abs{\gamma_n(\kappa)}\leq C_\Gamma g\inv\aleph^{-n} \abs{\aleph^{-n}\kappa}^5\qquad(\kappa\in D_n). 
\eeq
By virtue of $\abs{s_n(\kappa;0)}\leq \normw{r_n(\kappa)\pi_n(\kappa)}{n;-n}$, \eqref{eq:r_n_mod} gives 
\beq\label{eq:s_n-bound}
\abs{s_n(\kappa;0)}\leq  4C_{n}^2 C_\Gamma\abs{\epsilon}^2
\begin{cases}
g^3 & \text{if $n=0$},\\
A_n \, g\,\aleph^n e^{-c\mu^n} & \text{if $n\geq 1$},
\end{cases}
\eeq
in $D_n$.

{\bf (ii b.2) The Taylor expansion of $\bar p_{n+1}(\kappa)\equiv\bar P_{n+1}(\kappa;0)$.} Let us abbreviate
\beqn
\sigma_n(\kappa)\equiv \bar p_n(\kappa)-\bar p_n'(0)\kappa,
\eeqn
for each natural number $n$. The objective is to show that the estimates
\beq\label{eq:goal1}
\abs{\bar p_{n+1}'(0)}\leq\left(1-\frac{1}{n+3}\right)\frac{C_{n+1}\abs{\epsilon}^{3/2} g}{2},
\eeq
\ie, the iterate of \eqref{eq:taylor_it}, and
\beq\label{eq:goal2}
\sup_{\kappa\in D_n}\abs{(\sigma_{n+1}-\sigma_n)(\kappa)}\leq C_{n+1}\abs{\epsilon}^{7/4}g\, \aleph^{n+1}
\eeq
hold. Indeed, with the aid of such bounds together with \eqref{eq:taylor_it2}, \eqref{eq:bar_p} follows from
\beqn
\sup_{x\geq 0}{(x+\abs{\kappa})^k e^{-\alpha x}}=\Bigl(\frac{k}{\alpha}\Bigr)^k e^{\alpha\abs{\kappa}-k}\qquad(\alpha>0)
\eeqn
for $k=1,2$ and $\epsilon$ suitably small. Moreover, the Cauchy estimate
\begin{align*}
\abs{\sigma_{n+1}(\kappa)} \leq \abs{\sigma_n(\kappa)}+b^{-2}\abs{\kappa}^2\,\frac{\aleph^{-2n}}{1-\aleph}\,\sup_{\zeta\in D_{n}}\abs{(\sigma_{n+1}-\sigma_n)(\zeta)} \qquad(\kappa\in D_{n+1}),
\end{align*}
implies that also \eqref{eq:taylor_it2} gets successfully iterated.

The bound in \eqref{eq:gamma_n-bound} implies
\beqn
\gamma_n(0)=\gamma_n'(0)=0,
\eeqn
such that $\bar p_{n+1}'(0)=\bar p_n'(0)+s_n'(0;0)$ according to \eqref{eq:bar_p-flow}, and hence
\beqn
\bar p_{n+1}'(0)-\bar p_{n}'(0)=\frac{1}{2\pi i}\oint_{\de D_{n}}\frac{s_{n}(\zeta;0)}{\zeta^2}\,d\zeta.
\eeqn
Thus, resorting to \eqref{eq:s_n-bound}, 
\begin{align*}
\abs{\bar p_{n+1}'(0)-\bar p_{n}'(0)} & \leq \aleph^{-n} b^{-1}\sup_{\kappa\in D_{n}}\abs{s_n(\kappa,0)}
 \leq \frac{C_{n+1}\abs{\epsilon}^{3/2}g}{2(n+2)(n+3)},
\end{align*}
if the constant $C_{n+1}$ satisfies 
\beq\label{eq:C_n3}
C_{n+1}\geq
8(n+2)(n+3)C_\Gamma  b\inv A_n \tilde B_n \abs{\epsilon}^{1/2}C_n^2.
\eeq
The bound \eqref{eq:goal1} now follows, assuming also $C_n\leq C_{n+1}$.

We still need to demonstrate \eqref{eq:goal2}. This will be provided by \eqref{eq:bar_p-flow}, since then
\beqn
\sigma_{n+1}(\kappa)-\sigma_n(\kappa)= (p_n\,\gamma_n\,a_n\,p_n)(\kappa)+s_n(\kappa;0)-\sum_{l=0,1} s_n^{(l)}(0;0)\frac{\kappa^l}{l!},
\eeqn
such that \eqref{eq:phyp}, \eqref{eq:gamma_n-bound} and \eqref{eq:s_n-bound} yield \eqref{eq:goal2} if 
\beq\label{eq:C_n4}
C_{n+1}\geq 4C_\Gamma\aleph\inv\bigl(B_n^2 b^5+6 A_n\tilde B_n\bigr)\abs{\epsilon}^{1/4}C_n^2.
\eeq

{\bf Intuition behind (ii b.1--2).}
Due to the super-exponential decay of $s_n(\kappa;0)$ in \eqref{eq:s_n-bound} and the strong induction hypothesis $\abs{\delta_n}\leq C_0\abs{\epsilon}g\,\aleph^{2n}$ on the constant part of $p_n$, the flow of the remainder $\bar p_n=p_n-\delta_n$ reads roughly
\beq\label{eq:approximate_flow}
\bar p_{n+1}\approx (1-\bar p_n\gamma_n)\inv \bar p_n,
\eeq
by \eqref{eq:pflow}. Hence, the \textit{a priori} bound $\abs{(1-\bar p_n\gamma_n)\inv}\leq 1+C\abs{\epsilon}$ yields a sequence diverging in $n$, with very little hope of proving bounds such as \eqref{eq:tildep}---see \eqref{eq:bar_p}. However, the support of $\gamma_k$ is highly concentrated on the annulus $\aleph^{k+1}b \leq\abs{\kappa}\leq \aleph^k b$. Iterating for $n\geq k$ steps, with $\kappa$ on the latter interval,
\beqn
\bar p_{n+1}(\kappa)\approx (1-\bar p_1(\kappa)\gamma_k(\kappa))\inv \bar p_1(\kappa)=(1+\order{\epsilon})\,\bar p_1(\kappa).
\eeqn
That is, $\bar p_n$ remains close to $\bar p_1$, which enables proving \eqref{eq:tildep} through \eqref{eq:bar_p}.

In fact, our argument is different still: since $\chi_{n}(\aleph\,\kappa)=\chi_{n-1}(\kappa)$ and $G(\aleph\,\kappa;0)\approx \aleph\inv G(\kappa;0)$, we have $\gamma_{n}(\aleph\,\kappa)\approx \aleph\inv \gamma_{n-1}(\kappa)$ for $n\geq 2$. Inserting this into \eqref{eq:approximate_flow}, we notice that the \emph{approximate scaling invariance}
\beqn
\bar p_{n+1}(\aleph\,\kappa)\approx \aleph\,\bar p_n(\kappa)
\eeqn
is consistent with the flow. This is what the bounds \eqref{eq:taylor_it}--\eqref{eq:taylor_it2} reflect.

{\bf (ii c) The constant part $\delta_{n+1}$ of the diagonal.} Recall that $\gamma$ may be viewed as a function of $\delta_n$ by the induction hypotheses; the identity $\delta_n=\Delta_n(\gamma)$ is bijective on $J_n$. The flow produces a \emph{near-identity analytic} function $\delta_{n+1}=\delta_{n}+d_{n}(\delta_{n})$ of $\delta_{n}$ on the disk $I_{n}$, such that, for $\epsilon$ small enough,
\beq\label{eq:I_nsubset}
\delta_{n+1}(I_{n})\supset I_{n+1}.
\eeq
The analyticity of the map $\delta_n\mapsto d_n$ can be read off \eqref{eq:d=rtildepi} and the expression of $r_n$. As far as estimates are concerned,
\beqn
\abs{d_{n}}\leq \normw{r_n(0)\tilde\pi_n(0)}{n;-n}\leq CC_n^2C_\Gamma\,\abs{\epsilon}^2g
\begin{cases}
g^2 & \text{if $n=0$},\\
\aleph^{n}e^{-2c\mu^n} & \text{if $n\geq 1$},
\end{cases}
\eeqn
in the complex neighbourhood $2I_n$ of $I_n$ of radius $\half\abs{I_n}$, where $\abs{I_n}$ is the diameter of the disk $I_n$. Consequently, a Cauchy estimate yields the bound
\beq\label{eq:lip}
\sup_{\delta_n\in I_n}\abs{\de d_{n}/\de\delta_n}\leq \frac{\sup_{\delta_n\in 2I_n}\abs{d_n}}{\half\abs{I_n}}\leq\half
\eeq
on the Lipschitz constant of $d_n$ on $I_n$, provided $\abs{\epsilon}\leq \epsilon_{n+1}$ with
\beq\label{eq:epsilon_cond}
\epsilon_{n+1}\inv\geq 2 C_0\inv CC_n^2C_\Gamma
\begin{cases}
g & \text{if $n=0$},\\
\aleph^{-n}e^{-2c\mu^n} & \text{if $n\geq 1$}.
\end{cases}
\eeq
In this case also
\beq\label{eq:d_nbound2}
\abs{d_{n}}\leq \half\abs{I_n}-\half\abs{I_{n+1}}
\eeq
holds, which validates \eqref{eq:I_nsubset}, considering how the boundary of $I_n$ is transformed under $\delta_{n+1}$. 

Notice that \eqref{eq:lip} implies
\beq\label{eq:rel_expansive}
\bigl|\delta_{n+1}(x)-\delta_{n+1}(y)\bigr|\geq \half \abs{x-y}\qquad(x,y\in I_n),
\eeq
meaning that $\delta_n\mapsto\delta_{n+1}$ is \emph{one-to-one}. By continuity and \eqref{eq:I_nsubset}, there exists a closed set $\tilde J_{n+1}\subset I_{n}$ that is bijectively and analytically mapped onto $I_{n+1}$: $\tilde J_{n+1}\defas \delta_{n+1}\inv(I_{n+1})$. We can backtrack with the aid of the map $\Delta_n$, obtaining a closed subset $J_{n+1}\subset I_\gamma$ (see \eqref{eq:I_gamma}) that is bijectively and analytically mapped onto $I_{n+1}$ by the map $\Delta_{n+1}\defas\delta_{n+1}\circ\Delta_n$:
\beqn
J_{n+1}\defas \Delta_{n+1}\inv(I_{n+1}).
\eeqn
It follows immediately that
\beqn
J_{n+1}\subset J_n.
\eeqn

{\bf (iii) Large values of $n$ and the limit $g\to 0$.}
Suppose $C_n$ is independent of $g$, which is the case for $C_0$. The recursive conditions \eqref{eq:C_n1}, \eqref{eq:C_n2}, \eqref{eq:C_n3}, and \eqref{eq:C_n4} can be summarized in bounds of the form 
\beqn
C_{n+1}\geq K_n(g)\,C_n\mathand C_{n+1}\geq L_n(g)\abs{\epsilon}^{1/4}\,C_n^2.
\eeqn
Choosing $b\defas g/3$ (due to \eqref{eq:C_n3}) and $\aleph\defas\min{\left(\tfrac{1}{8},b^2\right)}$ (due to \eqref{eq:C_n4}; see also \eqref{eq:aleph_b_g}), which is allowed, we may bound $K_n(g)$ and $L_n(g)$ \emph{uniformly in $g$}: \linebreak $\sup_{0<g<g_0} K_n(g)\leq K_n$ and $\sup_{0<g<g_0} L_n(g)\leq L_n$. This follows from the fact that $\aleph\inv e^{-c\mu}\to 0$ as $\aleph\to 0$. Moreover, $L_n\leq L$ for each $n$, such that we may choose
\beqn
C_{n+1}\defas\max\,(K_n,\,L\abs{\epsilon}^{1/4}C_n)\,C_n.
\eeqn
The numbers $K_n>1$ converge to unity so fast that the number
$
K\defas\prod_{n=0}^\infty K_n>1
$ 
is finite. Now choose $\epsilon$ so small that
$
L\abs{\epsilon}^{1/4}KC_0\leq 1.
$
In particular, $C_1=K_0C_0$, and inductively $C_n=K_0\cdots K_{n-1}C_0\leq KC_0$. We conclude that the sequences $(C_n)$ and $(\epsilon_n)$ (see \eqref{eq:lambda_n_cond} and \eqref{eq:epsilon_cond}) converge to positive numbers.

{\bf (iv) Fine-tuning the Lyapunov exponent $\gamma$.} 
The maps $\delta_n$ are relatively expansive; \eqref{eq:rel_expansive} holds, while the target $I_n$ contracts by a factor of $\aleph^2<\half$ at each step. Thus, demanding $\Delta_n(J_n)=I_n$ at each step for the map $\Delta_n=\delta_n\circ\dots\circ\delta_0$ amounts to
\beqn
\abs{x-y}\leq 2^n\bigl|\Delta_n(x)-\Delta_n(y)\bigr|\leq Cg\,(2\aleph^2)^{n}\qquad(x,y\in J_n),
\eeqn
or $\lim_{n\to\infty} \abs{J_n}=0$. Because the $J_n$'s form an ever decreasing chain of closed disks, their intersection consists of precisely one point:
\beqn
\{\gamma\}\defas \bigcap_{n=0}^\infty J_n\subset I_\gamma.
\eeqn
The value of $\gamma$ is an analytic function of $\epsilon$, because the sequence $\Delta_n\inv(0)$ converges uniformly to $\gamma$ with respect to $\epsilon$. For real values of $\epsilon$, $\Delta_n$ sends real numbers to real numbers, making $\gamma$ real.
\end{proof}

\section{Proof of Theorem \ref{thm:manifolds}}\label{sec:Z}
Let us summarize what we have learned thus far. The solution $X^0(z)$ to the equations of motion in the uncoupled case was found. In the coupled case we resolved KAM-type small denominator issues, which contributed the $t\to -\infty$ ($z=0$) asymptotic $X_0(\theta)$ of the general solution $X(z,\theta)$, as well as the linearization $X_1(\theta)\equiv \de_zX(0,\theta)$. 

We can now solve \eqref{xeq}, and thus find the unstable manifold $\W^u_\lambda$ also ``far away'' from the torus $\T_\lambda$, by a Contraction Mapping argument.

To begin with, we single out the uncoupled part $X^0$ of the complete solution $X$;
\beqn
X=X^0+\widetilde X\mathwith{\widetilde X}\unperte\equiv 0.
\eeqn
As $\el^2X^0=(\gamma^2\sin\Phi^0,0)$, \eqref{xeq} now becomes
$
\el^2\widetilde X=-(\gamma^2\sin\Phi^0,0)+\Omega(X^0+\widetilde X).
$
In other words, the map $\widetilde X$ has to satisfy
\beq\label{til}
\K\widetilde X=\widetilde W(\widetilde X),
\eeq
where we define the linear operator
\beq\label{K}
\K\defas
\begin{pmatrix} 
L & 0\\
0 & \el^2
\end{pmatrix}
\mathwith 
L\defas\el^2-\gamma^2\cos\Phi^0
\eeq
and the nonlinear operator $\widetilde W$ through the expression
\beq\label{tilW}
\widetilde W(\widetilde X)\defas(-\gamma^2\sin\Phi^0-\gamma^2(\cos\Phi^0)\widetilde\Phi,0)+\Omega(X^0+\widetilde X).
\eeq

Throughout the rest of the work, we shall refer to different parts of the Taylor expansion of a suitable function $h(z,\theta)$ around $z=0$ using the notation
\beqn
h_k(\theta)\defas \frac{\de_z^k h(0,\theta)}{k!},\quad h_{\leq k}(z,\theta)\defas \sum_{j=0}^k z^kh_k(\theta)\mathand \delta_k h\defas h-h_{\leq k-1}.
\eeqn

Observe that $X_0=\widetilde X_0$ and $X_1=(4,0)+\widetilde X_1$ exist. Setting
\beq\label{Zdef}
\widetilde X(z,\theta)\equiv X_{\leq 1}(z,\theta)-(4,0)z+Z(z,\theta),
\eeq
we may transform equation~\eqref{til} into the equation
\beq\label{Zeq}
\K Z=W(Z)
\eeq
for $Z=\delta_2\widetilde X$, where we define $W$ through
\beq\label{Wdef}
W(Z)\defas\delta_2\left[\widetilde W(\widetilde X)+
\begin{pmatrix}
\gamma^2(\cos\Phi^0)\widetilde\Phi_{\leq 1} \\
0
\end{pmatrix}
\right],
\eeq
taking now \eqref{Zdef} as the \emph{definition of $\widetilde X$}.

Let us consider the complex Banach space $\A$ of (bounded) analytic functions $Z$ on the compact set
\beqn
\Pi_\tau\defas\left\{(z,\theta)\;\Big|\;\repart{(z,\theta)}\in[-1-\tau,1+\tau]\times\torus,\;\impart{(z,\theta)}\in[-\tau,\tau]^{d+1}\right\},
\eeqn
$\tau\geq 0$, equipped with the supremum norm, and its closed subspace
\beq\label{eq:A1}
\A_1\defas\left\{Z\in\A\;|\;Z_{\leq 1}=0\right\}.
\eeq
For future use, let us also define the closed origin-centered balls
\beqn
B(R)\defas\left\{Z\in\A\;|\;\supnorm{Z}\leq R\right\}\mathand B_1(R)\defas B(R)\cap\A_1.
\eeqn
Any element of $\A$ extends analytically to $\Pi_{\tau'}$ for some $\tau'>\tau$, allowing uniform estimates on its derivatives \emph{on} $\Pi_\tau$.

\begin{remark}
Whereas equation~\eqref{til} is plagued by small denominators, equation~\eqref{Zeq} is not. This is so due to the decomposition \eqref{Zdef} which separates the previously solved ``KAM-asymptotics'' $X_{\leq 1}$ from $\widetilde X$ and enables reducing \eqref{til} to \eqref{Zeq} on the space $\A_1$, which one could well call the small-denominator-free subspace of $\A$.
\end{remark}

\subsection{Existence and uniqueness of $Z$}
Postponing the proofs until the end of this section, we make two observations, important in demonstrating that \eqref{Zeq} is solvable.  

\begin{lemma}\label{lem:W}
With sufficiently small $R$, $\tau$, and $\epsilon$ (depending on the analyticity region of $f$), the operator $W:\A\to\A_1$ maps the ball $B(R)$ in $\A$ into a ball $B_1(R')$ in $\A_1$ with $R'=Cg^2(R^2+\abs{\epsilon})$, and $W|_{\A_1}$ is Lipschitz continuous on $B_1(R)$ with a Lipschitz constant proportional to $g^2(R+\abs{\epsilon})$. If the restriction of $Z\in\A$ to a \emph{real} neighbourhood of $\bigunit\times\torus$ has the \emph{real} range $\R\times\R^d$ and $\epsilon$ is real, then the same is true of $W(Z)$.
\end{lemma}

\begin{lemma}\label{lem:inverse}
If $0<\tau<1$, the linear operator $\K:\A_1\to\A_1$ has a bounded inverse $\K^{-1}\in\el(\A_1)$ obeying ${\norm{\K^{-1}}}_{\mathcal{L}(\A_1)}\leq C\gamma^{-2}\tau\inv(1-\tau^2)^{-2}$. It preserves analyticity in $\epsilon$. If the restriction of $Z\in\A$ to a \emph{real} neighbourhood of $\bigunit\times\torus$ has the \emph{real} range $\R\times\R^d$, the same is true of $\K\inv Z$.
\end{lemma}

We have developed enough machinery to extract a solution from $\eqref{Zeq}$:
\begin{theorem}\label{thm:Zexists}
For sufficiently small $R$, $\epsilon_0<R/2$, and $\tau$ (depending on the analyticity regions of $f$ and $X_{\leq 1}$), equation~\eqref{Zeq} has a unique solution $Z\in B_1(R)$. It is continuous on $D$, analytic in $\epsilon$, and bounded uniformly by $C\abs{\epsilon}$. The restriction $Z|_{\bigunit\times\torus}$ takes values in $\R\times\R^d$, provided $\epsilon$ is real.
\end{theorem}
\begin{proof}
We know by Lemmata~\ref{lem:W} and \ref{lem:inverse} that $\K^{-1}W$ maps $B_1(R)$ into itself. We may furthermore choose $\epsilon_0$ and $R$ such that the operator $\K^{-1}W$ becomes contractive on $B_1(R)$. The Banach Fixed Point Theorem implies that $\K^{-1}W$ has a unique fixed point $Z$ in the ball $B_1(R)$.

The theorem also implies that $Z$ is analytic in $\epsilon$. Namely, Lemma~\ref{lem:inverse} says that $\K^{-1}$ preserves such a property. Furthermore, the $\epsilon$-dependence of $W$ comes solely from $\gamma$, $X_0$, $X_1$, and $\Omega$, making it analytic. Hence, the uniformly convergent sequence ${((\K^{-1}W)^k(0))}_{k\in\N}$ reveals the analyticity of the limit $Z$. The latter is also $\R\times\R^d$-valued on $\bigunit\times\torus$ if $\epsilon$ is real.
Finally,
{\small
\beqn
\supnorm{Z}\leq
\supnorm{(\K^{-1}W)(Z)-(\K^{-1}W)(0)}+\supnorm{(\K^{-1}W)(0)}
\leq L\supnorm{Z}+C\abs{\epsilon}
\eeqn
}%
yields $\supnorm{Z}\leq C\abs{\epsilon}/(1-L)$. Here $(\K^{-1}W)(0)$ was bounded using $R'$ of Lemma~\ref{lem:W} at $R=0$.
\end{proof}

\subsection{Putting it all together}
To reach the statement of Theorem~\ref{thm:manifolds} about $X^u$, we glue together the pieces provided by Theorems~\ref{thm:tori}, \ref{thm:linearization}, and \ref{thm:Zexists}.

Assuming $\average{\Psi_0}=0$, we have constructed analytic maps $\gamma$ and 
\beqn
X(z,\theta)=X_0(\theta)+zX_1(\theta)+\delta_2 X(z,\theta)\mathwith\delta_2X=Z+\delta_2X^0
\eeqn
that solve \eqref{xeq} in a complex neighbourhood of $\bigunit\times\torus$ and satisfy the \emph{physical constraint} $\Phi_1\unperte=4$. Recall now \eqref{xtr}. Since \eqref{norm} is not automatically satisfied, we are required to pinpoint specific values of $\alpha$ and $\beta$ so as to fulfill $X_{\alpha,\beta}(1,0)=(\pi,0)$. To this end, we utilize the Implicit Function Theorem.

Consider the implicit equation $\mathfrak{X}(\epsilon,g;\alpha,\beta)\defas X(\alpha,\beta)+(0,\beta)-(\pi,0)=0$. Both $\mathfrak{X}$ and $\frac{\de \mathfrak{X}}{\de(\alpha,\beta)}$ are continuous, and we get from $X=(\Phi^0,0)+\order{\epsilon}$ that
\beqn
\mathfrak{X}(0,g;1,0)=0\mathand \det\left(\frac{\de\mathfrak{X}(\epsilon,g;\alpha,\beta)}{\de(\alpha,\beta)}\right)=\frac{4}{1+\alpha^2}+\order{\epsilon}
\eeqn
for $(\epsilon,g)\in D$ and for whichever values of $\alpha$ and $\beta$ the map $\mathfrak{X}$ is well-defined. Hence, if we choose $\epsilon_0$ small enough, there exist unique continuous functions $\alpha$ and $\beta$ on $D$, analytic with respect to $\epsilon$, such that $\alpha(0,g)=1$, $\beta(0,g)=0$, and
\beqn
\mathfrak{X}(\epsilon,g;\alpha(\epsilon,g),\beta(\epsilon,g))=0.
\eeqn 
Moreover, $\alpha(\epsilon,g)\in\R$ and $\beta(\epsilon,g)\in\R^d$ for $\epsilon$ real, as $\mathfrak{X}$ is then real-valued. A good reference here is \cite{ChierchiaBook}.

\subsection{Proofs of Lemmata~\ref{lem:W} and \ref{lem:inverse}}
We conclude the section by presenting the proofs of Lemmata~\ref{lem:W} and \ref{lem:inverse} used in the proof of Theorem~\ref{thm:Zexists}. 
\begin{proof}[Proof of Lemma~\ref{lem:W}]
Given $Z\in\A$ with $\supnorm{Z}\leq R$, we study $W(Z)$---defined in \eqref{Wdef}, and clearly an element of $\A_1$. Notice that in the relation \eqref{Zdef}, expressing $\widetilde X$ in terms of $Z$, the maps $X_0$ and $X_1$ were previously determined and are independent of $Z$. Furthermore, taking advantage of \eqref{Zdef} and Theorems~\ref{thm:tori} and \ref{thm:linearization}, we deduce
\beq\label{tilX}
\supnorm{\widetilde X}\leq C(\abs{\epsilon}+R).
\eeq

With the aid of \eqref{eq:Omega}, cast equation~\eqref{tilW} as
\beqn
\widetilde W(\widetilde X)\defas(g^2\sin(\Phi^0+\widetilde\Phi)-\gamma^2\sin\Phi^0-\gamma^2\cos(\Phi^0)\widetilde\Phi,0)+\lambda\,\widetilde\Omega(X^0+\widetilde X).
\eeqn
Recall that $f$ is analytic on the strip $\abs{\impart\phi},\abs{\impart\psi}\leq \eta$. Also, $\impart \Phi^0(z)= \order{\tau}$ on $\Pi_\tau$, when $\tau\ll 1$. Hence, owing to \eqref{tilX}, our function $\widetilde\Omega(X^0+\widetilde X)$ is well-defined for $\lambda$ and $R$ sufficiently small and the strip $\Pi_\tau$ about $\bigunit\times\torus$ narrow enough.

Since $\sin(\Phi^0+\widetilde\Phi)=\sin\Phi^0+\cos(\Phi^0)\widetilde\Phi+\order{\widetilde\Phi^2}$, in a neighbourhood of $\Pi_\tau$
{\small
\beqn
\supnorm{\widetilde W(\widetilde X)}\leq\abs{g^2-\gamma^2}\,\supnorm{\sin\Phi^0+\cos(\Phi^0)\widetilde\Phi}+Cg^2\supnorm{\widetilde\Phi}^2+\abs{\lambda}\,\supnorm{\widetilde\Omega(X^0+\widetilde X)}.
\eeqn
}%
The factor $g^2-\gamma^2$ is the reason we chose to subtract $\gamma^2\cos(\Phi^0)\widetilde\Phi$ from both sides in equation~\eqref{til}. Namely, $\abs{g^2-\gamma^2}=\abs{2g+\gamma-g}\abs{g-\gamma}\leq Cg^2\abs{\epsilon}$. Terms proportional to $\widetilde\Phi$ are dominated by \eqref{tilX}. Thus, for $\epsilon$ and $R$ small (independently of $g$ and each other),
\beqn
\supnorm{W(Z)}\leq Cg^2(R^2+\abs{\epsilon}).
\eeqn

In order to obtain the Lipschitz continuity of $W|_{\A_1}$, it suffices to show that $Z\overset{\eqref{Zdef}}{\mapsto}\widetilde X\mapsto\widetilde W(\widetilde X)$ is Lipschitz, as neither $(\widetilde W(\widetilde X))_{\leq 1}$ nor $\widetilde X_{\leq 1}$ depend on $Z\defasr\delta_2 \widetilde X$. To that end, we use the Mean Value Theorem, see \cite{Chae}, and conclude that for some $Z\defasr\delta_2\widetilde X$ on the line segment between two points $Z'\defasr\delta_2\widetilde X'$ and $Z''\defasr\delta_2\widetilde X''$ 
\beqn
\supnorm{\widetilde W(\widetilde X')-\widetilde W(\widetilde X'')}\leq\norm{D\widetilde W(\widetilde X)}\,\supnorm{Z'-Z''}.
\eeqn
The derivative is bounded by $Cg^2(R+\abs{\epsilon})$ given \eqref{tilX}, in particular when $\supnorm{Z}\leq R$.

From its explicit expression, one immediately recognizes that $W$ preserves the class of functions whose restriction to $\bigunit\times\torus$ has the \emph{real} range $\R\times\R^d$, if $\epsilon$ is real.
\end{proof}

\begin{proof}[Proof of Lemma~\ref{lem:inverse}]
$\el$ maps $\A_1$ into itself, and $\K$ in \eqref{K} inherits this feature.

Let us start with the ``pendulum part'' of $\K$, and solve
\beqn
Lf=g
\eeqn
resorting to the method of characteristics; we write $(z,\theta)=(\zeta e^{\gamma t},\vartheta+\omega t)$ in order to obtain an ordinary differential equation (ODE). Recalling the identity \eqref{Lid}, we see that
\beq\label{Lf=g}
\left(\de_t^2-\gamma^2\cos\Phi^0(\zeta e^{\gamma t})\right)f(\zeta e^{\gamma t},\vartheta+\omega t)=g(\zeta e^{\gamma t},\vartheta+\omega t),
\eeq
and our task reduces to studying $L_t\defas\de_t^2-\gamma^2\cos\Phi^0(\zeta e^{\gamma t})$. Since a translation in $t$ and $\vartheta$ eliminates $\zeta$, we can just as well set $\zeta=1$.

We proceed in the Fourier language. The function $f$ solves equation~\eqref{Lf=g} if and only if for all $q\in\Z^d$ the functions $u(t)\defas e^{iq\cdot\omega t}\hat f(e^{\gamma t},q)$ and $v(t)\defas e^{iq\cdot\omega t}\hat g(e^{\gamma t},q)$ satisfy
\beqn
L_t u=v.
\eeqn

Noticing that $\cos\Phi^0(e^{\gamma t})=2\tanh^2\gamma t-1$, we see that $L_t$ has got the zero mode
\beqn
u_1(t)\defas (\cosh\gamma t)^{-1},
\eeqn
\ie, $L_t u_1=0$. Since $L_t u=0$ is a linear second order ODE, there exists precisely one other zero mode $u_2$ of $L_t$ that is linearly independent of $u_1$. Because $u_1(t)\neq 0$ for any $t\in \R$, $u_2$ may be found by a standard procedure:
\beqn
u_2(t)\defas u_1(t)\int\frac{dt}{u_1^2(t)}=\frac{t}{2\cosh\gamma t}+\frac{\sinh\gamma t}{2\gamma},
\eeqn
omitting any additive constant emerging from the integral. Let us express the linear homogeneous equation $L_t u=0$ as the first order system $\dot U=AU$ with $U\defas (u,\dot u)^T$ and $A(t)\defas \bigl(\begin{smallmatrix} 0 & 1 \\ \gamma^2\cos\Phi^0(e^{\gamma t}) & 0 \end{smallmatrix}\bigr)$. Then 
$
w\defas 
\bigl(
\begin{smallmatrix}
u_1 & u_2 \\
\dot u_1 & \dot u_2
\end{smallmatrix}
\bigr)
$
is a fundamental matrix solution of the system (\ie, $\dot w=Aw$) with $\det w=1$ and thus
{\small
\beqn
w^{-1}=
\begin{pmatrix}
 \dot u_2 & -u_2 \\
-\dot u_1 & u_1
\end{pmatrix}
\mathand w(t)w^{-1}(s)=
\begin{pmatrix}
\,* & u_2(t)u_1(s)-u_1(t)u_2(s) \\
\,* & *
\end{pmatrix}.
\eeqn
}%

In terms of a first order system, the complete equation $L_t u=v$ reads $\dot U=AU+V$, $V\defas (0,v)^T$. Varying constants,
\beqn
U(t)=w(t)\left(w^{-1}(t_0)U(t_0)+\int_{t_0}^tw^{-1}(s)V(s)\,ds\right).
\eeqn

Next, we take $t_0\to-\infty$. In that limit $u(t_0)=\mathcal{O}(e^{2\gamma t_0})$, such that
\begin{align*}
&u(t)=\int_{-\infty}^t\left[u_2(t)u_1(s)-u_1(t)u_2(s)\right]v(s)\,ds.
\end{align*}
Equivalently,
\beq\label{eq:f-transformed}
\hat f(e^{\gamma t},q)=\int_{-\infty}^0 \widetilde K_\Phi(s;e^{\gamma t})\,\hat g(e^{\gamma t}e^{\gamma s},q)\,e^{iq\cdot\omega s}\,ds
\eeq
in terms of the kernel
\beqn
\widetilde K_\Phi(s;z)\defas \W_{\Phi 2}(z)\W_{\Phi 1}(z e^{\gamma s})-\W_{\Phi 1}(z)\W_{\Phi 2}(z e^{\gamma s}),
\eeqn
defined (by analytic continuation) on $\{(s,z)\in\R\times\C\,|\,z\notin\{\pm i,\pm i e^{-\gamma s}\}\}$, where
\beqn
\W_{\Phi 1}\defas 2 P\mathand \W_{\Phi 2}\defas \gamma\inv P\,\ln+\frac{1}{4}\gamma\inv Q,
\eeqn
and
\beq\label{eq:P&Q}
P(z)\defas (z^2+1)\inv z\mathand Q(z)\defas z\inv(z^2-1).
\eeq
This is so, because $\W_{\Phi j}(e^{\gamma t})\equiv u_j(t)$.

In a complex strip $\abs{\impart z}\leq \tau<1$, the inequality $\abs{z^2+1}\geq 1-\tau^2$ yields
\beq\label{Kbound}
\abs{\widetilde K_\Phi(s;z)}\leq C(1-\tau^2)^{-2} \gamma\inv e^{\gamma \abs{s}},\quad s\leq 0.
\eeq

Since $\hat f(0,q)=\hat g(0,q)=0$, we find that \eqref{eq:f-transformed} remains true if 0 replaces $e^{\gamma t}$. Inserting all this into the Fourier series of $f(z,\theta)$ leads to
\beq\label{f}
f(z,\theta)=\int_{-\infty}^0 \widetilde K_\Phi(s;z)\, g(ze^{\gamma s},\theta+\omega s)\,ds,\quad (z,\theta)\in\bigunit\times\torus.
\eeq
Here Fubini's Theorem was used, taking advantage of the bound \eqref{Kbound}. Indeed, we may express $g(z,\theta)=z^2 h(z,\theta)$, where $h$ is analytic in the same region as $g$. Since $\abs{\hat h(z,q)}\leq \sup_{\Pi_{\tau'}}\abs{h(z,\theta)}e^{-\tau'\abs{q}}\leq Ce^{-\tau'\abs{q}}$ for some $\tau'>\tau$, we have on $\Pi_\tau$ that
{\small
\begin{align*}
&\sum_{q\in\Z^d}\int_{-\infty}^0\left|{\widetilde K_\Phi(s;z)\,\hat g(ze^{\gamma s},q)\,e^{iq\cdot(\theta+\omega s)}}\right|ds\leq C(1-\tau^2)^{-2}\gamma^{-2}\abs{z}^2\sum_{q\in\Z^d} e^{-(\tau'-\tau)\abs{q}} <\infty.
\end{align*} 
}

Following the line of reasoning above, solving the ``rotator part''
\beqn
\el^2f=g
\eeqn
amounts to integrating $\ddot u=v$ and results in an expression like \eqref{f} with the kernel 
\beqn
\widetilde K_\Psi(s;z)\defas\W_{\Psi 2}(z)\W_{\Psi 1}(z e^{\gamma s})-\W_{\Psi 1}(z)\W_{\Psi 2}(z e^{\gamma s})\equiv -s,
\eeqn
introducing
\beqn
\W_{\Psi 1}\defas 1 \mathand \W_{\Psi 2}\defas \gamma\inv\ln.
\eeqn

For each index $n\in\N\cup\{\infty\}$ define now
\beqn
I_n(z,\theta)\defas\int_{-n}^0
\widetilde K(s;z)\,Z(ze^{\gamma s},\theta+\omega s)\,ds\mathwith
\widetilde K\defas
\begin{pmatrix}
\widetilde K_\Phi & 0 \\
0 & \widetilde K_\Psi
\end{pmatrix},
\eeqn
where $(z,\theta)\in\Pi_\tau$ and $Z\in\A_1$ are arbitrary. Also denote
\beqn
\widetilde \K Z\defas I_\infty.
\eeqn
Since the integrand here is an analytic function of $(z,\theta)$ on the compact region $\Pi_\tau$ and continuous in $s\in[-n,0]$, it follows from an exercise in function theory that $I_n$ \emph{with $n<\infty$} is analytic on $\Pi_\tau$; see p.~123 of \cite{Ahlfors}. As an element of $\A_1$, $Z(z,\theta)$ has the representation $z^2\widetilde Z(z,\theta)$, where $\widetilde Z$ is analytic on $\Pi_\tau$. Accordingly, \eqref{Kbound} implies
\begin{align}\label{IntoI}
\left|{\widetilde\K Z(z,\theta)-I_n(z,\theta)}\right| \leq C\gamma\inv\int_{-\infty}^{-n}e^{-\gamma s}\,\abs{Z(ze^{\gamma s},\theta+\omega s)}\,ds
\leq C\abs{z}^2\frac{\supnorm{\widetilde Z}}{\gamma^2\, e^{\gamma n}},
\end{align}
showing that $I_n\to \widetilde\K Z$ uniformly on $\Pi_\tau$ as $n\to\infty$. Hence, also $\widetilde\K Z$ is analytic on the latter region. Moreover, $I_n(z,\theta)=\order{z^2}$ as $z\to 0$, which by virtue of \eqref{IntoI} yields $\widetilde \K:\A_1\to\A_1$.

We showed above that if $Z\in\A_1$ and $\K Z=Z'$ (thus $Z'\in\A_1$), then $Z=\widetilde\K Z'$ holds on $\bigunit\times\torus\subset\Pi_\tau$. But each side of the latter equation is analytic on $\Pi_\tau$ and hence agree there, meaning that $\widetilde\K$ is the left inverse of $\K$: $\widetilde\K\K=\one_{\A_1}$. A direct computation shows that it is also the right inverse. In other words,
\beqn
\text{$\widetilde \K=\K\inv$ on $\A_1$.}
\eeqn

$K(s;z)\in\R$, provided $z\in\R$. Thus, should the restriction $Z|_{\bigunit\times\torus}$ be real-valued, so is $(\K\inv Z)|_{\bigunit\times\torus}$.

The integrals $I_n$ also depend analytically on $\gamma$. Thus, according to Theorem~\ref{thm:linearization}, they are analytic functions on the domain $\abs{\epsilon}<\epsilon_0$. Since $\abs{\gamma-g}<Cg\abs{\epsilon}$, the trivia $\gamma>\half g>0$ and \eqref{IntoI} guarantee that the convergence $I_n\to\K\inv Z$ takes place uniformly on compact subsets of $D$ defined in \eqref{eq:D} ($g$ bounded away from zero).

It remains to be checked that $\K^{-1}$ is bounded. For $Z\in\A_1$,
$
Z(z,\theta)=\sum_{k=2}^\infty \frac{1}{k!}\,Z_k(\theta)\,z^k
$
converges in the disk $\mathbb{\bar D}(0,\tau)\defas\left\{z\in\C\;\big|\;\abs{z}\leq \tau\right\}$. Using the Cauchy inequalities
$
\left|{Z_k(\theta)}\right|\leq k!\,\tau^{-k}\,\supnorm{Z}
$
we deduce the bound
\beqn
\abs{Z(z,\theta)}\leq 2(\abs{z}/\tau)^2\supnorm{Z} \quad\text{if}\quad z\in\mathbb{\bar D}(0,\tau/2)
\eeqn
In $\Pi_\tau$, $\abs{z}\leq R$ for a certain $R=1+\order{\tau}$, such that $ze^{\gamma s}\in\mathbb{\bar D}(0,\tau/2)$ whenever $s\leq S\defas-\gamma\inv\ln(2R/\tau))$. The bound \eqref{Kbound} for $\widetilde K_\Phi$ applies to $\widetilde K_\Psi$ as well. Summarizing,
\begin{align*}
\supnorm{\K^{-1}Z} \leq \frac{C\supnorm{Z}}{\gamma(1-\tau^2)^2}\left(\int_S^0 e^{-\gamma s}\,ds+\int_{-\infty}^S\frac{e^{\gamma s}R^2}{\tau^2(1+\tau)^2}\,ds\right)
\leq \frac{C\supnorm{Z}}{\gamma^2\tau(1-\tau^2)^2},
\end{align*}
which finishes the proof.
\end{proof}


\psset{treemode=R,treefit=tight,radius=2pt,tnpos=a,npos=.5,nrot=:U,labelsep=3pt
,treenodesize=-1pt,levelsep=1.2cm}
\newcommand{\rad}{7pt}


\section{Analytic Continuation of the Solution}\label{sec:continuation}
Here we present the proof of Theorem~\ref{thm:extension}. In the notation of Section~\ref{sec:Z}, the \emph{existing} map $Z=\delta_2\widetilde X$ solves \eqref{Zeq} and, by virtue of $\widetilde W$'s analyticity, admits the representation
\beq\label{eq:Zrec}
\begin{split}
\delta_2\widetilde X=\K\inv \delta_2 \biggl[
\begin{pmatrix}
\gamma^2 \cos\Phi^0 & 0 \\
0 & 0
\end{pmatrix}
\widetilde X_{\leq 1}+\sum_{k=0}^\infty w^{(k)}\bigl(\widetilde X_{\leq 1}\bigr)^{\otimes k}
\biggr]+ \\
+\,\K\inv\sum_{k=1}^\infty \Bigl[w^{(k)}\bigl(\widetilde X_{\leq 1}+\delta_2\widetilde X\bigr)^{\otimes k}-w^{(k)}\bigl(\widetilde X_{\leq 1}\bigr)^{\otimes k}\Bigr]
\end{split}
\eeq
on the set $\Pi_\tau$, taking $\epsilon$ small enough, and denoting 
\beq\label{eq:wdefinition}
w^{(k)}\defas \frac{1}{k!}D^k\widetilde W(0)
\eeq
and a repeated argument of such a symmetric $k$-linear operator by
\beqn
(x)^{\otimes k}\defas  (\underset{\text{$k$ times}}{x,\dots,x}),
\eeqn
for the sake of brevity. Observe that we have omitted a $\delta_2$ in front of the square brackets on the second line of \eqref{eq:Zrec} as redundant. 

Equation~\eqref{eq:Zrec} may be viewed as a recursion relation for $\delta_2\widetilde X$. It is crucial that 
\beq\label{eq:nodeorder}
w^{(0)},\,w^{(1)}=\order{\epsilon g^2},
\eeq
when $(\epsilon,g)\in D$; see \eqref{eq:D}. Namely, any given order $\delta_2\widetilde X^\ell$ in the convergent expansion
\beqn
\delta_2\widetilde X=\sum_{\ell=1}^\infty \epsilon^\ell \,\delta_2\widetilde X^\ell
\eeqn
is then completely determined by $\widetilde X_{\leq 1}$ and the \emph{lower orders} $\delta_2\widetilde X^l$ ($1\leq l\leq\ell-1$) through the right-hand side of \eqref{eq:Zrec}. Moreover, since $\widetilde X_{\leq 1}=\order{\epsilon}$, only \emph{finitely many} terms in the sum over the index $k$ are involved. Together these facts imply that only \emph{finitely many} recursive steps using \eqref{eq:Zrec} are needed to completely describe any given order $\delta_2\widetilde X^\ell$ in terms of $\widetilde X_{\leq 1}$ alone and that, at each such step, only \emph{finitely many} terms from the $k$-sum contribute. 

It is important to understand that $\widetilde X_{\leq 1}$ is a predetermined function. As we shall see, the recursion procedure will then provide the analytic continuation of each $X^{u,\ell}=\widetilde X^\ell_{\leq 1}+\delta_2\widetilde X^\ell$ ($\ell\geq 1$) to the large region $\mathbb{U}_{\tau,\vartheta}\times\{\abs{\impart{\theta}}\leq\sigma\}$ of Theorem~\ref{thm:extension}. 

\subsection{Tree expansion}
We next give a pictorial representation of the above recursion. It involves tree diagrams similar to those of Gallavotti, \textit{et al.} (see, \eg, \cite{GallavottiTwistless,ChierchiaGallavotti}), with one difference: there will be no resummations nor cancellations, as the expansion in \eqref{eq:Zrec} contains no resonances and is instead well converging. This so-called tree expansion is needed for bookkeeping and pedagogical purposes; we simply choose to draw a tree instead of spelling out a formula. 

Let us first define the auxiliary functions
\beqn
h^{(k)}\defas 
\begin{cases}
 w^{(0)}+\bigl[\bigl(\begin{smallmatrix} 
\gamma^2 \cos\Phi^0 & 0 \\
0 & 0
\end{smallmatrix}\bigr)
+ w^{(1)}
\bigr]\widetilde X_{\leq 1} & \text{if $k=1$,} 

\vspace{1mm}\\	

w^{(k)}\bigl(\widetilde X_{\leq 1}\bigr)^{\otimes k} & \text{if $k=2,3,\dots$},
\end{cases}
\eeqn
and make the identifications
\beq\label{eq:endnode}
\raisebox{3pt}{
\pstree{\Tp}{\Tcircle{\tiny $k$}}
}
\defas\K\inv\delta_2 h^{(k)} 
\mathand 
\raisebox{3pt}{
\pstree{\Tp}{\Tc[fillstyle=hlines,fillcolor=black,hatchsep=2.5pt]{\rad}}
}
\defas\K\inv\delta_2 \,\sum_{k=0}^\infty h^{(k)}.
\eeq
Furthermore, let
\beqn
\raisebox{3pt}{
\pstree{\Tp}{\Tc{\rad}}
}
\defas\delta_2\widetilde X,\quad 
\raisebox{3pt}{
\pstree{\Tp}{\TC*}
} 
\defas \widetilde X_{\leq 1},
\eeqn
and, for $k\geq 1$,
\beqn
\raisebox{3pt}{
\pstree[treesep=1.2cm]{\Tp}{\pstree{\TC*~[tnpos=r,tnsep=7.5pt]{\raisebox{1pt}{\tiny $k$ lines}}}{\Tp[name=top] \Tp[name=bot]}}
\ncarc[linestyle=dotted,nodesep=5pt]{top}{bot}  
}
\defas \K\inv w^{(k)}.
\eeqn
In the diagram representing the $k$-linear $w^{(k)}$, the $k$ ``free'' lines to the right of the node stand for the arguments. We say that these lines \emph{enter} the \emph{internal} node, whereas the single line to the left of the node \emph{leaves} it. For instance,
\beqn
\raisebox{3pt}{
\pstree[treesep=0.5cm]{\Tp}{\pstree{\TC*}{\TC* \Tc[bbh=12pt]{\rad} \Tcircle[]{\tiny 4}}}
}
=\K\inv w^{(3)}\bigl(\widetilde X_{\leq 1},\,\delta_2\widetilde X,\,\K\inv\delta_2 h^{(4)}\bigr).
\eeqn
Notice that, as $w^{(k)}$ is symmetric, permuting the lines entering a node does not change the resulting function. We emphasize that all of the functions introduced above are analytic on $\Pi_\tau$ and $\abs{\epsilon}<\epsilon_0$. 

In terms of such \emph{tree diagrams}, or simply \emph{trees}, equation~\eqref{eq:Zrec} reads
\beq\label{eq:tree-rec}
\begin{split}
\raisebox{3pt}{
\pstree{\Tp}{\Tc{\rad}}
}
= \, &
\raisebox{3pt}{
\pstree{\Tp}{\Tc[fillstyle=hlines,fillcolor=black,hatchsep=2.5pt]{\rad}}
}
+
\raisebox{3pt}{
\pstree{\Tp}{\pstree{\TC*}{\Tc{\rad}}}
}
+
\raisebox{3pt}{
\pstree{\Tp}{\pstree{\TC*}{\Tc{\rad} \TC*}}
}
+
\\
& +
\raisebox{3pt}{
\pstree{\Tp}{\pstree{\TC*}{\TC* \Tc{\rad}}}
}
+
\raisebox{3pt}{
\pstree{\Tp}{\pstree{\TC*}{\Tc{\rad} \Tc{\rad}}}
}
+\,\cdots,
\end{split}
\eeq
using multilinearity to split the sums $\widetilde X_{\leq 1}+\delta_2\widetilde X$ into pieces. Above, the sum after the first tree consists of \emph{all} trees having one internal node and an arbitrary number of \emph{end nodes, at least one of which, however, is a white circle}. This rule encodes the fact that on the second line of \eqref{eq:Zrec} the summation starts from $k=1$ and that the contributions with only $\widetilde X_{\leq 1}$ in the argument (\ie, trees with only black dots as end nodes) are cancelled.

Using \eqref{eq:Zrec} recursively now amounts to replacing each of the lines with a white-circled end node by the complete expansion of such a tree above. This is to be understood additively, so that replacing one end node, together with the line leaving it, by a sum of two trees results in a sum of two new trees. For example, such a replacement in the third tree on the right-hand side of \eqref{eq:tree-rec} by the first two trees gives the sum
\beqn
\raisebox{3pt}{
\pstree{\Tp}{\pstree{\TC*}{\Tc[fillstyle=hlines,fillcolor=black,hatchsep=2.5pt]{\rad} \TC*}}
}
+
\raisebox{3pt}{
\pstree{\Tp}{\pstree{\TC*}{\pstree{\TC*}{\Tc{\rad}}  
\TC*}}
}.
\eeqn

Before proceeding, we introduce a little bit of terminology. The leftmost line in a tree is called the \emph{root line}, whereas the node it leaves (\ie, the uniquely defined leftmost node) is called the \emph{root}. A line leaving a node $v$ and entering a node $v'$ can always be interpreted as the root line of a \emph{subtree}, the maximal tree consisting of lines and nodes in the original tree with $v$ as its root. We call $v$ a (not necessarily unique) \emph{successor} of $v'$, whereas $v'$ is the unique \emph{predecessor} of $v$.

The recursion \eqref{eq:tree-rec} can be repeated on a given tree if it has at least one white circle left. Otherwise, the tree in question must satisfy 
\begin{itemize}
\item[{\bf (R$\mathbf 1'$)}] The tree has only filled circles (\raisebox{3pt}{\Tc[fillstyle=hlines,fillcolor=black,hatchsep=2.5pt]{\rad}}) and black dots (\raisebox{3pt}{\TC*}) as its end nodes,
\end{itemize}
together with
\begin{itemize}
\item[{\bf (R$\mathbf 2'$)}] Any internal node has an entering (line that is the root line of a) subtree containing at least one filled circle as an end node.
\end{itemize}
After all, the recursion can only stop by replacing an existing white circle with a filled one. Continuing \textit{ad infinitum} yields the expansion
\beq\label{eq:tree-exp1}
\raisebox{3pt}{
\pstree{\Tp}{\Tc{\rad}}
}
=\,\sum{\bigl(\text{Trees satisfying (R$1'$) and (R$2'$)}\bigr)}=\, \sideset{}{'}\sum_{\text{trees $T$}}T,
\eeq
where the prime restricts the summation to trees $T$ satisfying (R$1'$) and (R$2'$). We point out that each admissible tree appears precisely once in this sum, considering different two trees that can be superposed by a (nontrivial) permutation of subtrees that enter the same node.

The earlier discussion concerning the description of $\delta_2\widetilde X^\ell$ in terms of a finite sum involving only $\widetilde X_{\leq 1}$ translates to the language of trees in a straightforward fashion. First, the second part of \eqref{eq:nodeorder} and $\widetilde X_{\leq 1}=\order{\epsilon}$ amount pictorially to
\beqn
\raisebox{3pt}{
\pstree{\Tp}{\pstree{\TC*}{\Tp}}
}
=\order{\epsilon}
\mathand
\raisebox{3pt}{
\pstree{\Tp}{\TC*}
} 
=\order{\epsilon}.
\eeqn
Second, $w^{(k)}=\order{g^2}$ and the first part of \eqref{eq:nodeorder} yield
\beqn
\raisebox{3pt}{
\pstree{\Tp}{\Tcircle{\tiny $k$}}
} 
=\order{\epsilon^k}\qquad(k=1,2,\dots)
\eeqn
and
\beqn
\raisebox{3pt}{
\pstree[treesep=1.2cm]{\Tp}{\pstree{\TC*~[tnpos=r,tnsep=7.5pt]{\raisebox{1pt}{\tiny $k$ lines}}}{\Tp[name=top] \Tp[name=bot]}}
\ncarc[linestyle=dotted,nodesep=5pt]{top}{bot}  
}
=\order{1}\qquad(k=2,3,\dots).
\vspace{2mm}
\eeqn
Expanding the filled end nodes
\beq\label{eq:node-exp}
\raisebox{3pt}{
\pstree{\Tp}{\Tc[fillstyle=hlines,fillcolor=black,hatchsep=2.5pt]{\rad}}
}
=\,\sum_{k=1}^\infty 
\raisebox{3pt}{
\pstree{\Tp}{\Tcircle{\tiny $k$}}
},
\eeq
according to \eqref{eq:endnode}, on the right-hand side of \eqref{eq:tree-exp1}, we get a new version of the latter by replacing the rules (R$1'$) and (R$2'$), respectively,  with
\begin{itemize}
\item[{\bf (R1)}] The tree has only numbered circles (\raisebox{3pt}{{\tiny \Tcircle{$k$}}} with arbitrary values of $k$) and black dots (\raisebox{3pt}{\TC*}) as its end nodes,
\end{itemize}
and
\begin{itemize}
\item[{(\bf R2)}] Any internal node has an entering (line that is the root line of a) subtree containing at least one numbered circle as an end node.
\end{itemize}
Let us define the \emph{degree} of a tree as the positive integer
\beq\label{eq:deg}
\deg{T}\defas\#(\!
\raisebox{3pt}{
\pstree{\Tp}{\pstree{\TC*}{\Tp}}
} 
\!)\,+\,\#(\!
\raisebox{3pt}{
\pstree{\Tp}{\TC*}
}
\!)\,+\,\sum_{k=1}^\infty \,k\,\#(\!
\raisebox{3pt}{
\pstree{\Tp}{\Tcircle{\tiny $k$}}
}
\!)
\eeq
for any tree $T$ satisfying (R1) and (R2). By $\#(G)$ we mean the number of occurrences of the graph $G$ in the tree $T$. That is, the degree of a tree is the number of its end nodes with suitable weights plus the number of nodes with precisely one entering line. Since a tree has finitely many nodes, its degree is well-defined. Then a rearrangement of the sum arising from \eqref{eq:node-exp} being inserted into \eqref{eq:tree-exp1} yields formally
\beq\label{eq:tree-exp2}
\raisebox{3pt}{
\pstree{\Tp}{\Tc{\rad}}
}
=\,\sum_{l=1}^\infty\, \sideset{}{^*}\sum_{\substack{\text{trees $T$} \vspace{1pt} \\ \deg{T}=l}}T,
\eeq
where the asterisk reminds us that the rules (R1) and (R2) are being respected.

According to the analysis above, the particular graphs appearing in the definition of $\deg{T}$ are the only possible single-node subgraphs of $T$ proportional to a positive power of $\epsilon$. Since each tree is an analytic function of $\epsilon$, writing again $(\piste)^k$ for the $k$th coefficient of the power series in $\epsilon$, we have
\beqn
T=\sum_{k=\deg T}^\infty \epsilon^k \,T^k=\epsilon^{\deg T}\,\sum_{k=0}^\infty \epsilon^k \,T^{k+\deg T}.
\eeqn
Hence, only trees with degree \emph{at most} equal to $\ell$ can contribute to $\delta_2\widetilde X^\ell$: 
\beq\label{eq:tree-exp-order}
\delta_2\widetilde X^\ell=\,\sum_{l=1}^\ell\, \sideset{}{^*}\sum_{\deg{T}=l}T^\ell=\, {\Biggl(\;
\sideset{}{^*}\sum_{\deg{T}\leq \ell}T
\Biggr)}^{\!\!\ell}
\eeq
 or, alternatively,
\beq\label{eq:tree-exp3}
\delta_2\widetilde X=\, \sideset{}{^*}\sum_{\deg{T}\leq \ell}T+\order{\epsilon^{\ell+1}}\qquad(\epsilon\to 0)
\eeq
for \emph{each and every} $\ell=1,2,\dots$. The expansion in \eqref{eq:tree-exp2} is in fact just a compact way of writing \eqref{eq:tree-exp3}. We emphasize that the latter can be derived completely rigorously, for each value of $\ell$ separately, but resorting to the use of formal series allowed us to treat all orders of $\delta_2\widetilde X$ at once. We call the series \eqref{eq:tree-exp2} an \emph{asymptotic expansion} of $\delta_2\widetilde X$; the partial sums ${\sum_{\deg{T}\leq \ell}^*}\,T$ need not converge to $\delta_2\widetilde X$ for any fixed $\epsilon$ as $\ell\to\infty$, but for a fixed $\ell$ the error is bounded by an $\ell$-dependent constant times $\abs{\epsilon}^{\ell+1}$ on the mutual domain of analyticity, $\abs{\epsilon}<\epsilon_0$.
\begin{example}
The beginning of the asymptotic expansion \eqref{eq:tree-exp3} reads
\vspace{1mm}
\beqn
\psset{levelsep=1cm}
\begin{split}
\delta_2\widetilde X
\quad = \quad \, & \raisebox{3pt}{
\pstree{\Tp}{\Tcircle{\tiny 1}}
}
+\order{\epsilon^2} \quad = \quad
\raisebox{3pt}{
\pstree{\Tp}{\Tcircle{\tiny 1}}
}
+
\raisebox{3pt}{
\pstree{\Tp}{\Tcircle{\tiny 2}}
}
+
\raisebox{3pt}{
\pstree{\Tp}{\pstree{\TC*}{\Tcircle{\tiny 1}}}
}
+  \\[2mm] 
& \, +
\raisebox{3pt}{
\pstree{\Tp}{\pstree{\TC*}{\Tcircle{\tiny 1} \TC*}}
}
+
\raisebox{3pt}{
\pstree{\Tp}{\pstree{\TC*}{\TC* \Tcircle{\tiny 1}}}
}
+
\raisebox{3pt}{
\pstree{\Tp}{\pstree{\TC*}{\Tcircle{\tiny 1} \Tcircle{\tiny 1}}}
}
+\order{\epsilon^3}.
\end{split}
\eeqn
\end{example}

\subsection{Analyticity domain of trees}
As already pointed out, all trees $T$ above are analytic functions of $(z,\theta,\epsilon)$ on $\Pi_\tau\times\{\abs{\epsilon}<\epsilon_0\}$. Due to the projections $\delta_2$ appearing in \eqref{eq:endnode}, they also satisfy $T|_{z=0}=\de_zT|_{z=0}=0$, \ie, are elements of the space $\A_1$ defined in \eqref{eq:A1}. On this space, the inverse of $\K=\bigl(\begin{smallmatrix} L & 0 \\ 0 & \el^2 \end{smallmatrix}\bigr)$ (see \eqref{K}) constructed in the proof of Lemma~\ref{lem:inverse} satisfies
\beq\label{eq:inverse}
\K\inv h(z,\theta)=\int_{-\infty}^0 \widetilde K(s;z)\,h(z e^{\gamma s},\theta+\omega s)\,ds.
\eeq
Consequently, we will now show that the analyticity domain of a tree in the $z$-variable is in fact much larger than the neighbourhood of $[-1,1]$ that is included in $\Pi_\tau$; namely it includes the wedgelike region
\beqn
\mathbb{U}_{\tau,\vartheta}\defas \bigl\{\abs{z}\leq \tau\bigr\}\,\bigcup\,\bigl\{\arg{z}\in [-\vartheta,\vartheta]\cup [\pi-\vartheta,\pi+\vartheta]\bigr\}\subset\C
\eeqn
(with a new $\tau$ and ``small'' $\vartheta$).
\begin{lemma}[Analytic continuation of trees]\label{lem:an-cont}
Without affecting the analyticity domain with respect to $\epsilon$, there exist numbers $0<\tau<1$, $0<\vartheta<\pi/2$, and $0<\sigma<\eta$ such that each tree in the sums \eqref{eq:tree-exp2} and \eqref{eq:tree-exp3} extends to an analytic function of $(z,\theta)$ on $\mathbb{U}_{\tau,\vartheta}\times\{\abs{\impart{\theta}}\leq\sigma\}$.
\end{lemma}
\begin{proof}
Observe that, as a polynomial, $\widetilde X_{\leq 1}$ is an entire function of $z$. On the other hand, $\Phi^0(z)=4\arctan z=2i\bigl(\log(1-iz)-\log(1+iz)\bigr)$, implying that $\abs{\impart\Phi^0(z)}\leq\eta$ in $\mathbb{U}_{\tau,\vartheta}$ with $\tau$ and $\vartheta$ sufficiently small. By Remark~\ref{rem:manifolds}, $f(\Phi^0(z),\theta)$ is analytic, making the maps $h^{(k)}$ and $\widetilde X_{\leq 1}$ analytic on $\mathbb{U}_{\tau,\vartheta}\times\{\abs{\impart{\theta}}\leq\sigma\}$ for some $0<\sigma<\eta$, where $\eta$ is determined by $f$ and $\sigma$ by $\widetilde X_{\leq 1}$ (ultimately by $f$ and $\omega$).

Suppose $h=\delta_2 h$ is a map analytic on $\mathbb{U}_{\tau,\vartheta}\times\{\abs{\impart{\theta}}\leq\sigma\}$. Then the integrand in \eqref{eq:inverse} is analytic in a neighbourhood of the latter set. By virtue of Fubini's theorem,
\beqn
\oint_{\Gamma}\K\inv h(\zeta,\theta)\,d\zeta=\int_{-\infty}^0 \oint_{\Gamma} \widetilde K(s;\zeta)\,h(\zeta e^{\gamma s},\theta+\omega s)\,d\zeta\,ds=0
\eeqn
for any closed contour $\Gamma$ inside a sufficiently small neighbourhood of $\mathbb{U}_{\tau,\vartheta}$ and enclosing $z$. Hence, Morera's theorem yields analyticity of $\K\inv h$ with respect to $z$. As always, analyticity with respect to $\theta$ follows from an exponentially decaying bound on the Fourier coefficients. Applying this argument at each node of a tree proves the claim.  
\end{proof}

\begin{proof}[Proof of Theorem~\ref{thm:extension}]
Since the number of terms in the sum in \eqref{eq:tree-exp-order} is finite and the functions $\widetilde X_{\leq 1}$ and $X^0$ in $X=X^0+\widetilde X_{\leq 1}+\delta_2\widetilde X$ are analytic on $\mathbb{U}_{\tau,\vartheta}\times\{\abs{\impart{\theta}}\leq\sigma\}$, the analyticity of $X^\ell$ follows from Lemma~\ref{lem:an-cont}.

From the equations of motion, \eqref{xeq}, a Taylor expansion yields
\beqn
\el^2\widetilde X=-\el^2 X^0+\Omega(X^0)+\sum_{m=1}^\infty \frac{1}{m!}D^m\Omega(X^0)(\widetilde X)^{\otimes m},
\eeqn
where the trigonometric degree of $D^m\Omega(X^0)$ is $N$ for $\epsilon\neq 0$ but vanishes at $\epsilon=0$ because $X^0$ does not depend on $\theta$. For each $k\geq 1$, let $n_k$ stand for the trigonometric degree of $\widetilde X^k$.  Equating like powers of $\epsilon$ in the expansion above, we infer two things. First, $n_1=N$. Second, we must have, for each $\ell\geq 2$,
\beqn
n_\ell\leq
\begin{cases}
n_{k_1}+\dots+ n_{k_m} & \text{where $k_1+\dots+ k_m=\ell$,} \\
N+n_{k_1}+\dots+ n_{k_m} & \text{where $k_1+\dots+ k_m=\ell-1$},
\end{cases}
\eeqn
because the trigonometric degree of a product is at most the sum of the trigonometric degrees of the factors; $e^{iq\cdot\theta}e^{iq\cdot\theta}=e^{i2q\cdot\theta}$ and $e^{iq\cdot\theta} e^{-iq\cdot\theta}=1$.

Next, assume that $n_k\leq kN$ holds for each $1\leq k\leq \ell-1$, recalling that this is the case if $k=1$. Subsequently, the estimate for $n_\ell$ above becomes $n_\ell\leq \ell N$.
\end{proof}


\bibliographystyle{amsalpha}
\bibliography{References}

\end{document}